\documentclass[12pt]{article}
\usepackage{float}
\usepackage{authblk}
\usepackage{amssymb,amsmath}
\usepackage{graphicx}
\usepackage{caption} 
\usepackage{subcaption}
\usepackage{cite}
\usepackage{xcolor}
\usepackage{setspace}
\usepackage{appendix}
\usepackage{makecell}
\usepackage{array}
\usepackage{parskip}
\usepackage{tabularx}
\newcolumntype{M}[1]{>{\centering\arraybackslash}m{#1}}
\usepackage[colorlinks=true,urlcolor=blue,citecolor=blue,linkcolor=red,linktocpage=true,pdfproducer=medialab]{hyperref}
\usepackage[a4paper,width=17cm,height=25cm]{geometry}
\usepackage{morefloats}
\usepackage{siunitx}
\makeatletter \renewcommand{\@dotsep}{10000} \makeatother
\title{ Effects of Family Non-universal $Z^{\prime}$ Model in the angular observables of  $B\to(\rho,a_{1})\mu^{+}\mu^{-}$  decays}
\author{Nimra Farooq%
	\thanks{\texttt nimrafarooq\_1995@hotmail.com }} 
\author{Marwah Zaki%
	\thanks{\texttt marwahiiui@gmail.com}} 
\author {M. Ali Paracha%
	\thanks{\texttt aliparacha@sns.nust.edu.pk}}
\author{Faisal Munir Bhutta%
	\thanks{\texttt faisal.munir@sns.nust.edu.pk}}
\affil{Department of Physics, School of Natural Sciences, National University of Sciences and Technology, H-12, Islamabad, 44000, Pakistan}
\date{\today}
\begin{document}
\maketitle
 \begin{abstract}
We present the angular distribution of the four-fold $B\to\rho (\to\pi\pi)\mu^{+}\mu^{-}$ and $B\to a_{1}(\to\rho_{\parallel, \perp}\pi)\mu^{+}\mu^{-}$ decays both in the Standard Model and the family non-universal $Z^{\prime}$ model. At the quark level, these decays are governed by the $b\to d\mu^{+}\mu^{-}$ transition. Along with different angular observables, we also give predictions of differential branching ratios, forward-backward asymmetry, longitudinal polarization fraction of $\rho$, and $a_{1}$ mesons. Our analysis shows that the signatures of family non-universal $Z^{\prime}$ model are more distinct in the observables associated with the $B\to\rho(\to\pi\pi)\mu^{+}\mu^{-}$ decay, compared to that of the $B\to a_{1}(\to\rho_{\parallel, \perp}\pi)\mu^{+}\mu^{-}$ decay. Future measurements of the predicted angular observables, both at current and future high energy colliders, will add to the useful complementary data required to clarify the structure of the family non-universal $Z^{\prime}$ model in $|\Delta b|$=$|\Delta d|=1$ processes.
\end{abstract}
\section{Introduction}
Flavor physics plays a pivotal role not only in testing the Standard Model (SM) parameters but also in tracing out the signatures of New Physics (NP). Much of the efforts in this area focus on the detailed study of B meson decays as they are rich in phenomenology. Furthermore, from the NP point of view the rare decays of $B$ meson, in particular the decays induced by flavor changing neutral current (FCNC) transitions $b\to q$ with $q=d,s$ are of great interest, as these decays are loop suppressed and only allowed via GIM mechanism in the SM \cite{Glashow:1970gm}. Specifically, the observables involved in the rare radiative $b\to q\gamma$ and the rare semileptonic $b\to q\ell^{+}\ell^{-}$ decays allow to explore the structure of NP. Semileptonic decays which involve $b\to s$ current have been analyzed rigorously over the past many years and they have shown discrepancies from the SM predictions both in the lepton flavor-dependent (LFD) observables and the lepton flavor-independent observables defined as lepton flavor universality (LFU) ratios.

Among LFD ($b\to s\mu\mu$) observables, deviations are observed from the SM predictions, in the branching fractions of $B\to K\mu^{+}\mu^{-}$ \cite{LHCb:2014cxe}, $B\to K^{\ast}\mu^{+}\mu^{-}$ \cite{LHCb:2014cxe,LHCb:2013zuf,LHCb:2016ykl}, and $B_{s}\to\phi\mu^{+}\mu^{-}$ \cite{LHCb:2013tgx,LHCb:2015wdu} decays. The observed branching fractions suggest lower values as compared to their SM predictions. Also, in the $B^{0}\to K^{\ast 0}\mu^{+}\mu^{-}$ decay, the angular observable, $P_{5}^{\prime}$, has shown mismatch from the SM values \cite{Descotes-Genon:2012isb,Descotes-Genon:2013vna}. Interestingly, global fits predict the NP effects being present in the LFD observables, involving $b\to s\mu\mu$ transition. Following it, NP effects in different complementary decay modes, driven by the same quark level transition $b\to s\mu\mu$, such as $B\to K_{1}\mu^{+}\mu^{-}$ \cite{Ishaq:2013toa,Munir:2015gsp,Huang:2018rys,MunirBhutta:2020ber}, $B\to K_{2}^{\ast}\mu^{+}\mu^{-}$ \cite{Das:2018orb,Mohapatra:2021izl}, $B_s\to f_{2}^{\prime}\mu^{+}\mu^{-}$ \cite{Rajeev:2020aut,Mohapatra:2021izl}, and $B_{c}\to D_{s}^{(\ast)}\mu^{+}\mu^{-}$ \cite{Dutta:2019wxo,Mohapatra:2021ynn,Zaki:2023mcw,Li:2023mrj} have been investigated both in model-independent approach and the specific NP models.
%\cite{Bifani:2018zmi, Albrecht:2021tul, London:2021lfn, Munir:2015gsp, Ishaq:2013toa, Zaki:2023mcw, MunirBhutta:2020ber}.

LFU ratios in $b\to s$ sector have been measured by LHCb collaboration \cite{LHCb:2017avl,LHCb:2019hip,LHCb:2021trn}, defined as $R_{K^{(\ast)}}=\frac{\mathcal{B}(B\to K^{(\ast)}\mu^{+}\mu^{-})}{\mathcal{B}(B\to K^{(\ast)}e^{+}e^{-})}$, in different $q^{2}$ bins, and their analysis showed a $3\sigma$ deviation from the SM prediction. A similar analysis has been done by BELLE collaboration \cite{BELLE:2019xld,Belle:2019oag}, for the same ratios $R_{K^{(\ast)}}$ in the $q^{2}\in (1-6)$ $\text{GeV}^{2}$ bin and it shows consistency with the SM predictions but with large experimental uncertainties. Furthermore, the recent measurements of the ratios  $R_{K^{(\ast)}}$ in the low and central $q^{2}$ region of the spectrum by LHCb collaboration \cite{LHCb:2022qnv,LHCb:2022vje}, have shown an agreement with the SM predictions.
 
Apart from the LFU violation in $R_{K^{(\ast)}}$, LFU violation has also been studied in flavor changing charged current (FCCC) semileptonic $B\to D^{(\ast)}\ell\nu_{\ell}$ decay mode via the ratios $R_{D^{(\ast)}}=\frac{\mathcal{B}(B\to D^{(\ast)}\tau^{+}\nu_{\tau})}{\mathcal{B}(B\to D^{(\ast)}\ell\nu_{\ell})}$ with $(\ell=e, \mu)$ \cite{LHCb:2015gmp, BaBar:2013mob, Belle:2016dyj, LHCb:2017rln}. However the recent analysis of $R_{D^{\ast}}$ and $R_{D}$ by LHCb\cite{HFLAG:2022} and Belle collaboration\cite{Belle:2019rba}, shows a good agreement with the SM predictions. To draw any conclusion regarding the status of NP, one must exploit the other sectors.

In this work, we consider FCNC processes governed by $b\to d\ell^{+}\ell^{-}$ transition, as these modes are CKM suppressed compared to that of $b\to s\ell^{+}\ell^{-}$ transitions. The typical branching ratios that belong to $b\to d\ell^{+}\ell^{-}$ processes are of order $10^{-8}$, and hence the measurements of these modes are considered to be challenging. Up to now only the branching ratios of rare $b\to d\mu^{+}\mu^{-}$ and $b\to d\gamma$ decays have been measured, and the observed decay modes are (i) $B^{+}\to\pi^{+}\mu^{+}\mu^{-}$\cite{LHCb:2015hsa} (ii) $B^{0}\to\mu^{+}\mu^{-}$\cite{LHCb:2021awg} (iii) $B^{0}_{s}\to K^{\ast 0}\mu^{+}\mu^{-}$\cite{LHCb:2018rym} (iv) $B\to X_{d}\gamma$\cite{Misiak:2015xwa,BaBar:2010vgu}. In literature, researchers use the data set to extract information on NP Wilson coefficients from the global fit analysis, see Refs.\cite{Bause:2022rrs,Du:2015tda,Ali:2013zfa,Rusov:2019ixr}. Furthermore, the experimental data can be used to extract the information on the Wilson coefficients of different NP models such as family non-universal $Z^{\prime}$ Model\cite{Paracha:2020gws,Nayek:2018rcq}, supersymmetric models\cite{Choudhury:2002fk} and Two Higgs doublet Models \cite{Aliev:1998sk}.

The goal of our study is to use the family non-universal $Z^{\prime}$ effective Hamiltonian and perform the four-fold angular analysis of $B\to\rho(\to\pi\pi)\mu\mu$ and $B\to a_{1}(\to\rho_{\parallel,\perp}\pi)\mu\mu$ decays. For $B\to\rho$ decay we use the fit results for simplified series expansion (SSE) coefficients in the fit to Light cone sum rules (LCSR) form factors\cite{Bharucha:2015bzk}, and for the $B\to a_{1}$ decay we use the perturbative QCD (pQCD) form factors\cite{Li:2009tx}. Both the decays are analyzed in the low $q^{2}$ region of the spectrum. For the decay channel $\rho\to\pi\pi$ the probability is $100\%$, while for the decays $a_{1}\to\rho_{\parallel}\pi$, and $a_{1}\to\rho_{\perp}\pi$ the probability is  and $17\%$ and $43\%$, respectively. In our study we choose the values of Wilson coefficients from\cite{Nayek:2018rcq} and give the predictions of different physical observables such as differential branching fractions, forward-backward asymmetry, longitudinal helicity fraction of $\rho$ and $a_{1}$ mesons, and the individual normalized angular observables within the SM and the two scenarios of the family non-universal $Z^{\prime}$ model.

The organization of this paper is as follows: In section \ref{TH}, we present the theoretical framework which includes effective Hamiltonian of the family non-universal $Z^{\prime}$ model for $b\to d\mu^{+}\mu^{-}$ transition, where the basis operators remain the same as that of the SM. We then express the matrix elements for $B\to\rho\mu^{+}\mu^{-}$ and $B\to a_{1}\mu^{+}\mu^{-}$ decays in terms of form factors. Furthermore, using the helicity formalism, we derive the four-fold angular decay distribution of $B\to\rho(\to\pi\pi)\mu\mu$, and $B\to a_{1}(\to\rho_{\parallel,\perp}\pi)\mu\mu$ decays, which contains the angular coefficients given in terms of the helicity amplitudes. These angular coefficients are then used to construct various physical observables along with the normalized angular coefficients. In section \ref{NA}, we present the numerical analysis of the physical observables in the SM and the family non-universal $Z^{\prime}$ model, and finally, section \ref{Conc}, summarizes our work.

\section{Theoretical Framework}\label{TH}
In this section, we provide the effective electroweak (EW) Hamiltonian approach\cite{Buchalla:1995vs,Chetyrkin:1996vx}, where the SM heavy degrees of freedom such as $W^{\pm}, Z^{0}$ gauge bosons and the top quark are integrated out. The effective Hamiltonian is then used to calculate the full angular distribution of $B\to\rho(770)(\to\pi\pi)\mu^{+}\mu^{-}$ and $B\to a_{1}(1260)(\to\rho_{\parallel, \perp}\pi)\mu^{+}\mu^{-}$ decays. Using the form of four-fold angular distribution, we extract the $q^{2}$ dependent angular coefficients, which further will be used to analyze the signatures of the family non-universal $Z^{\prime}$ model.
\subsection{Highlights of the Family Non-universal \texorpdfstring{$Z^{\prime}$ Model}{Z'}}
A family non-universal $Z^{\prime}$ gauge boson can naturally be derived in many extensions of the SM. The simplest way to incorporate the $Z^{\prime}$ gauge boson is by incorporating extra $U^{\prime}(1)$ gauge symmetry. The model was formulated by Langacker and Plümacher\cite{Langacker:2000ju}. One of the features of this model is that FCNC transitions could be induced at tree level, due to the non-diagonal chiral coupling matrix. The signatures of $Z^{\prime}$ gauge boson arise in two different ways.
\begin{enumerate}
    \item[(i)] By introducing new Wilson coefficients only and the basis of operators remains the same as that of SM.
    \item[(ii)] In other approach, new Wilson coefficients and new operators are added to the SM effective Hamiltonian.
\end{enumerate}
In this work, we will analyze the family non-universal $Z^{\prime}$ model using the above mentioned $B$ meson decays, and the NP in this model arises due to the modification of the Wilson coefficients $C^{\text{eff}}_{9}$ and $C_{10}$, the structure of the operators remains the same as that of SM. The Wilson coefficients can get modified due to the off-diagonal couplings of quarks as well as leptons with $Z^{\prime}$ gauge boson. 
The  current due to extra $U^{\prime}(1)$ gauge symmetry in the SM eigenstate basis can be written as\cite{Langacker:2000ju,Barger:2009qs}
\begin{eqnarray}
 J_{\mu}=\sum_{i,j} \bar{\psi}_{i}\gamma_{\mu}[\epsilon_{\psi_{L_{ij}}}P_{L}+\epsilon_{\psi_{R_{ij}}}P_{R}]\psi_{j},
\end{eqnarray}
where the summation runs overall quarks and leptons field $\psi_{ij}$, $P_{R,L}=\frac{1}{2}(1\pm\gamma^{5})$ are the right-handed and left-handed projectors, and $\epsilon_{\psi_{R,L}}$ indicates the chiral couplings of the new gauge boson.

 As already discussed, in the family non-universal $Z^{\prime}$ model, FCNC transition arises at the tree level if the chiral coupling matrices $\epsilon_{\psi R, L}$  are non- diagonal, however if the couplings of $Z^{\prime}$ gauge bosons are diagonal but non-universal, flavor-changing couplings are generated by fermion mixing. The fermion Yukawa matrix $h_{\psi}$ can be diagonalized in the weak eigenstate basis through CKM unitary matrices $V^{\psi}_{R, L}$ and can be expressed as,
 \begin{eqnarray}
h_{\psi,diag}=V^{\psi}_{R}h_{\psi}V^{\dag\psi}_{L}.\label{hpsi}
\end{eqnarray}
Hence, the chiral $Z^{\prime}$ couplings in the fermion mass eigenstates can be expressed as \cite{Barger:2009qs}
\begin{eqnarray}
 B^{\psi_{L}}_{ij}=(V^{\psi}_{L}\epsilon_{\psi_{L}}V^{\dag\psi}_{L})_{ij}, & B^{\psi_{R}}_{ij}=(V^{\psi}_{R}\epsilon_{\psi_{R}}V^{\dag\psi}_{R})_{ij}.\label{NC}
\end{eqnarray}
In Eq. (\ref{NC}), the non-vanishing quark coupling matrices $B^{\psi_{L, R}}_{ij}$ represent the signature of the NP, and also two decades ago it was shown that the flavor non-universal $Z^{\prime}$ model can be used to improve the precision of electroweak data \cite{Erler:1999nx}.

At quark level the decay $B\to M\mu^{+}\mu^{-}$ ($M=\rho(770),a_{1}(1260)$) modes are governed by $b\to d$ transitions, hence the FCNC Lagrangian due to $Z^{\prime}$ model can be written as \cite{Nayek:2018rcq}
\begin{eqnarray}
 \mathcal{L}^{Z^{\prime}}_{\text{FCNC}}=-g^{\prime}(B^{L}_{db}\bar{d}_{L}\gamma_{\mu}b_{L}+B^{R}_{db}\bar{d}_{R}\gamma_{\mu}b_{R})Z^{\prime\mu}+h.c.,\label{Lag}
\end{eqnarray}
where $g^{\prime}$ is the gauge coupling associated with $U^{\prime}(1)$ gauge group.
\subsection{Effective Hamiltonian for \texorpdfstring{$b\to d\mu^{+}\mu^{-}$}{b->du+u-} Transition in the SM and the \texorpdfstring{$Z^{\prime}$ Model}{Z'}}
 The signatures of $Z^{\prime}$ gauge boson can also be analyzed through the decay modes of $B$ mesons within the framework of SM low energy effective field theory. As mentioned earlier that in this framework, the heavy degrees of freedom including the new particles are integrated out (Wilson coefficients), and the effective Hamiltonian appears in terms of four Fermi operators as well as the Wilson coefficients. 

The effective Hamiltonian for modes $B\to M\mu^{+}\mu^{-}$ ($M=\rho(770),a_{1}(1260)$) in the SM can be written as,
  \begin{eqnarray}
  H^{\text{SM}}_{\text{eff}}=-\frac{4G_{F}\alpha}{\sqrt{2}}V_{tb}V^{\ast}_{td}\left[\sum_{i=1}^{10}C_{i}O_{i}-\lambda_{u}\{C_{1}[O^{u}_{1}-O_{1}]
+C_{2}[O^{u}_{2}-O_{2}]\}\right],\label{Heff}
  \end{eqnarray}
where  $G_{F}$ is the Fermi coupling constant, $V_{ij}$, and $\lambda_{u}=\frac{V_{ub}V^{\ast}_{ud}}{V_{tb}V^{\ast}_{td}}$, are the corresponding CKM factors and their ratios, respectively. The explicit form of the four fermion operators that contribute to the said process in the SM can be written as,
 \begin{eqnarray}
 O_{7\gamma} &=&\frac{e}{16\pi ^{2}}m_{b}\left( \bar{d}\sigma _{\mu \nu }P_{R}b\right) F^{\mu \nu }\,,  \notag \\
O_{9} &=&\frac{e^{2}}{16\pi ^{2}}(\bar{d}\gamma _{\mu }P_{L}b)(\bar{\ell}\gamma^{\mu }\ell)\,,  \label{op-form} \\
O_{10} &=&\frac{e^{2}}{16\pi ^{2}}(\bar{d}\gamma _{\mu }P_{L}b)(\bar{\ell} \gamma ^{\mu }\gamma _{5} \ell)\,,  \notag
\end{eqnarray}
where $F^{\mu\nu}$ are the electromagnetic field strength tensor, $e$ is an electromagnetic coupling constant and $m_{b}$ appears in the electromagnetic dipole operator expression is the running $b$ quark mass in $\overline{\text{MS}}$ scheme.

In Eq. (\ref{Heff}), $C_{i}(\mu)$ represents the Wilson coefficients at the energy scale $\mu$. The form of the $C_{7}^{\text{eff}}(q^{2})$ and $C_{9}^{\text{eff}}(q^{2})$ Wilson coefficients \cite{Bobeth:1999mk,Beneke:2001at,Asatrian:2001de,Asatryan:2001zw,Greub:2008cy,Du:2015tda}, that contain the factorizable contributions from current-current, QCD penguins and chromomagnetic dipole operators $O_{1-6,8}$ are given in appendix \ref{append}.

At the quark level the decays $B\to M\mu^{+}\mu^{-}$ are governed by $b\to d\mu^{+}\mu^{-}$ transition and in the framework of SM its amplitude can be written as,
\begin{eqnarray}
\mathcal{M}^{\text{SM}}(b\to d\mu^{+}\mu^{-})&=&\frac{G_{F}\alpha V_{tb}V_{td}^{\ast}}{2\sqrt{2}\pi}\bigg\{C_{9}^{eff}\langle M(k,\epsilon)|\bar d\gamma^{\mu}(1-\gamma^{5})b|B(p)\rangle
\bar{\ell}\gamma_{\mu}\ell\notag\\
&+&C_{10}^{\text{SM}}\langle M(k,\epsilon)|\bar d\gamma^{\mu}(1-\gamma^{5})b|B(p)\rangle\bar{\ell}\gamma_{\mu}\gamma_{5}\ell\notag\\
&-&\frac{2m_{b}}{q^{2}}C_{7}^{\text{eff}}\langle M(k,\epsilon)|\bar{d}i\sigma^{\mu\nu}q_{\nu}(1+\gamma^{5})b|B(p)\rangle\bar{\ell}\gamma_{\mu}\ell\bigg\},\label{AMP}
\end{eqnarray}

 As discussed above that in the family non-universal $Z^{\prime}$ model, the FCNC transition arises at the tree level, hence by ignoring the $Z-Z^{\prime}$ mixing and assuming that the couplings
of right-handed quark flavors with $Z^{\prime}$ boson are diagonal\cite{Arhrib:2006sg,Cheung:2006tm,Chang:2009wt}, the 
effective Hamiltonian for $b\to d\mu^{+}\mu^{-}$ transition in the family non-universal $Z^{\prime}$ model can be straightforwardly written as,
\begin{eqnarray}
 \mathcal{H}^{Z^{\prime}}_{\text{eff}}=-\frac{2G_{F}}{\sqrt{2}}V_{tb}V^{\ast}_{td}\left[-\frac{B^{L}_{db}B^{L}_{\ell\ell}}{V_{tb}V^{\ast}_{td}}
 (\bar{d}b)_{V-A}(\bar{\ell}\ell)_{V-A}-\frac{B^{L}_{db}B^{R}_{\ell\ell}}{V_{tb}V^{\ast}_{td}}(\bar{d}b)_{V-A}(\bar{\ell}\ell)_{V+A}\right],\label{HZP}
\end{eqnarray}
where $B^{L}_{db}=|B^{L}_{db}|e^{-i\phi_{db}}$ represents the left-handed coupling of quarks with $Z^{\prime}$ gauge boson 
and $\phi_{db}$ is the new CP-violating phase which is not present in the SM. In condensed notation, one can write Eq. (\ref{HZP}) as
\begin{eqnarray}
 \mathcal{H}^{Z^{\prime}}_{\text{eff}}=-\frac{4G_{F}}{\sqrt{2}}V_{tb}V^{\ast}_{td}\left[\Lambda_{db}C^{Z^{\prime}}_{9}O_{9}
 +\Lambda_{db}C^{Z^{\prime}}_{10}O_{10}\right],\label{HZP1}
\end{eqnarray}
where
\begin{eqnarray}
\Lambda_{db}=\frac{4\pi e^{-i\phi_{db}}}{\alpha V_{tb}V^{\ast}_{td}},\label{coup}\\
 C^{Z^{\prime}}_{9}=|B^{L}_{db}|S_{LR}; &C^{Z^{\prime}}_{10}=|B^{L}_{db}|D_{LR},\label{WC1}
\end{eqnarray}
and 
\begin{eqnarray}
%\Lambda_{db}=\frac{4\pi e^{-i\phi_{db}}}{\alpha %V_{tb}V^{\ast}_{td}},\label{coup}\\
 S_{LR}=B^{L}_{\ell\ell}+B^{R}_{\ell\ell},\notag\\
 D_{LR}=B^{L}_{\ell\ell}-B^{R}_{\ell\ell}.\label{coup1}
\end{eqnarray}
In Eq. (\ref{coup1}), $S_{LR}$ and $D_{LR}$ constitutes the couplings of new $Z^{\prime}$ gauge boson with left and right-handed leptons. The total amplitude for the decay $B\to M\mu^{+}\mu^{-}$ in terms of SM and in $Z^{\prime}$ model can be written as,
\begin{eqnarray}
\mathcal{M}^{\text{tot}}(b\to d\ell^{+}\ell^{-})&=&\frac{G_{F}\alpha V_{tb}V_{td}^{\ast}}{2\sqrt{2}\pi}\bigg\{C_{9}^{\text{tot}}\langle M(k,\epsilon)|\bar d\gamma^{\mu}(1-\gamma^{5})b|B(p)\rangle
\bar{\ell}\gamma_{\mu}\ell\notag\\
&+&C_{10}^{\text{tot}}\langle M(k,\epsilon)|\bar d\gamma^{\mu}(1-\gamma^{5})b|B(p)\rangle\bar{\ell}\gamma_{\mu}\gamma_{5}\ell\notag\\
&-&\frac{2m_{b}}{q^{2}}C_{7}^{\text{eff}}\langle M(k,\epsilon)|\bar{d}i\sigma^{\mu\nu}q_{\nu}(1+\gamma^{5})b|B(p)\rangle\bar{\ell}\gamma_{\mu}\ell\bigg\}\,,\label{AMP1}
\end{eqnarray}
where in Eq. (\ref{AMP1})
\begin{eqnarray}
 \mathcal{M}^{\text{tot}}=\mathcal{M}^{\text{SM}}+\mathcal{M}^{\text{ZP}},
\end{eqnarray}
and
\begin{eqnarray}
 C_{9}^{\text{tot}}=C_{9}^{\text{eff}}+\Lambda_{db} C^{Z^{\prime}}_{9},\notag\\
 C_{10}^{\text{tot}}=C_{10}^{\text{SM}}+\Lambda_{db} C^{Z^{\prime}}_{10}.\label{WC2}
\end{eqnarray}
The amplitude for the decays $B\to M\ell^{+}\ell^{-}$ in the framework of SM and in the family non-universal $Z^{\prime}$ model can also be written as,
\begin{eqnarray}
\mathcal{M}^{\text{tot}}\left(B\to M\ell^{+}\ell^{-}\right)=\frac{G_{F}\alpha}{2\sqrt{2}\pi}V_{tb}V^{\ast}_{td}\Big\{T^{1,M}_{\mu}(\bar{\ell}\gamma^{\mu}\ell)
+T^{2,M}_{\mu}(\bar{\ell}\gamma^{\mu}\gamma_{5}\ell)\Big\},\label{Amp1}
\end{eqnarray}
where
\begin{eqnarray}
T^{1,M}_{\mu}&=&C_{9}^{\text{tot}}\langle M(k,\varepsilon)|\bar s\gamma_{\mu}(1-\gamma_{5})b|B(p)\rangle
-\frac{2m_{b}}{q^{2}}C_{7}^{\text{eff}}
\langle M(k,\varepsilon)|\bar s i\sigma_{\mu\nu}q^{\nu}(1+\gamma_{5})b|B(p)\rangle,\label{Amp1a}
\\
T^{2,M}_{\mu}&=&C_{10}^{\text{tot}}\langle M(k,\varepsilon)|\bar s\gamma_{\mu}(1-\gamma_{5})b|B(p)\rangle,
\end{eqnarray}
where $T^{i,M}_{\mu}$, $i=(1,2)$, contain the matrix elements of $B\to M$.
\subsection{Matrix Elements of \texorpdfstring{$B\to(\rho(770), a_{1}(1260))\mu^{+}\mu^{-}$}{} Decays}
The form factors for $B\to\rho$ and $B\to a_{1}$ decays can be expressed in terms of Lorentz invariant form factors as,
\begin{align}
\left\langle \rho(k,\overline\epsilon)\left\vert \bar{s}\gamma
_{\mu }b\right\vert B(p)\right\rangle &=\frac{2\epsilon_{\mu\nu\alpha\beta}}
{m_{B}+m_{\rho}}\overline\epsilon^{\,\ast\nu}p^{\alpha}k^{\beta}V(q^{2}),\label{2.13a}
\\
\left\langle \rho(k,\overline\epsilon)\left\vert \bar{s}\gamma_{\mu}\gamma_{5}b\right\vert
B(p)\right\rangle &=i\left(m_{B}+m_{\rho}\right)g_{\mu\nu}\overline\epsilon^{\,\ast\nu}A_{1}(q^{2})
\notag\\
&-iP_{\mu}(\overline\epsilon^{\ast}\cdot q)\frac{A_{2}(q^{2})}{\left(m_{B}+m_{\rho}\right)}\notag\\
&-i\frac{2m_{\rho}}{q^{2}}q_{\mu}(\overline\epsilon^{\,\ast}\cdot q)
\left[A_{3}(q^{2})-A_{0}(q^{2})\right],\label{2.13b}
\end{align}
where $P_{\mu}=p_{\mu}+k_{\mu}$, $q_{\mu}=p_{\mu}-k_{\mu}$, and 
with $A_3(0)=A_0(0)$. We have used $\epsilon_{0123}=+1$ convention throughout the study. The additional tensor form factors are expressed as,
\begin{align}
\left\langle \rho(k,\overline\epsilon)\left\vert \bar{s}i\sigma
_{\mu \nu }q^{\nu }b\right\vert B(p)\right\rangle
&=-2\epsilon _{\mu\nu\alpha\beta}\overline\epsilon^{\,\ast\nu}p^{\alpha}k^{\beta}T_{1}(q^{2}),\label{FF11}\\
\left\langle \rho(k,\overline\epsilon )\left\vert \bar{s}i\sigma
_{\mu \nu }q^{\nu}\gamma_{5}b\right\vert B(p)\right\rangle
&=i\Big[\left(m^2_{B}-m^2_{\rho}\right)g_{\mu\nu}\overline\epsilon^{\,\ast\nu}\notag\\
&-(\overline\epsilon^{\,\ast }\cdot q)P_{\mu}\Big]T_{2}(q^{2})+i(\overline\epsilon^{\,\ast}\cdot q)\notag\\
&\times\left[q_{\mu}-\frac{q^{2}}{m^2_{B}-m^2_{\rho}}P_{\mu}
\right]T_{3}(q^{2}).\label{F3}
\end{align}
The relations between the form factors in \cite{Bharucha:2015bzk}, and the form factors given in above matrix elements are
\begin{align}
A_{12}(q^{2})&=\frac{\left(m_{B}+m_{\rho}\right)^2(m^2_{B}-m^2_{\rho}-q^2)
A_{1}(q^{2})-\lambda A_{2}(q^{2})}{16m_{B}m^2_{\rho}\left(m_{B}+m_{\rho}\right)},\notag
\\
T_{23}(q^{2})&=\frac{\left(m^2_{B}-m^2_{\rho}\right)(m^2_{B}+3m^2_{\rho}-q^2)
T_{2}(q^{2})-\lambda T_{3}(q^{2})}{8m_{B}m^2_{\rho}\left(m_{B}-m_{\rho}\right)}.\label{relation}
\end{align}
and
\begin{eqnarray}
 \langle a_{1}(k,\overline\epsilon)|V_{\mu}|B(p)\rangle&=&-\overline\epsilon^{\ast}_{\mu}(m_{B}+m_{a_{1}})V_{1}(q^{2})+
 (p+k)_{\mu}(\overline\epsilon^{\ast}.q)\frac{V_{2}(q^{2})}{m_{B}+m_{a_{1}}}\notag\\
 &+&q_{\mu}(\overline\epsilon^{\ast}.q)\frac{2m_{a_{1}}}{q^{2}}[V_{3}(q^{2})-V_{0}(q^{2})]\,,\label{F1}\\
 \langle a_{1}(k,\overline\epsilon)|A_{\mu}|B(p)\rangle&=& \frac{2i\epsilon_{\mu\nu\alpha\beta}}{m_{B}+m_{a_{1}}}
 \overline\epsilon^{\ast\nu}p^{\alpha}k^{\beta}A(q^{2})\,,\label{F2}
\end{eqnarray}
where $V^{\mu}=\bar d\gamma^{\mu}b$ and $A^{\mu}=\bar d\gamma^{\mu}\gamma^{5}b$ are vector and axial vector currents, $\overline\epsilon^{\ast\nu}$
are polarization vectors of the axial vector meson. The relation for vector form factor $V_{3}(q^{2})$ given in Eq. (\ref{F1})
can be written as 
\begin{eqnarray}
 V_{3}(q^{2})&=&\frac{m_{B}+m_{a_{1}}}{2m_{a_{1}}}V_{1}(q^{2})-\frac{m_{B}-m_{a_{1}}}{2m_{a_{1}}}V_{2}(q^{2})\,,\label{V3}\\
 V_{3}(0)&=&V_{0}(0)\,.\notag
\end{eqnarray}
\begin{eqnarray}
 \langle a_{1}(k,\overline\epsilon)|\bar di\sigma_{\mu\nu}q^{\nu}b|B(p)\rangle&=&[(m_{B}^{2}-m_{a_{1}}^{2})\overline\epsilon^{\ast}_{\mu}
 -(\overline\epsilon^{\ast}.q)(p+k)_{\mu}]T_{2}(q^{2})\notag\\
 &+&(\overline\epsilon^{\ast}.q)\left[q_{\mu}-\frac{q^{2}}{m^2_{B}-m^2_{a_{1}}}(p+k)_{\mu}\right]T_{3}(q^{2})\,,\label{T1}\\
 \langle a_{1}(k,\overline\epsilon)|\bar di\sigma_{\mu\nu}q^{\nu}\gamma^{5}b|B(p)\rangle&=&2i\epsilon_{\mu\nu\alpha\beta}\overline\epsilon^{\ast\nu}p^{\alpha}k^{\beta}T_{1}(q^{2})\,.\label{T2}
\end{eqnarray} 
\subsection{Helicity Formalism of \texorpdfstring{$B\to(\rho(770), a_{1}(1260))\mu^{+}\mu^{-}$}{} Decays} 
\begin{figure}[ht]
    \centering
    \includegraphics[scale=0.45]{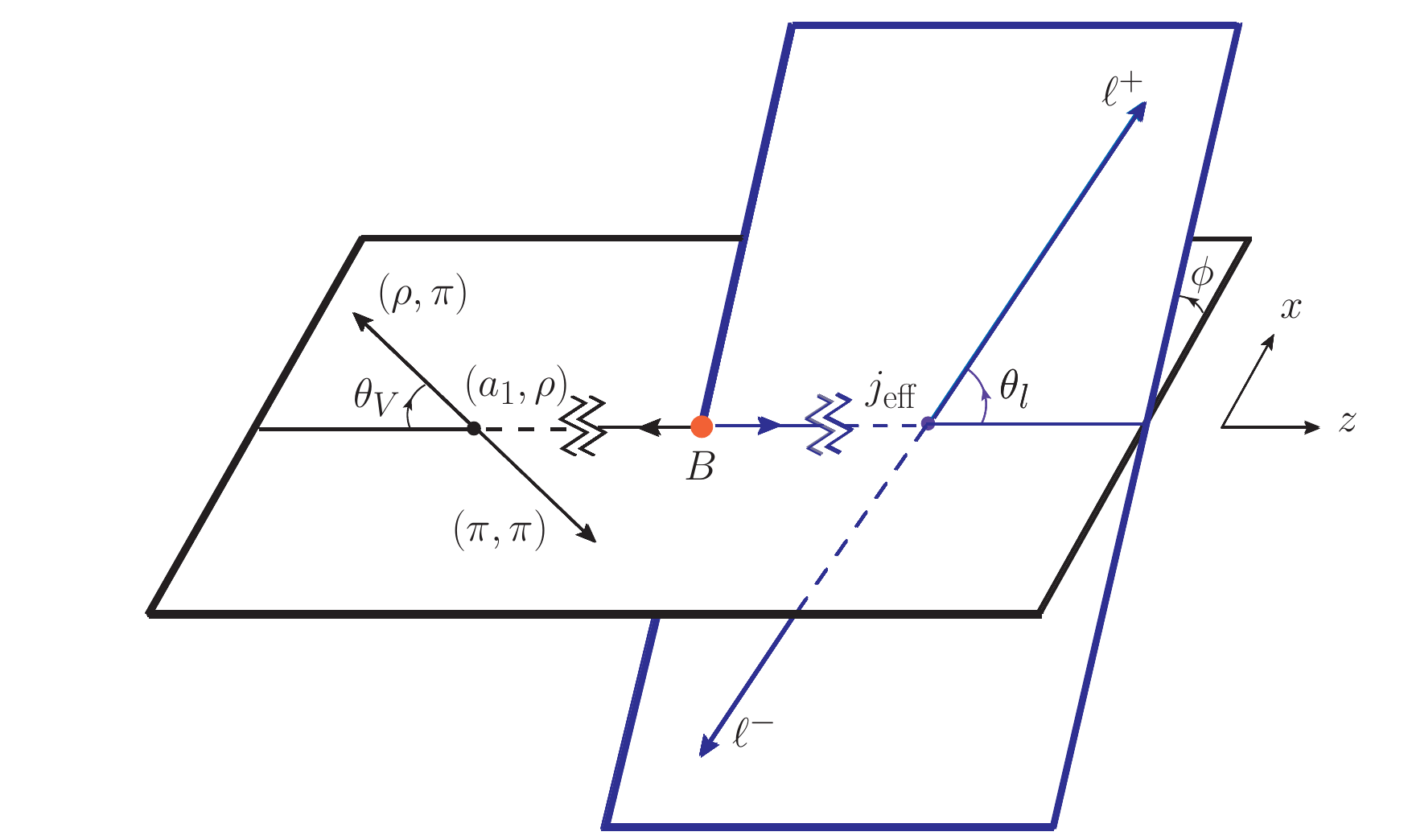}
    \caption{Kinematics of the $B\rightarrow \rho(\rightarrow\pi\pi)l^{+}l^{-}$, and $B\rightarrow a_{1}(\rightarrow\rho\pi)l^{+}l^{-}$ decays.}
    \label{fig:my_label}
\end{figure}

To calculate the angular distribution of the four-fold $B\to\rho (\to\pi\pi)\mu^{+}\mu^{-}$ and $B\to a_{1}(\to\rho_{\parallel, \perp}\pi)\mu^{+}\mu^{-}$ decays, we use the helicity formalism and follow \cite{Faessler:2002ut}. The kinematics of the four-fold decays under consideration are shown in Fig. \ref{fig:my_label}. The completeness and orthogonality properties of helicity basis can read as follows,
\begin{eqnarray}
\varepsilon^{\ast\alpha}(n)\varepsilon_{\alpha}(l)=g_{nl}, \qquad\quad \sum_{n, l=t, +, -, 0}\varepsilon^{\ast\alpha}(n)\varepsilon^{\beta}(l)g_{nl}=g^{\alpha\beta},\label{C22}
\end{eqnarray}
with $g_{nl}=\text{diag}(+, -, -, -)$. From the completeness relation given in Eq. (\ref{C22}), the contraction of leptonic tensors $L^{(k)\alpha\beta}$ and hadronic tensors $H^{ij}_{\alpha\beta}=T^{i,M}_{\alpha}\overline{T}^{\,j,M}_{\beta}$ $(i, j=1, 2)$, can be written as
\begin{eqnarray}
L^{(k)\alpha\beta}H^{ij}_{\alpha\beta}=\sum_{n, n^{\prime}, l, l^{\prime}}L^{(k)}_{nl}g_{nn^{\prime}}g_{ll^{\prime}}H^{ij}_{n^{\prime}l^{\prime}},\label{LH}
\end{eqnarray}
where the leptonic and hadronic tensors can be written in the helicity basis as follows
\begin{eqnarray}
L^{(k)}_{nl}=\varepsilon^{\alpha}(n)\varepsilon^{\ast\beta}(l)L^{(k)}_{\alpha\beta}, &&\qquad H^{ij}_{nl}=\varepsilon^{\ast\alpha}(n)\varepsilon^{\beta}(l)H^{ij}_{\alpha\beta}.\label{LHT}
\end{eqnarray}
Both leptonic and hadronic tensors shown in Eq. (\ref{LHT}), can be evaluated in two different frames of reference. The lepton tensor $L^{(k)}_{nl}$ is evaluated in the dimuon center of mass (CM) frame, and the hadronic tensor $H^{ij}_{nl}$ is evaluated in the rest frame of $B$ meson. For the above mentioned decays, one can write the hadronic tensor as follows,
\begin{eqnarray}
H^{ij}_{nl}&=&\big(\varepsilon^{\ast\alpha}(n)T^{i,M}_{\alpha}\big)\cdot\big(\overline{\varepsilon^{\ast\beta}(l)T^{j,M}_{\beta}}\big)\notag
\\
&=&\big(\varepsilon^{\ast\alpha}(n)\overline\epsilon^{\ast\mu}(r)T^{i,M}_{\alpha,\mu}\big)\cdot\big(\overline{\varepsilon^{\ast\beta}(l)
\overline\epsilon^{\ast\nu}(s)T^{j,M}_{\beta,\nu}}\big)\delta_{rs}\equiv H^{i,M}_n \,\overline{H}^{\, j,M_l}_l.\label{HA5}
\end{eqnarray}
%\begin{figure}[h]
%    \centering
%    \includegraphics[scale=0.80]{Btorho.png}
%    \caption{Kinematics of decay mode %$B\rightarrow \rho(\rightarrow\pi\pi)l^{+}l^{-}$ }
%    \label{fig:my_label}
%\end{figure}
\subsection{Helicity Amplitudes for \texorpdfstring{$B\to\rho\mu^{+}\mu^{-}$}{} and \texorpdfstring{$B\to a_{1}\mu^{+}\mu^{-}$}{} Decays}
The explicit expressions of the helicity amplitudes for $B\to\rho$ and $B\to a_{1}$, are given as
%obtained in terms of the Wilson coefficients of the SM and the family non-universal $Z^{\prime}$ model as, 
\begin{align}
H^{1, \rho}_t&=-i\sqrt{\frac{\lambda}{q^2}}C_{9}^{\text{tot}}A_0(q^{2}),\notag
\\
H^{2, \rho}_t&=-i\sqrt{\frac{\lambda}{q^2}}C_{10}^{\text{tot}}A_0(q^{2}),\notag
\\
H^{1, \rho}_{\pm}&=-i\left(m^2_{B}-m^2_{\rho}\right)\Big[C_{9}^{\text{tot}}
\frac{A_{1}(q^{2})}{\left(m_{B}-m_{\rho}\right)}\notag
\\
&+\frac{2m_{b}}{q^{2}}C_{7}^{\text{eff}}T_{2}(q^{2})\Big]
\pm i\sqrt{\lambda}\Big[C_{9}^{\text{tot}}
\frac{V(q^{2})}{\left(m_{B}+m_{\rho}\right)}+\frac{2m_{b}}{q^{2}}C_{7}^{\text{eff}}T_{1}(q^{2})\Big],\notag
\\
H^{2, \rho}_{\pm}&=-iC_{10}^{\text{tot}}\left(m_{B}+m_{\rho}\right)
A_{1}(q^{2})\pm i\sqrt{\lambda}C_{10}^{\text{tot}}
\frac{V(q^{2})}{\left(m_{B}+m_{\rho}\right)},\notag
\\
H^{1, \rho}_0&=-\frac{8im_{B}m_{\rho}}{\sqrt{q^2}}\Bigg[C_{9}^{\text{tot}}
A_{12}(q^{2})+m_b C_{7}^{\text{eff}}\frac{T_{23}(q^{2})}{m_{B}+m_{\rho}}\Bigg],\notag
\\
H^{2, \rho}_0&=-\frac{8im_{B} m_{\rho}}{\sqrt{q^2}}\Bigg[C_{10}^{\text{tot}}
A_{12}(q^{2})\Bigg].\label{heamprho}
\end{align}

%\\
%H^{1,\rho}_0&=-\frac{i}{2m_{\rho}\sqrt{q^2}}\Bigg[C_{9}^{\text{tot}}
%\Big\{(m^2_{B}-m^2_{\rho}-q^2)\left(m_{B}+m_{\rho}\right)A_{1}(q^{2})\notag
%\\
%&-\frac{\lambda}{m_{B}+m_{\rho}}A_{2}(q^{2})\Big\}+2m_b C_{7}^{\text{eff}}\Big\{(m^2_{B}+3m^2_{\rho}-q^2)T_{2}(q^{2})
%-\frac{\lambda}{m^2_{B}-m^2_{\rho}}T_{3}(q^{2})\Big\}
%\Bigg],\notag
%\\
%H^{2, \rho}_0&=-\frac{i}{2m_{\rho}\sqrt{q^2}}C_{10}^{\text{tot}}
%\Bigg[(m^2_{B}-m^2_{\rho}-q^2)\left(m_{B}+m_{\rho}\right)A_{1}(q^{2})-%\frac{\lambda}{m_{B}+m_{\rho}}A_{2}(q^{2})\Bigg].\label{heamprho}
%\end{align}
and
\begin{align}
H^{1,a_{1}}_t&=-\sqrt{\frac{\lambda}{q^2}}C_{9}^{\text{tot}}V_0(q^{2}),\notag
\\
H^{2,a_{1}}_t&=-\sqrt{\frac{\lambda}{q^2}}C_{10}^{\text{tot}}V_0(q^{2}),\notag
\\
H^{1,a_{1}}_{\pm}&=-\left(m^2_{B}-m^2_{a_{1}}\right)\Big[C_{9}^{\text{tot}}
\frac{V_{1}(q^{2})}{\left(m_{B}-m_{a_{1}}\right)}+\frac{2m_{b}}{q^{2}}C_{7}^{\text{eff}}T_{2}(q^{2})\Big]\notag
\\
&\pm \sqrt{\lambda}\Big[C_{9}^{\text{tot}}
\frac{A(q^{2})}{\left(m_{B}+m_{a_{1}}\right)}+\frac{2m_{b}}{q^{2}}C_{7}^{\text{eff}}T_{1}(q^{2})\Big],\notag
\\
H^{2,a_{1}}_{\pm}&=-C_{10}^{\text{tot}}\left(m_{B}+m_{a_{1}}\right)
V_{1}(q^{2})\pm \sqrt{\lambda}C_{10}^{\text{tot}}
\frac{A(q^{2})}{\left(m_{B}+m_{a_{1}}\right)},\notag
\\
H^{1,a_{1}}_0&=-\frac{1}{2m_{a_{1}}\sqrt{q^2}}\Bigg[C_{9}^{\text{tot}}
\Big\{(m^2_{B}-m^2_{a_{1}}-q^2)\left(m_{B}+m_{a_{1}}\right)V_{1}(q^{2})\notag
\\
&-\frac{\lambda}{m_{B}+m_{a_{1}}}V_{2}(q^{2})\Big\}+2m_b C_{7}^{\text{eff}}\Big\{(m^2_{B}+3m^2_{a_{1}}-q^2)T_{2}(q^{2})
-\frac{\lambda}{m^2_{B}-m^2_{a_{1}}}T_{3}(q^{2})\Big\}
\Bigg],\notag
\\H^{2, a_1}_0&=-\frac{1}{2m_{a_{1}}\sqrt{q^2}}C_{10}^{\text{tot}}
\Bigg[(m^2_{B}-m^2_{a_{1}}-q^2)\left(m_{B}+m_{a_{1}}\right)V_{1}(q^{2})-\frac{\lambda}{m_{B}+m_{a_{1}}}V_{2}(q^{2})\Bigg].\label{heampa1}
\end{align}
\subsection{Four fold distribution of \texorpdfstring{$B\to\rho(\to\pi\pi)\mu^{+}\mu^{-}$}{} and \texorpdfstring{$B\to a_{1}(\to\rho\pi)\mu^{+}\mu^{-}$}{} Decays}
%The NP effects in effective theory arise due to Wilson coefficients and a set of new operators. However for the present work, we investigate the family non-universal $Z^{\prime}$ model where the effects of NP arise due to the Wilson coefficients only; these NP effects are contained in
The four-fold decay distribution depends on the square of the dilepton invariant mass $q^{2}$, angles $\theta_{\ell}$, $\theta_{V}$ and $\phi$ as shown in Fig. \ref{fig:my_label}. For $B\to\rho$ decay mode, the four-fold distribution can be written as,
\begin{eqnarray}
 \frac{d^4\Gamma\left(B\to\rho\,(\to \pi\pi)\mu^+\mu^-\right)}{dq^2 \ d\cos{\theta_{l}} \ d\cos {\theta}_{V} \ d\phi} &=& \frac{9}{32 \pi} \mathcal{B}(\rho\to \pi\pi)\notag
 \\
&\times&\bigg[I^{\rho}_{1s}\sin^2\theta_{V}+I^{\rho}_{1c}\cos^2\theta_{V}\notag\\
&+&\Big(I^{\rho}_{2s}\sin^2\theta_{V}+I^{\rho}_{2c}\cos^2\theta_{V}\Big)\cos{2\theta_{l}}
\notag\\
&+&\Big(I^{\rho}_{6s}\sin^2\theta_{V}+I^{\rho}_{6c}\cos^2\theta_{V}\Big)\cos{\theta_{l}}\notag\\
&+&\Big(I^{\rho}_{3}\cos{2\phi}
+I^{\rho}_{9}\sin{2\phi}\Big)\sin^2\theta_{V}\sin^2\theta_{l}\notag
\\
&+&\Big(I^{\rho}_{4}\cos{\phi}+I^{\rho}_{8}\sin{\phi}\Big)\sin2\theta_{V}\sin2\theta_{l}\notag
\\
&+&\Big(I^{\rho}_{5}\cos{\phi}+I^{\rho}_{7}\sin{\phi}\Big)\sin2\theta_{V}\sin\theta_{l}\bigg].\notag\\
\label{fullad}
\end{eqnarray}
The explicit expressions of $I^{\rho}_{n\lambda}$ in terms of the helicity amplitudes are obtained as,
\begin{eqnarray}\label{IsDpirho}
I^{\rho}_{1s} &=& \frac{(2+\beta_l^2)}{2}N^2\left(|H_+^1|^2+|H_+^2|^2+|H_-^1|^2+|H_-^2|^2\right)\notag
\\&+&\frac{4m_l^2}{q^2}N^2\left(|H_+^1|^2-|H_+^2|^2+|H_-^1|^2-|H_-^2|^2\right),\\
I^{\rho}_{1c} &=& 2N^2\left(|H_0^1|^2+|H_0^2|^2\right)+\frac{8m_l^2}{q^2}N^2\left(|H_0^1|^2-|H_0^2|^2+2|H_t^2|^2\right),\\
I^{\rho}_{2s} &=& \frac{\beta_l^2}{2}N^2\left(|H_+^1|^2+|H_+^2|^2+|H_-^1|^2+|H_-^2|^2\right),\\
I^{\rho}_{2c} &=& -2\beta_l^2N^2\left(|H_0^1|^2+|H_0^2|^2\right),\\
I^{\rho}_{3}&=&-2\beta_l^2N^2\bigg[\mathcal{R}e\left(H_+^{1}H_-^{1\ast}+H_+^{2}H_-^{2\ast}\right)\bigg],\\
I^{\rho}_{4}&=&\beta_l^2N^2\bigg[\mathcal{R}e\left(H_+^{1}H_0^{1\ast}+H_-^{1}H_0^{1\ast}\right)
+\mathcal{R}e\left(H_+^{2}H_0^{2\ast}+H_-^{2}H_0^{2\ast}\right)\bigg],\\
I^{\rho}_{5}&=&-2\beta_lN^2\bigg[\mathcal{R}e\left(H_+^{1}H_0^{2\ast}-H_-^{1}H_0^{2\ast}\right)
+\mathcal{R}e\left(H_+^{2}H_0^{1\ast}-H_-^{2}H_0^{1\ast}\right)\bigg],\\
I^{\rho}_{6s}&=&-4\beta_lN^2\bigg[\mathcal{R}e\left(H_+^{1}H_+^{2\ast}-H_-^{1}H_-^{2\ast}\right)\bigg],\\
I^{\rho}_{6c}&=&0,\\
I^{\rho}_{7}&=&-2\beta_lN^2\bigg[\mathcal{I}m\left(H_0^{1}H_+^{2\ast}+H_0^{1}H_-^{2\ast}\right)
+\mathcal{I}m\left(H_0^{2}H_+^{1\ast}+H_0^{2}H_-^{1\ast}\right)\bigg],\\
I^{\rho}_{8}&=&\beta_l^2N^2\bigg[\mathcal{I}m\left(H_0^{1}H_+^{1\ast}-H_0^{1}H_-^{1\ast}\right)
+\mathcal{I}m\left(H_0^{2}H_+^{2\ast}-H_0^{2}H_-^{2\ast}\right)\bigg],\\
I^{\rho}_{9}&=&2\beta_l^2N^2\bigg[\mathcal{I}m\left(H_+^{1}H_-^{1\ast}+H_+^{2}H_-^{2\ast}\right)\bigg],
\end{eqnarray}
For the decay $B\to a_{1}\,(\to \rho_{\|(\perp)}\pi)\mu^+\mu^-$ the four fold distribution can be written as,
%\begin{eqnarray}
% \frac{d^4\Gamma\left(B\to a_{1}\,(\to \rho_{\perp(\|)}\pi)\mu^+\mu^-\right)}{dq^2 \ d\cos{\theta_{l}} \ d\cos {\theta}_{V} \ d\phi} &=& \frac{9}{32 \pi} \mathcal{B}(a_{1}\to \rho_{\perp(\|)}\pi)\notag
% \\
%&\times&\bigg[I^{a_{1}}_{1s,\perp}(I^{a_{1}}_{1s,\|})\sin^2\theta_{V}+I^{a_{1}}_{1c,\perp}(I^{a_{1}}_{1c,\|})\cos^2\theta_{V}\notag\\
%&+&\Big(I^{a_{1}}_{2s,\perp}(I^{a_{1}}_{2s,\|})\sin^2\theta_{V}+I^{a_{1}}_{2c,\perp}(I^{a_{1}}_{2c,\|})\cos^2\theta_{V}\Big)\cos{2\theta_{l}}
%\notag\\
%&+&\Big(I^{a_{1}}_{6s,\perp}(I^{a_{1}}_{6s,\|})\sin^2\theta_{V}+I^{a_{1}}_{6c,\perp}(I^{a_{1}}_{6c,\|})\cos^2\theta_{V}\Big)\cos{\theta_{l}}\notag\\
%&+&\Big(I^{a_{1}}_{3,\perp}(I^{a_{1}}_{3,\|})\cos{2\phi}
%+I^{a_{1}}_{9,\perp}(I^{a_{1}}_{9,\|})\sin{2\phi}\Big)\sin^2\theta_{V}\sin^2\theta_{l}\notag
%\\
%&+&\Big(I^{a_{1}}_{4,\perp}(I^{a_{1}}_{4,\|})\cos{\phi}+I^{\rho}_{8,\perp}(I^{a_{1}}_{8,\|})\sin{\phi}\Big)\sin2\theta_{V}\sin2\theta_{l}\notag
%\\
%&+&\Big(I^{a_{1}}_{5,\perp}(I^{a_{1}}_{5,\|})\cos{\phi}+I^{\rho}_{7,\perp}(I^{a_{1}}_{7,\|})\sin{\phi}\Big)\sin2\theta_{V}\sin\theta_{l}\bigg],\notag\\
%\label{fullad1}
%\end{eqnarray}
\begin{eqnarray}
 \frac{d^4\Gamma\left(B\to a_{1}\,(\to \rho_{\|(\perp)}\pi)\mu^+\mu^-\right)}{dq^2 \ d\cos{\theta_{l}} \ d\cos {\theta}_{V} \ d\phi} &=& \frac{9}{32 \pi} \mathcal{B}(a_{1}\to \rho_{\|(\perp)}\pi)\notag
 \\
&\times&\bigg[I^{a_{1}}_{1s,\|(\perp)}\sin^2\theta_{V}+I^{a_{1}}_{1c,\|(\perp)}\cos^2\theta_{V}\notag\\
&+&\Big(I^{a_{1}}_{2s,\|(\perp)}\sin^2\theta_{V}+I^{a_{1}}_{2c,\|(\perp)}\cos^2\theta_{V}\Big)\cos{2\theta_{l}}
\notag\\
&+&\Big(I^{a_{1}}_{6s,\|(\perp)}\sin^2\theta_{V}+I^{a_{1}}_{6c,\|(\perp)}\cos^2\theta_{V}\Big)\cos{\theta_{l}}\notag\\
&+&\Big(I^{a_{1}}_{3,\|(\perp)}\cos{2\phi}
+I^{a_{1}}_{9,\|(\perp)}\sin{2\phi}\Big)\sin^2\theta_{V}\sin^2\theta_{l}\notag
\\
&+&\Big(I^{a_{1}}_{4,\|(\perp)}\cos{\phi}+I^{a_{1}}_{8,\|(\perp)}\sin{\phi}\Big)\sin2\theta_{V}\sin2\theta_{l}\notag
\\
&+&\Big(I^{a_{1}}_{5,\|(\perp)}\cos{\phi}+I^{a_{1}}_{7,\|(\perp)}\sin{\phi}\Big)\sin2\theta_{V}\sin\theta_{l}\bigg].\notag\\
\label{fullad1}
\end{eqnarray}

where, $I^{a_{1}}_{n\lambda,\|}$ and $I_{n\lambda,\perp}^{a_{1}}$ are the angular coefficients. The explicit expressions of $I^{a_{1}}_{n\lambda,\|}$ in terms of the helicity amplitudes are written as,
\begin{eqnarray}\label{IsDpia1}
I^{a_{1}}_{1s,\|} &=& \frac{(2+\beta_l^2)}{2}N^2\left(|H_+^1|^2+|H_+^2|^2+|H_-^1|^2+|H_-^2|^2\right)\notag
\\&+&\frac{4m_l^2}{q^2}N^2\left(|H_+^1|^2-|H_+^2|^2+|H_-^1|^2-|H_-^2|^2\right),\\
I^{a_{1}}_{1c,\|} &=& 2N^2\left(|H_0^1|^2+|H_0^2|^2\right)+\frac{8m_l^2}{q^2}N^2\left(|H_0^1|^2-|H_0^2|^2+2|H_t^2|^2\right),\\
I^{a_{1}}_{2s,\|} &=& \frac{\beta_l^2}{2}N^2\left(|H_+^1|^2+|H_+^2|^2+|H_-^1|^2+|H_-^2|^2\right),\\
I^{a_{1}}_{2c,\|} &=& -2\beta_l^2N^2\left(|H_0^1|^2+|H_0^2|^2\right),\\
I^{a_{1}}_{3,\|}&=&-2\beta_l^2N^2\bigg[\mathcal{R}e\left(H_+^{1}H_-^{1\ast}+H_+^{2}H_-^{2\ast}\right)\bigg],\\
I^{a_{1}}_{4,\|}&=&\beta_l^2N^2\bigg[\mathcal{R}e\left(H_+^{1}H_0^{1\ast}+H_-^{1}H_0^{1\ast}\right)
+\mathcal{R}e\left(H_+^{2}H_0^{2\ast}+H_-^{2}H_0^{2\ast}\right)\bigg],\\
I^{a_{1}}_{5,\|}&=&-2\beta_lN^2\bigg[\mathcal{R}e\left(H_+^{1}H_0^{2\ast}-H_-^{1}H_0^{2\ast}\right)
+\mathcal{R}e\left(H_+^{2}H_0^{1\ast}-H_-^{2}H_0^{1\ast}\right)\bigg],\\
I^{a_{1}}_{6s,\|}&=&-4\beta_lN^2\bigg[\mathcal{R}e\left(H_+^{1}H_+^{2\ast}-H_-^{1}H_-^{2\ast}\right)\bigg],\\
I^{a_{1}}_{6c,\|}&=&0,\\
I^{a_{1}}_{7,\|}&=&-2\beta_lN^2\bigg[\mathcal{I}m\left(H_0^{1}H_+^{2\ast}+H_0^{1}H_-^{2\ast}\right)
+\mathcal{I}m\left(H_0^{2}H_+^{1\ast}+H_0^{2}H_-^{1\ast}\right)\bigg],\\
I^{a_{1}}_{8,\|}&=&\beta_l^2N^2\bigg[\mathcal{I}m\left(H_0^{1}H_+^{1\ast}-H_0^{1}H_-^{1\ast}\right)
+\mathcal{I}m\left(H_0^{2}H_+^{2\ast}-H_0^{2}H_-^{2\ast}\right)\bigg],\\
I^{a_{1}}_{9,\|}&=&2\beta_l^2N^2\bigg[\mathcal{I}m\left(H_+^{1}H_-^{1\ast}+H_+^{2}H_-^{2\ast}\right)\bigg],
\end{eqnarray}
whereas  the expressions of $I_{n\lambda,\perp}^{a_{1}}$ in terms of the helicity amplitudes are written as,
\begin{eqnarray}\label{30df}
I^{a_{1}}_{1s,\perp} &=&\frac{(2+\beta_l^2)}{4}N^2\left(|H_+^1|^2+|H_+^2|^2+|H_-^1|^2+|H_-^2|^2\right)+\left(|H_0^1|^2+|H_0^2|^2\right)\notag\\
&+&\frac{2m_l^2}{q^2}N^2\bigg[\left(|H_+^1|^2-|H_+^2|^2+|H_-^1|^2-|H_-^2|^2\right)
+2\left(|H_0^1|^2-|H_0^2|^2+2|H_t^2|^2\right)\bigg],\\
I^{a_{1}}_{1c,\perp} &=& \frac{(2+\beta_l^2)}{2}N^2\left(|H_+^1|^2+|H_+^2|^2+|H_-^1|^2+|H_-^2|^2\right)
\notag\\
&+&\frac{4m_l^2}{q^2}N^2\left(|H_+^1|^2-|H_+^2|^2+|H_-^1|^2-|H_-^2|^2\right),\\
I^{a_{1}}_{2s,\perp} &=& -\beta_l^2N^2\bigg[\left(|H_0^1|^2+|H_0^2|^2\right)-\frac{1}{4}\left(|H_+^1|^2+|H_+^2|^2+|H_-^1|^2+|H_-^2|^2\right)\bigg]
,\\
I^{a_{1}}_{2c,\perp} &=& \frac{\beta_l^2}{2}N^2\left(|H_+^1|^2+|H_+^2|^2+|H_-^1|^2+|H_-^2|^2\right),\\
I^{a_{1}}_{3,\perp}&=&\beta_l^2N^2\bigg[\mathcal{R}e\left(H_+^{1}H_-^{1\ast}+H_+^{2}H_-^{2\ast}\right)\bigg],\\
I^{a_{1}}_{4,\perp}&=&-\frac{\beta_l^2}{2}N^2\bigg[\mathcal{R}e\left(H_+^{1}H_0^{1\ast}+H_-^{1}H_0^{1\ast}\right)
+\mathcal{R}e\left(H_+^{2}H_0^{2\ast}+H_-^{2}H_0^{2\ast}\right)\bigg],
\end{eqnarray}
\begin{eqnarray}
I^{a_{1}}_{5,\perp}&=&\beta_lN^2\bigg[\mathcal{R}e\left(H_+^{1}H_0^{2\ast}-H_-^{1}H_0^{2\ast}\right)
+\mathcal{R}e\left(H_+^{2}H_0^{1\ast}-H_-^{2}H_0^{1\ast}\right)\bigg],\\
I^{a_{1}}_{6s,\perp}&=&-2\beta_lN^2\bigg[\mathcal{R}e\left(H_+^{1}H_+^{2\ast}-H_-^{1}H_-^{2\ast}\right)\bigg],\label{6sperp}\\
I^{a_{1}}_{6c,\perp}&=&-4\beta_lN^2\bigg[\mathcal{R}e\left(H_+^{1}H_+^{2\ast}-H_-^{1}H_-^{2\ast}\right)\bigg],\label{6cperp}\\
I^{a_{1}}_{7,\perp}&=&\beta_lN^2\bigg[\mathcal{I}m\left(H_0^{1}H_+^{2\ast}+H_0^{1}H_-^{2\ast}\right)
+\mathcal{I}m\left(H_0^{2}H_+^{1\ast}+H_0^{2}H_-^{1\ast}\right)\bigg],\\
I^{a_{1}}_{8,\perp}&=&-\frac{\beta_l^2}{2}N^2\bigg[\mathcal{I}m\left(H_0^{1}H_+^{1\ast}-H_0^{1}H_-^{1\ast}\right)
+\mathcal{I}m\left(H_0^{2}H_+^{2\ast}-H_0^{2}H_-^{2\ast}\right)\bigg],\\
I^{a_{1}}_{9,\perp}&=&-\beta_l^2N^2\bigg[\mathcal{I}m\left(H_+^{1}H_-^{1\ast}+H_+^{2}H_-^{2\ast}\right)\bigg],
\end{eqnarray}
where
\begin{eqnarray}\label{24abc}
N=V_{tb}V^{\ast}_{td}\Bigg[\frac{G_{F}^2\alpha^2}{3.2^{10} \pi^5 m_{B}^{3}} q^2\sqrt{\lambda}\beta_l\Bigg]^{1/2},
\end{eqnarray}
with $\lambda\equiv \lambda(m^2_{B}, m^2_{M}, q^2)$ and $\beta_l=\sqrt{1-4m_l^2/q^2}$.
%\begin{sloppypar}
%\subsection{Physical Observables for \texorpdfstring{$B\to\rho(\to\pi\pi)\mu^{+}\mu^{-}$}{} and \texorpdfstring{$B\to a_{1}(\to\rho_{\|,\perp}\pi)\mu^{+}\mu^{-}$}{} Decays }
%\end{sloppypar}
%In this section, we write an expressions for physical observables such as branching ratio, lepton forward-backward asymmetry, longitudinal helicity fractions and lepton flavor non-universality ratio for $B\to\rho(\to\pi\pi)\mu^{+}\mu^{-}$ and $B\to a_{1}(\to\rho_{\|,\perp}\pi)\mu^{+}\mu^{-}$ decays in terms of angular coefficients $I^{\rho}_{n\lambda,\|}$ and  $I^{\rho}_{n\lambda,\perp}$.
\subsection {Physical Observables for \texorpdfstring{$B\to\rho(\to\pi\pi)\mu^{+}\mu^{-}$}{}Decay} 
In this section, we give the expressions of the physical observables such as the differential decay rate, lepton forward-backward asymmetry, longitudinal helicity fraction of $\rho$ and the normalized angular observables $\langle I^{\rho}_{n\lambda}\rangle$, for $B\to\rho(\to\pi\pi)\mu^{+}\mu^{-}$ decay.\\
\textbf{(i) Differential decay rate:}
From the full angular distribution Eq. \eqref{fullad}, integration over $\cos\theta_l = [-1,1]$, $\cos\theta_V = [-1,1]$, and $\phi = [0,2\pi]$ yields the $q^{2}$ dependent differential decay rate expression, which in terms of the angular coefficients is as follows,
\begin{eqnarray}
\frac{d\Gamma \left(B\to\rho(\to \pi\pi)\mu^{+}\mu^{-}\right)}{dq^{2}}=\mathcal{B}(\rho\to \pi\pi)\frac{1}{4}(3I^{\rho}_{1c}+6I^{\rho}_{1s}-I^{\rho}_{2c}-2I^{\rho}_{2s}).\label{DBRrho}
\end{eqnarray}
\textbf{(ii) Lepton forward-backward asymmetry:}
From the full angular distribution Eq. \eqref{fullad}, the integration over $\cos\theta_V = [-1,1]$, and $\phi = [0,2\pi]$, gives the double differential decay rate $\left(\frac{d^{2}\Gamma}{dq^{2}d\cos\theta_{\ell}}\right)$. The lepton forward-backward asymmetry corresponding to $\theta_\ell$ is $A_{\text{FB}} = (F-B)/(F+B)$, where $F$ and $B$ are the forward and backward hemispheres. The forward backward asymmetry for $B\to\rho\mu^{+}\mu^{-}$ decay can be obtained by integrating $\frac{d^{2}\Gamma}{dq^{2}d\cos\theta_{\ell}}$, and is defined as,
\begin{eqnarray}
A_{\text{FB}}^{\rho}(q^{2})=\frac{\int_{0}^{1}\frac{d^{2}\Gamma}{dq^{2}d\cos\theta_{\ell}}d\cos\theta_{\ell}-\int_{-1}^{0}\frac{d^{2}\Gamma}{dq^{2}d\cos\theta_{\ell}}d\cos\theta_{\ell}}{\int_{-1}^{1}\frac{d^{2}\Gamma}{dq^{2}d\cos\theta_{\ell}}d\cos\theta_{\ell}}.
\end{eqnarray}
In terms of the angular coefficients $I's$ the lepton forward-backward asymmetry for $B\to\rho(\to\pi\pi)\mu^{+}\mu^{-}$  as a function of $q^{2}$ can be expressed as,
\begin{eqnarray}
A_{\text{FB}}^{\rho}(q^{2})=\frac{6I^{\rho}_{6s}}{2(3I^{\rho}_{1c}+6I^{\rho}_{1s}-I^{\rho}_{2c}-2I^{\rho}_{2s})}.\label{AFB}
\end{eqnarray}
\textbf{(iii) Longitudinal helicity fraction:}
From the full angular distribution Eq. \eqref{fullad}, the integration over $\cos\theta_l = [-1,1]$, and $\phi = [0,2\pi]$, gives the double differential decay rate $\left(\frac{d^{2}\Gamma}{dq^{2}d\cos\theta_{V}}\right)$. The longitudinal helicity fraction of the decay $B\to \rho(\to \pi\pi)\mu^{+}\mu^{-}$, when $\rho$ meson is longitudinally polarized can be defined as,
\begin{eqnarray}
f_{L}^{\rho}(q^{2})=\frac{\int_{-1}^{1}\frac{d^{2}\Gamma}{dq^{2}d\cos\theta_{V}}\left(\frac{5}{2}\cos^2\theta_V-\frac{1}{2}\right)d\cos\theta_{V}}{{d\Gamma \left(B\to\rho(\to \pi\pi)\mu^{+}\mu^{-}\right)}/{dq^{2}}}.
\end{eqnarray}
In terms of the angular coefficients $I's(q^{2})$ the longitudinal helicity fraction for the decay $B\to \rho(\to \pi\pi)\mu^{+}\mu^{-}$ can be written as,
\begin{eqnarray}
f_{L}^{\rho}(q^{2})=\frac{3I^{\rho}_{1c}-I^{\rho}_{2c}}{3I^{\rho}_{1c}+6I^{\rho}_{1s}-I^{\rho}_{2c}-2I^{\rho}_{2s}}.
\end{eqnarray}
\textbf{(iv) Normalized angular observables:}
\begin{eqnarray}
\langle I_{n\lambda}^{\rho}\rangle=\frac{\mathcal{B}(\rho\to \pi\pi)I_{n\lambda}^{\rho}}{d\Gamma \left(B\to\rho(\to \pi\pi)\mu^{+}\mu^{-}\right)/dq^{2}}.
\end{eqnarray}
\textbf{(v) Binned normalized angular observables:}
\begin{eqnarray}
\langle I_{n\lambda}^{\rho}\rangle_{\left[q^{2}_{\text{min}},q^{2}_{\text{max}}\right]}=\frac{\int^{q^{2}_{\text{max}}}_{q^{2}_{\text{min}}}\mathcal{B}(\rho\to \pi\pi)I_{n\lambda}^{\rho}\,dq^2}{\int^{q^{2}_{\text{max}}}_{q^{2}_{\text{min}}}(d\Gamma \left(B\to\rho(\to \pi\pi)\mu^{+}\mu^{-}\right)/dq^{2})dq^2}.
\end{eqnarray}
\subsection {Physical Observables for \texorpdfstring{$B\to a_{1}(\to\rho_{||,\perp}\pi)\mu^{+}\mu^{-}$}{} Decay}
The formulas of physical observables for $B\to a_{1}(\to\rho_{||,\perp}\pi)\mu^{+}\mu^{-}$ decay can be expressed as,\\
\textbf{(i) Differential decay rates:}
\begin{eqnarray}
\frac{d\Gamma \left(B\to a_{1}(\to\rho_{||}\pi)\mu^{+}\mu^{-}\right)}{dq^{2}}=\mathcal{B}(a_{1}\to\rho_{||}\pi)\frac{1}{4}(3I^{a_{1}}_{1c,||}+6I^{a_{1}}_{1s,||}-I^{a_{1}}_{2c,||}-2I^{a_{1}}_{2s,||}).\label{DBR1}
\end{eqnarray}
\begin{eqnarray}
\frac{d\Gamma \left(B\to a_{1}(\to\rho_{\perp}\pi)\mu^{+}\mu^{-}\right)}{dq^{2}}=\mathcal{B}(a_{1}\to\rho_{\perp}\pi)\frac{1}{4}(3I^{a_{1}}_{1c,\perp}+6I^{a_{1}}_{1s,\perp}-I^{a_{1}}_{2c,\perp}-2I^{a_{1}}_{2s,\perp}).\label{DBR2}
\end{eqnarray}
\begin{eqnarray}
\frac{d\Gamma \left(B\to a_{1}(\to\rho\pi)\mu^{+}\mu^{-}\right)}{dq^{2}}&=&
\frac{d\Gamma \left(B\to a_{1}(\to\rho_{||}\pi)\mu^{+}\mu^{-}\right)}{dq^{2}}+\frac{d\Gamma \left(B\to a_{1}(\to\rho_{\perp}\pi)\mu^{+}\mu^{-}\right)}{dq^{2}}.\notag
\\
\label{DBRtotal}
\end{eqnarray}
\textbf{(ii) Lepton forward-backward asymmetry:}
From the full angular distribution Eq. \eqref{fullad1}, the integration over $\cos\theta_V = [-1,1]$, and $\phi = [0,2\pi]$, gives the double differential decay rate $\left(\frac{d^{2}\Gamma_{\parallel(\perp)}}{dq^{2}d\cos\theta_{\ell}}\right)$, where $\Gamma_{\parallel(\perp)}\equiv\Gamma \left(B\to a_{1}(\to\rho_{\parallel(\perp)}\pi)\mu^{+}\mu^{-}\right)$. The lepton forward-backward asymmetry corresponding to $\theta_\ell$ can be obtained from these polarized double differential decay rates as
\begin{eqnarray}
A_{\text{FB}}^{a_{1}}\left(q^2\right) &\equiv& \left[\int_{0}^1 d\cos\theta_\ell\frac{d^2\Gamma_{\parallel(\perp)}}{dq^2d\cos\theta_{\ell}}-\int_{-1}^0 d\cos\theta_\ell\frac{d^2\Gamma_{\parallel(\perp)}}{dq^2d\cos\theta_{\ell}}\right]\bigg/{\frac{d\Gamma_{\parallel(\perp)}}{dq^2}},\label{FBexp}
\end{eqnarray} 
which in terms of the angular coefficient functions is given by,
\begin{eqnarray}
A_{\text{FB}}^{a_{1}}(q^{2})=\frac{3\left(I^{a_{1}}_{6c,\perp}+2I^{a_{1}}_{6s,\perp}\right)}{2(3I^{a_{1}}_{1c,\perp}+6I^{a_{1}}_{1s,\perp}-I^{a_{1}}_{2c,\perp}-2I^{a_{1}}_{2s,\perp})}=\frac{6I^{a_{1}}_{6s,||}}{2(3I^{a_{1}}_{1c,||}+6I^{a_{1}}_{1s,||}-I^{a_{1}}_{2c,||}-2I^{a_{1}}_{2s,||})}.\label{FB}
\end{eqnarray}
\textbf{(iii) Longitudinal helicity fraction:}
From the full angular distribution Eq. \eqref{fullad1}, the integration over $\cos\theta_l = [-1,1]$, and $\phi = [0,2\pi]$, gives the double differential decay rate $\left(\frac{d^{2}\Gamma_{\parallel(\perp)}}{dq^{2}d\cos\theta_{V}}\right)$. The longitudinal helicity fraction of the decay  $B\to a_{1}\mu^{+}\mu^{-}$, when $a_{1}$ meson is longitudinally polarized can be defined as,
\begin{eqnarray}
f_{L}^{a_1}(q^{2})=\frac{\int_{-1}^{1}\frac{d^{2}\Gamma_{||}}{dq^{2}d\cos\theta_{V}}\left(\frac{5}{2}\cos^2\theta_V-\frac{1}{2}\right)d\cos\theta_{V}}{{d\Gamma \left(B\to a_{1}(\to\rho_{||}\pi)\mu^{+}\mu^{-}\right)}/{dq^{2}}}=\frac{\int_{-1}^{1}\frac{d^{2}
\Gamma_{\perp}}{dq^{2}d\cos\theta_{V}}\left(2-5\cos^2\theta_V\right)d\cos\theta_{V}}{{d\Gamma \left(B\to a_{1}(\to\rho_{\perp}\pi)\mu^{+}\mu^{-}\right)}/{dq^{2}}},
\end{eqnarray}
which in terms of the angular coefficient functions is given by,
\begin{eqnarray}
f_{L}^{a_1}(q^{2})=\frac{3I^{a_{1}}_{1c,||}-I^{a_{1}}_{2c,||}}{3I^{a_{1}}_{1c,||}+6I^{a_{1}}_{1s,||}-I^{a_{1}}_{2c,||}-2I^{a_{1}}_{2s,||}}=\frac{(6I^{a_{1}}_{1s,\perp}-2I^{a_{1}}_{2s,\perp})-(3I^{a_{1}}_{1c,\perp}-I^{a_{1}}_{2c,\perp})}{3I^{a_{1}}_{1c,\perp}+6I^{a_{1}}_{1s,\perp}-I^{a_{1}}_{2c,\perp}-2I^{a_{1}}_{2s,\perp}}.
\end{eqnarray}
\textbf{(iv) Normalized angular observables:}
We introduce the normalized angular observables
\begin{eqnarray}
    \langle \widehat{I}^{a_{1}}_{n\lambda, \parallel\left(\perp\right)}\rangle  =  \frac{\mathcal{B}(a_{1}\to \rho_{\|(\perp)}\pi)I^{a_1}_{n\lambda,\parallel\left(\perp\right)}}{d\Gamma\left(B\to a_{1}(\to\rho\pi)\mu^{+}\mu^{-}\right)/dq^2}.\label{A-Coeffincients}
\end{eqnarray}
\textbf{(v) Binned normalized angular observables:}
\begin{eqnarray}
\langle \widehat{I}^{a_{1}}_{n\lambda, \parallel\left(\perp\right)}\rangle_{\left[q^{2}_{\text{min}},q^{2}_{\text{max}}\right]}=\frac{\int^{q^{2}_{\text{max}}}_{q^{2}_{\text{min}}}\mathcal{B}(a_{1}\to \rho_{\|(\perp)}\pi)I^{a_1}_{n\lambda,\parallel\left(\perp\right)}\,dq^2}{\int^{q^{2}_{\text{max}}}_{q^{2}_{\text{min}}}(d\Gamma\left(B\to a_{1}(\to\rho\pi)\mu^{+}\mu^{-}\right)/dq^2)dq^2}.\label{BRa1perp}
\end{eqnarray}
To compute the branching ratios $\mathcal{B}(a_{1}\to \rho_{\|(\perp)}\pi)$ given in Eq. (\ref{BRa1perp}), one needs the amplitude of the decay whose expression is given as follows \cite{Colangelo:2019axi}
\begin{eqnarray}
\langle\rho(p_{\rho},\eta)\pi(p_{\pi})|a_{1}(k,\overline\epsilon)\rangle=g_{1}(\overline\epsilon.\eta)(k.p_{\rho})+g_{2}(\overline\epsilon.p_{\rho})(k.\eta),\label{SAmp}
\end{eqnarray}
where $g_{1},g_{2}$ are strong coupling constants and $\overline\epsilon$, $\eta$ are the polarization of $a_{1}$ and $\rho$ meson. 

The form of $\mathcal{B}(a_{1}\to \rho_{\|(\perp)}\pi)$ for longitudinal and transverse $\rho$ meson can be written as,
\begin{eqnarray}
\mathcal{B}(a_{1}\to \rho_{\|(\perp)}\pi)=\frac{1}{\Gamma_{a_{1}}}\frac{|\vec{p}_{\rho}|}{24\pi m^{2}_{a_{1}}}\Gamma_{\|(\perp)},\label{decay2}
\end{eqnarray}
where $|\vec{p}_{\rho}|=\frac{1}{2m_{a_{1}}}\sqrt{\lambda(m^{2}_{a_{1}},m^{2}_{\rho},m^{2}_{\pi})}$, and
\begin{eqnarray}
\Gamma_{\|}=\frac{m^{2}_{a_{1}}}{m^{2}_{\rho}}[(m^{2}_{\rho}+|\vec{p}_{\rho}|^{2})g_{1}+|\vec{p}_{\rho}|^{2}g_{2}]^{2},\label{gammapar}\\
\Gamma_{\perp}=2g^{2}_{1}m^{2}_{a_{1}}|\vec{p}_{\rho}|^{2}\left(1+\frac{m^{2}_{\rho}}{|\vec{p}_{\rho}|^{2}}\right).\label{gammaperp}
\end{eqnarray}
The coupling constants $g_{1}$ and $g_{2}$ can be related through the amplitude given in \cite{Roca:2003uk}
\begin{eqnarray}
\langle\rho(p_{\rho},\eta)\pi(p_{\pi})|a_{1}(k,\overline\epsilon)\rangle=-\frac{2\lambda_{a_{1}\rho\pi}}{m_{\rho}m_{a_{1}}}\left[(k.p_{\rho})(\overline\epsilon.\eta)-(\overline\epsilon.p_{\rho})(k.\eta)\right]\label{amp3}
\end{eqnarray}
Comparing Eq. (\ref{SAmp}), and (\ref{amp3}) gives $g_{1}=-g_{2}=\frac{2\lambda_{a_{1}\rho\pi}}{m_{\rho}m_{a_{1}}}$. Using the numerical values of masses from\cite{ParticleDataGroup:2022pth}, one gets $\mathcal{B}(a_{1}\to \rho_{\|}\pi)=17.2\%$, and $\mathcal{B}(a_{1}\to \rho_{\perp}\pi)=43\%$. These values of branching ratios are used to analyze the above mentioned physical observables.
\section{Numerical Analysis}\label{NA}
\subsection{Input Parameters}
To analyze the signatures of family non-universal $Z^{\prime}$ gauge boson in the observables that belongs to $B\to(\rho,a_{1})\mu^{+}\mu^{-}$ decays, we use the input parameters such as hadronic transition form factors, which are calculated in the framework of Light cone sum rules (LCSR)  for the case of $B\to\rho(\pi\pi)\mu^{+}\mu^{-}$ decay \cite{Bharucha:2015bzk}, and in perturbative QCD (pQCD) approach for the case of $B\to a_{1}(\to\rho\pi)\mu^{+}\mu^{-}$ decay \cite{Li:2009tx}.

\begin{table*}[!htbp]
\begin{center}
\captionsetup{margin=2.0cm}
\caption{Masses of resonances of quantum numbers $J^{P}$ as represented for the parametrization of the form factors $F_{i}$ for $b\to d$ transition.}\label{tab:bestfitWC}
			\begin{tabular}{|clc|}
				\hline
				$F_{i}$  & $J^{P}$ &  $m^{b\to d}_{R,i}$ \\
				\hline
                $A_{0}$ &  $0^{-}$                                         &$5.279$       \\
                $T_{1},V$ & $1^{-}$                  &$5.325$        \\
                $T_{2},T_{23},A_{1},A_{12}$ & $1^{+}$      & $5.274$        \\
                \hline
            \end{tabular}
	\end{center}
 
\end{table*}
The combined fit of the simplified series expansion (SSE) parametrization to LCSR results for $B\to\rho(\pi\pi)\mu^{+}\mu^{-}$ are given as follows\cite{Bharucha:2015bzk}
\begin{eqnarray}
z(t)=\frac{\sqrt{t_{+}-t}-\sqrt{t_{+}-t_{0}}}{\sqrt{t_{+}-t}+\sqrt{t_{+}-t_{0}}},\label{SSE}
\end{eqnarray}
where, $t_{\pm}\equiv (m_{B}\pm m_{\rho})^{2}$ and $t_{0}\equiv t_{+}\left(1-\sqrt{1-\frac{t_{-}}{t_{+}}}\right)$.
We can write the expressions of the transition form factors for the decay $B\to\rho$ as,
\begin{eqnarray}
F_{i}(q^{2})=P_{i}(q^{2})\sum_{k}\alpha^{i}_{k}[z(q^{2})-z(0)]^{k},\label{FF}
\end{eqnarray}
where $P_{i}(q^{2})=\frac{1}{(1-\frac{q^{2}}{m^{2}_{R,i}})}$, is a simple pole corresponding to the first resonance in the spectrum. The resonance masses and the fit results for the SSE expansion coefficients in the fit to the LCSR computation for the decay $B\to\rho$  are presented in Table \ref{tab:bestfitWC} and Table \ref{FF table1}.
\begin{table*}[!htbp]
\centering
%\footnotesize
\captionsetup{margin=0.5cm}
 \caption{\small Fit results for the SSE expansion coefficients in the fit to the LCSR computation for the decay $B\to\rho$ \cite{Bharucha:2015bzk} decay.}\label{FF table1}
 \renewcommand{\arraystretch}{1.5}
    \scalebox{0.87}{ 
\begin{tabular}{|c|c|c|c|c|c|c|c|}
\hline
&$A_{0}$&$A_{1}$&$A_{12}$&$V$&$T_{1}$&$T_{2}$&$T_{23}$
\\ \hline
$\alpha_{0}$&$0.36\pm 0.04$&$0.26\pm 0.03$&$0.30\pm 0.03$&$0.33\pm 0.03$&$0.27\pm 0.03$&$0.27\pm 0.03$&$0.75\pm 0.08$\\ \hline
$\alpha_{1}$&$-0.83\pm 0.20$&$0.39\pm 0.14$&$0.76\pm 0.20$&$-0.86\pm 0.18$&$-0.74\pm 0.14$&$0.47\pm 0.13$&$1.90\pm 0.43$\\ \hline
$\alpha_{2}$&$1.33\pm 1.05$&$0.16\pm 0.41$&$0.46\pm 0.76$&$1.80\pm 0.97$& $1.45\pm 0.77$&$0.58\pm 0.46$&$2.93\pm 1.81$ \\
 \hline
\end{tabular}}
\end{table*}

 For $B\to a_{1}$ decay, the transition form factors were calculated in the framework of pQCD approach. The form factors that are involved in $B\to a_{1}$ decay, can be parametrized in the whole kinematical $q^{2}$ region as follows\cite{Li:2009tx},
 \begin{eqnarray}
 F(q^{2})=\frac{F(0)}{1-a(q^{2}/m^{2}_{B})+b(q^{2}/m^{2}_{B})^{2}}.
 \end{eqnarray}
 The numerical results for $B\to a_{1}$ decay at $q^{2}=0$, in the pQCD approach are presented in Table \ref{FF table}.
\begin{table*}[!htbp]
\centering
\captionsetup{margin=0.8cm}
\caption{\small The numerical values of transition form factors for $B\to a_{1}$ decay at $q^{2}=0$, and the fitted parameters $a$ and $b$ \cite{Li:2009tx}.}\label{FF table}
 \renewcommand{\arraystretch}{1.5}
    \scalebox{1.0}{
\begin{tabular}{|c|c|c|c|c|c|c|c|}
\hline
&$A$&$V_{0}$&$V_{1}$&$V_{2}$&$T_{1}$&$T_{2}$&$T_{3}$
\\ \hline
$F(0)$ & $0.26^{+0.09}_{-0.09}$ &   $0.34^{+0.16}_{-0.17}$  & $0.43^{+0.15}_{-0.15}$  &   $0.13^{+0.03}_{-0.04}$  & $0.34^{+0.13}_{-0.13}$ & $0.34^{+0.13}_{-0.13}$ & $0.37^{+0.17}_{-0.12}$\\ \hline
$a$ &  $1.72^{+0.05}_{-0.05}$  &  $1.73^{+0.05}_{-0.06}$   &   $0.75^{+0.05}_{-0.05}$  &  $--$  & $1.60^{+0.06}_{-0.05}$ & $0.71^{+0.07}_{-0.05}$ & $1.60^{+0.06}_{-0.05}$\\ \hline
 $b$ &   $0.66^{+0.07}_{-0.06}$ &  $0.66^{+0.06}_{-0.08}$ & $-0.12^{+0.05}_{-0.02}$ & $--$ & $0.53^{+0.06}_{-0.04}$ & $-0.16^{+0.03}_{-0.02}$ & $0.53^{+0.06}_{-0.04}$ \\
 \hline
 \end{tabular}}
\end{table*}
The numerical values of Wilson coefficients in the SM, evaluated at the renormalization scale $\mu\sim m_{b}$ \cite{Blake:2016olu}, are presented in Table \ref{wc table}.
%The other input parameters are the Wilson coefficients, the various masses of quarks and leptons, the branching fractions of $\mathcal{B}(D_{s}^{\ast}\to D_{s}\gamma)$, $\mathcal{B}(D_{s}^{\ast}\to D_{s}\pi)$, the values of CKM matrix and the life time of $B_{c}$ meson. The numerical values of these parmeters are presented in Table \ref{wc table} and Table \ref{input}.
\begin{table*}[!htbp]
\centering
\captionsetup{margin=0.30cm}
\caption{\small The numerical values of the SM Wilson coefficients up to NNLL accuracy, evaluated at the renormalization scale $\mu\sim m_{b}$ \cite{Blake:2016olu}.}
\label{wc table}
\begin{tabular}{|c|c|c|c|c|c|c|c|c|c|}
\hline
$C_{1}$&$C_{2}$&$C_{3}$&$C_{4}$&$C_{5}$&$C_{6}$&$C_{7}$&$C_{8}$&$C_{9}$&$C_{10}$
\\ \hline
  $-0.294$ &   $1.017$  & $-0.0059$  &   $-0.087$  &
  $0.0004$  &  $0.0011$   &   $-0.324$  &  $-0.176$  &
    $4.114$  &  $-4.193$ \\
\hline
\end{tabular}
\end{table*}
In order to analyze the normalized angular observables and the other observables such as differential branching ratios, forward-backward asymmetry, longitudinally polarized final state vector and axial vector mesons in $B\to\rho(\to\pi\pi)\mu^{+}\mu^{-}$ and $B\to a_{1}(\to\rho_{\perp,\parallel})\mu^{+}\mu^{-}$ decays, respectively, in the framework of the family non-universal $Z^{\prime}$ model, the numerical values of $Z^{\prime}$ model parameters are collected in Table \ref{TZP1}.
\begin{table}[tbh]
\captionsetup{margin=3.0cm}
\caption{The numerical values of lepton and quark coupling in $Z^{\prime}$ model are collected from Ref. \cite{Nayek:2018rcq}.}\label{TZP1}
\centering
\begin{tabular}{|c|c|c|c|c|}
\hline
Scenarios & $|B^{L}_{db}|\times 10^{-3}$ & $\phi_{db}$ in degrees &  $S_{LR}$  & $D_{LR}$  \\ \hline
S1 &$0.16\pm 0.08$ & $-33\pm 45$ & $0.08$  & $-0.02$\\ \hline
S2 &$0.12\pm 0.03$ & $-23\pm 21$ & $0.08$ & $-0.02$ \\ \hline
\end{tabular}
%\begin{tabular}{|c|c|c|}
%\hline
%Scenarios & $|B^{L}_{db}|\times 10^{-3}$ & $\phi_{db}$ in degrees  \\ \hline
%S1 &$0.16\pm 0.08$ & $-33\pm 45$ \\ \hline
%S2 &$0.12\pm 0.03$ & $-23\pm 21$\\ \hline
%\end{tabular}
\end{table}

To gauge the effects of family non-universal $Z^{\prime}$ model in the above mentioned physical observables, the numerical values of coupling $B^{L}_{db}$ and the weak phase $\phi_{db}$ presented in Table \ref{TZP1} are fixed, and are constrained from $B^{0}_{q}-\bar{B}^{0}_{q}$ mixing\cite{Chang:2009tx}. The scenarios S1(S2) given in Table \ref{TZP1} represents the constraints from UTfit collaboration on the parameters $C_{B_{q}}$ and $\phi_{B_{q}}$\cite{Bona:2009tn}. The explicit form of $C_{B_{q}}$ and $\phi_{B_{q}}$  is given by,
\begin{eqnarray}
C_{B_{q}}e^{2i\phi_{B_{q}}}\equiv\frac{\langle B_{q}|H^{\text{full}}_{\text{eff}}|\bar{B}_{q}\rangle}{\langle B_{q}|H^{\text{SM}}_{\text{eff}}|\bar{B}_{q}\rangle}
\end{eqnarray}
Using the maximum allowed values of the coupling constants given in Table \ref{TZP1}, the numerical values of the non-universal family $Z^{\prime}$ model Wilson coefficients $C_{9}^{Z^{\prime}}$ and $C_{10}^{Z^{\prime}}$ in scenarios S1 and S2 are presented in Table \ref{WCZP}.
\begin{table}[tbh]
\captionsetup{margin=3.0cm}
\caption{The  numerical values of family non-universal $Z^{\prime}$ model Wilson coefficients $C_{9}^{Z^{\prime}}$ and $C_{10}^{Z^{\prime}}$ in scenarios S1 and S2.}\label{WCZP}
\centering
%\begin{tabular}{|c|c|c|}
%\hline
% & S1 & S2  \\ \hline
%$C_{9}^{Z^{\prime}}$ &3.606 - 0.766$i$ & 2.303 + 0.080$i$ \\ \hline
%$C_{10}^{Z^{\prime}}$ &-0.901 + 0.191$i$ & -0.575 - 0.020$i$\\ \hline
%\end{tabular}
\begin{tabular}{|c|c|c|}
\hline
 & S1 & S2  \\ \hline
$C_{9}^{Z^{\prime}}$ &0.0000192 & 0.000012 \\ \hline
$C_{10}^{Z^{\prime}}$ &$-4.8\times 10^{-6}$ & $-3.0\times 10^{-6}$\\ \hline
\end{tabular}
\end{table}

\subsection{Phenomenological Analysis of the Physical Observables in \texorpdfstring{$B\to\rho (\to\pi\pi)\mu^{+}\mu^{-}$}{} and \texorpdfstring{$B\to a_{1}(\to\rho_{\parallel, \perp}\pi)\mu^{+}\mu^{-}$}{} Decays}
\begin{figure*}[t!]
\begin{tabular}{cc}
\hspace{0.0cm}($\mathbf{a}$)&\hspace{-0.9cm}($\mathbf{b}$)\\
\includegraphics[scale=0.54]{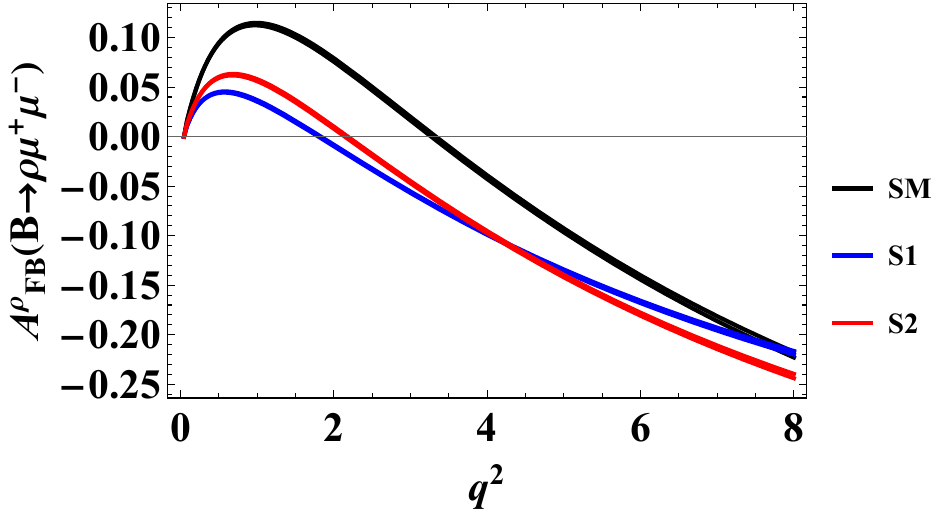}\ \ \
 &\hspace{-1.2cm}\ \ \ \includegraphics[scale=0.50]{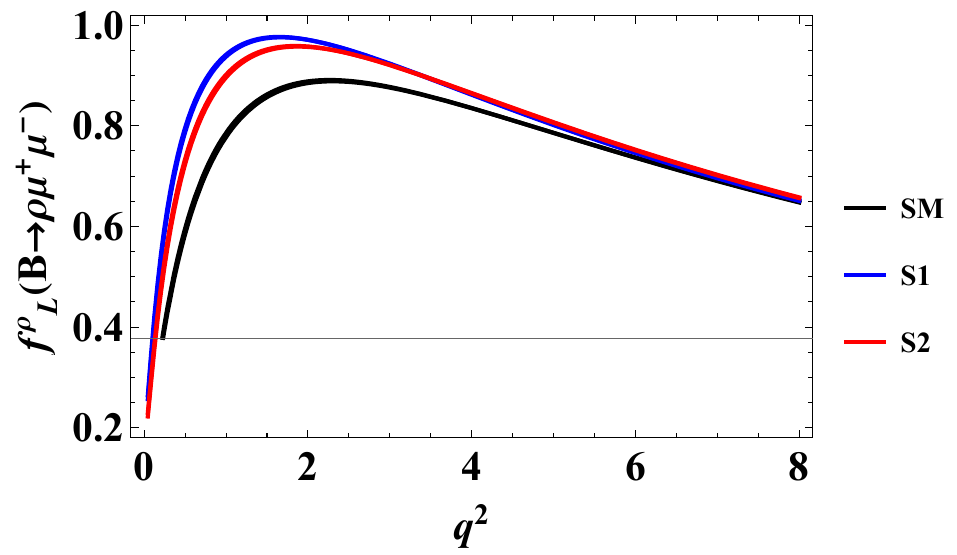}\\
\hspace{0.0cm}($\mathbf{c}$)&\hspace{-0.9cm}($\mathbf{d}$)\\
\includegraphics[scale=0.49]{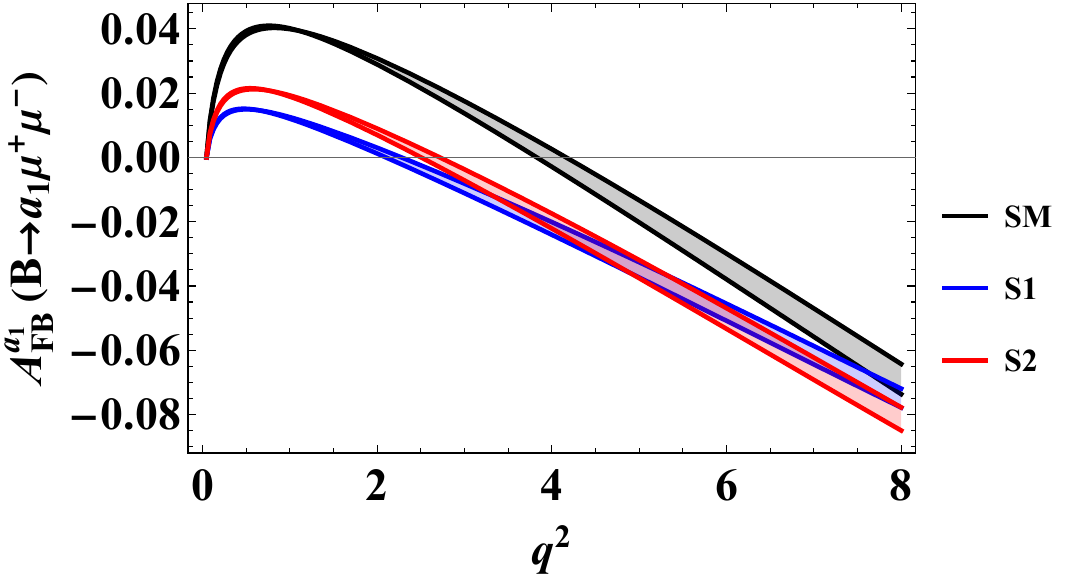}\ \ \
&\hspace{-1.2cm}\ \ \ \includegraphics[scale=0.46]{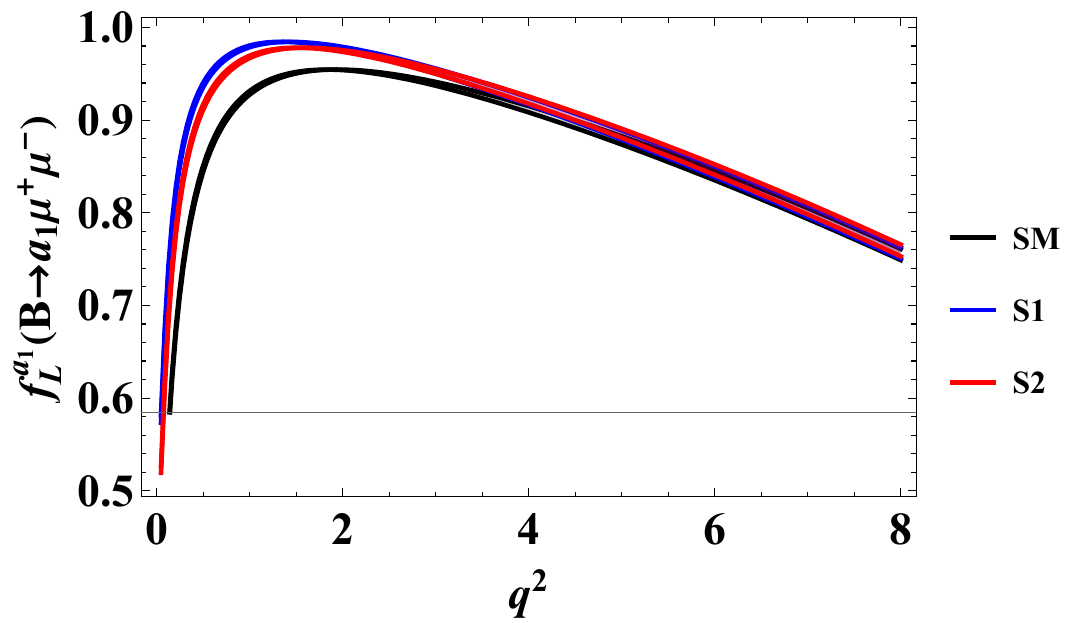}\\\
\end{tabular}
\caption{ (a) Lepton forward-backward asymmetry $A_{\text{FB}}^{\rho}$, (b) longitudinal polarization fraction of $\rho$ meson $f_{L}^{\rho}$, for the $B\to\rho\mu^{+}\mu^{-}$ decay, while (c) Lepton forward-backward asymmetry $A_{\text{FB}}^{a_1}$, (d) longitudinal polarization fraction of $a_1$ meson $f_{L}^{a_1}$, for the $B\to a_{1}\mu^{+}\mu^{-}$ decay in the SM and the two scenarios of the family non-universal $Z^{\prime}$ model.}
\label{FBAfL}
\end{figure*}

In this section, we present our phenomenological analysis of the family non-universal $Z^{\prime}$ model via physical observables constructed from the combination of different angular coefficients such as the differential branching ratios $\left(d\mathcal{B}/dq^{2}\right)$, lepton forward-backward asymmetry $(A_{\text{FB}})$, longitudinal polarization fraction $(f_L)$ of $\rho$ and $a_{1}$ mesons, in the $B\to\rho(\to\pi\pi)\mu^{+}\mu^{-}$ and $B\to a_{1}(\to\rho_{\parallel, \perp})\mu^{+}\mu^{-}$ decays, respectively.
\begin{figure*}[t!]
%\centering
\begin{tabular}{cc}
\hspace{0.6cm}($\mathbf{a}$)&\hspace{0.1cm}($\mathbf{b}$)\\
\includegraphics[scale=0.45]{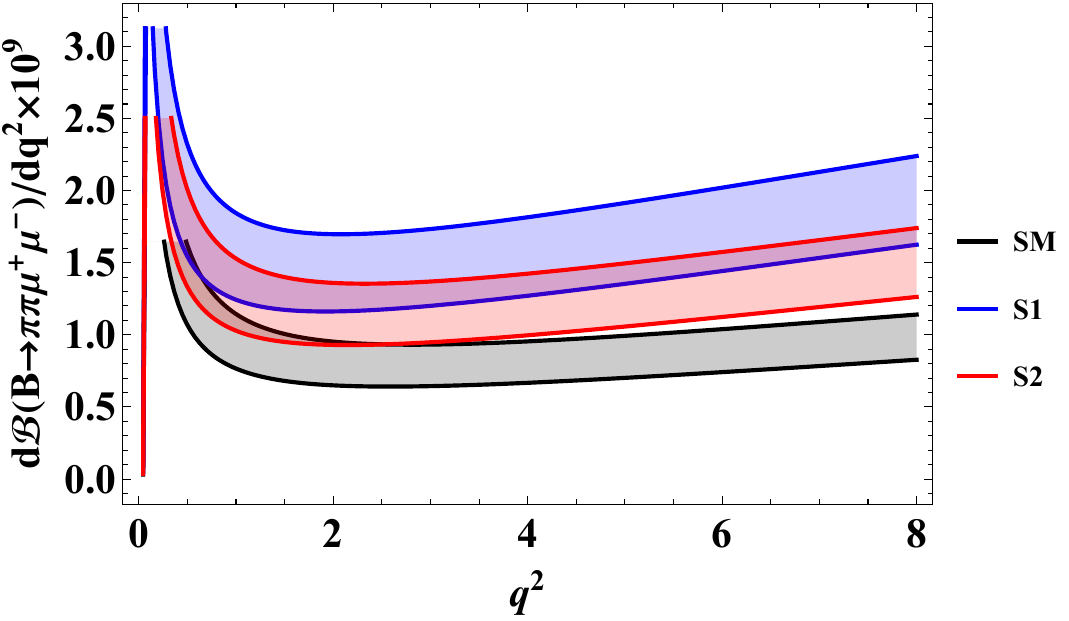}\ \ \
& \hspace{-1.1cm}\ \ \ \includegraphics[scale=0.44]{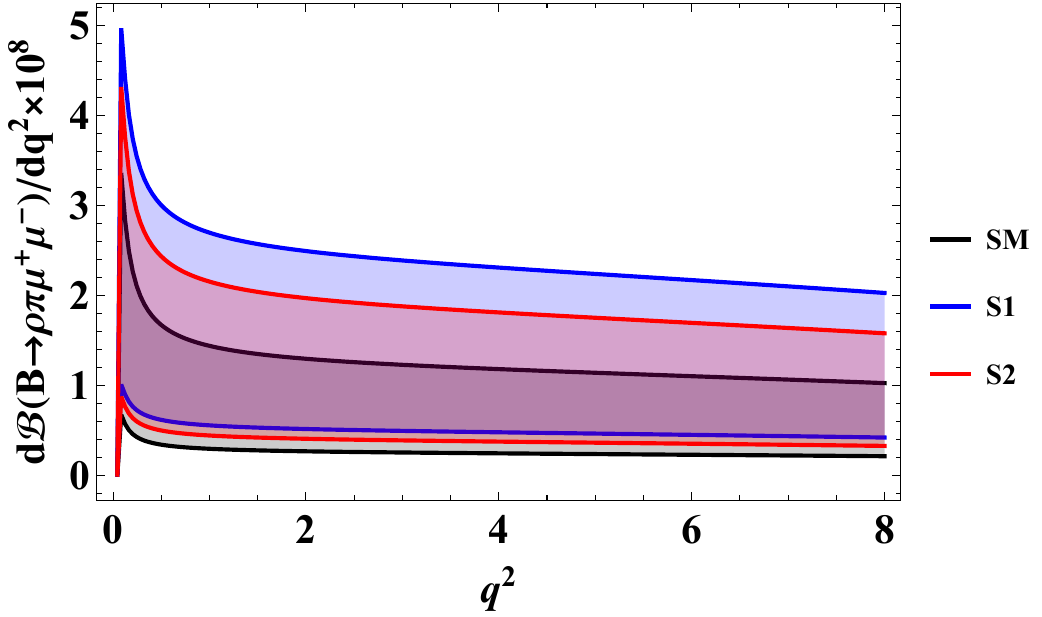}\\\
\hspace{0.6cm}($\mathbf{c}$)&\hspace{0.1cm}($\mathbf{d}$)\\
\includegraphics[scale=0.45]{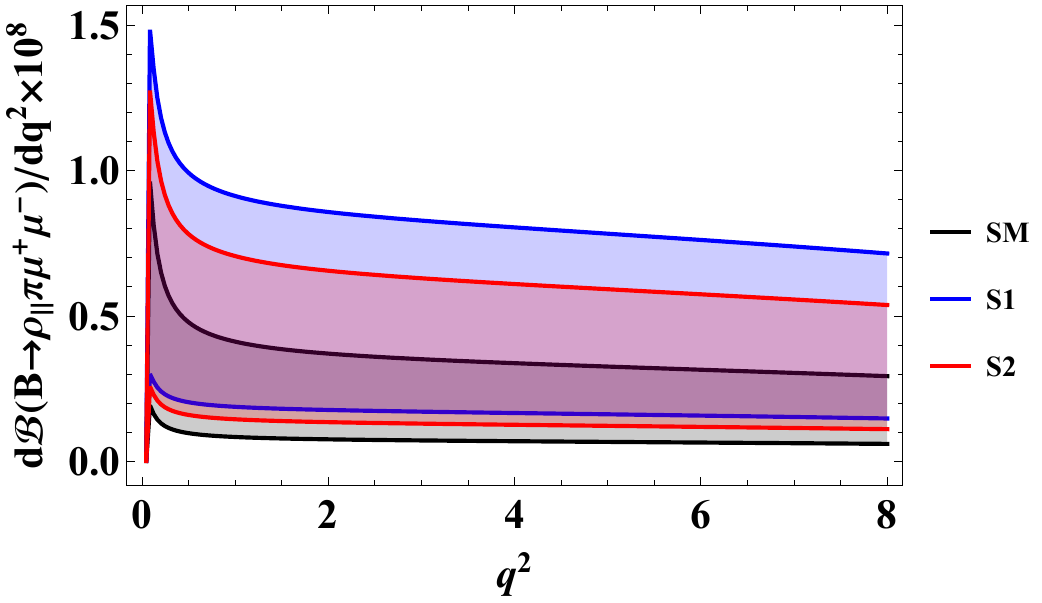}\ \ \
&\hspace{-1.1cm} \ \ \ \includegraphics[scale=0.43]{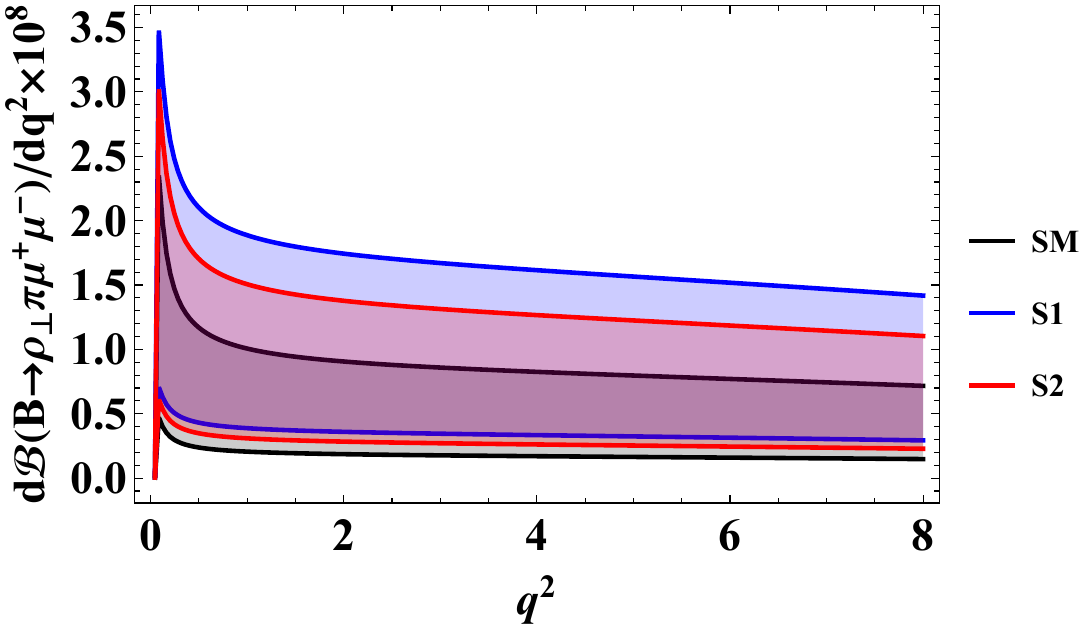}\\\
\end{tabular}
\caption{Differential branching ratio of the decay (a) $B\to\rho(\to\pi\pi)\mu^{+}\mu^{-}$, (b) $B\to a_{1}(\to\rho\pi)\mu^{+}\mu^{-}$, (c) $B\to a_{1}(\to\rho_{\parallel}\pi)\mu^{+}\mu^{-}$, and (d) $B\to a_{1}(\to\rho_{\perp}\pi)\mu^{+}\mu^{-}$, in the SM and the two scenarios of the family non-universal $Z^{\prime}$ model.}
\label{DBR}
\end{figure*}
The predicted numerical values of these observables, in different $q^2$ bins, for the SM as well as for the two different scenarios of family non-universal $Z^{\prime}$ model are given in Tables \ref{table:1}-\ref{table:8}, of appendix \ref{append1}. The listed errors in these tables originate mainly from the uncertainties of the form factors. Furthermore, in Figs. \ref{DBR}-\ref{FBAfL}, we have plotted the above mentioned physical observables as a function of $q^{2}$. Following are our predictions regarding the physical observables.
\begin{itemize}
    \item Figs. \ref{DBR}(a), \ref{DBR}(b), \ref{DBR}(c), and \ref{DBR}(d) depict the differential branching ratios $\frac{d\mathcal{B}}{dq^{2}}$ of $B\to\rho(\to\pi\pi)\mu^{+}\mu^{-}$, $B\to a_{1}(\to\rho\pi)\mu^{+}\mu^{-}$, $B\to a_{1}(\to\rho_{\parallel}\pi)\mu^{+}\mu^{-}$ and $B\to a_{1}(\to\rho_{\perp}\pi)\mu^{+}\mu^{-}$ decays, respectively, in the framework of the SM and the scenarios S1 and S2 of the family non-universal $Z^{\prime}$ model. Fig. \ref{DBR}(a) indicates that, after including uncertainties of the form factors, the predictions of differential branching ratio in two scenarios of the family non-universal $Z^{\prime}$ model deviate from the SM predictions such that they show a tendency towards higher values of differential branching ratios as compared to the SM expectations, which is more dominant in Scenario S1. In Figs. \ref{DBR}(b), \ref{DBR}(c), and \ref{DBR}(d), our results show similar trend of higher values of differential branching ratios in two scenarios of the family non-universal $Z^{\prime}$ model, however in this case SM predictions largely overlap with the scenarios of the family non-universal $Z^{\prime}$ model due to the larger uncertainties originating from the form factors. At very low $q^2 < 1$ GeV$^2$, these differential branching ratios are dominated by the SM magnetic dipole Wilson coefficient $C_{7}$, so at $q^2\to 0$, the results of differential branching ratios indicate singularity corresponding to photon pole.   
\item Figs. \ref{FBAfL} represents the lepton forward-backward asymmetry and longitudinally polarized final state mesons as a function of $q^{2}$ in the framework of the SM and the family non-universal $Z^{\prime}$ model for $B\to\rho\mu^{+}\mu^{-}$ and $B\to a_{1}\mu^{+}\mu^{-}$ decays. Figs. \ref{FBAfL}(a) and fig. \ref{FBAfL}(c) show the zero position of the $A_{\text{FB}}(q^{2})$ for $B\to\rho\mu^{+}\mu^{-}$ and $B\to a_{1}\mu^{+}\mu^{-}$ decays in the framework of SM and two scenarios S1 and S2 of family non-universal $Z^{\prime}$ model, respectively. Both scenarios S1 and S2 of the family non-universal $Z^{\prime}$ are shifted towards left as compared to the SM prediction, as a result, the zero crossing of $A_{\text{FB}}^{\rho}(q^{2})$ and $A_{\text{FB}}^{a_1}(q^{2})$ for both scenarios is distinguishable from the SM expectation. Also overall $q^2$ predictions of $A_{\text{FB}}^{\rho}(q^{2})$ and $A_{\text{FB}}^{a_1}(q^{2})$ in the S1 and S2 scenarios for the whole $q^2$ range show discrimination from the SM results.
%On the other hand, in Fig. \ref{FBAfL}(c), forward-backward asymmetry $A_{\text{FB}}^{a_1}(q^{2})$, predictions in scenario S1 largely overlap with the SM results, due to large error bands originating from the uncertainties of the form factors, whereas, scenario S2 is distinct due to the smaller error band.
%Another interesting observable that is used to analyze the signature of the family non-universal $Z^{\prime}$ model is longitudinally polarized final state meson as a function of $q^{2}$.
In Fig. \ref{FBAfL}(b), we have plotted the longitudinal helicity fraction $f_{L}^{\rho}$, for the decay $B\to\rho\mu^{+}\mu^{-}$ decay in the SM framework and the two scenarios S1, S2 of the family non universal $Z^{\prime}$ model. The predictions in S1 and S2 scenarios, clearly show a departure from the SM result in the region $q^{2}=(0.1-5)$ 
$\text{GeV}^{2}$. However, for the case of $B\to a_{1}\mu^{+}\mu^{-}$ decay, longitudinal helicity fraction $f_{L}^{a_1}$, shows deviation from the SM result in the region $q^{2}=(0.1-3)$ $\text{GeV}^{2}$, only as shown in fig. \ref{FBAfL}(d). From the values of the NP Wilson coefficients reported in Table \ref{WCZP}, it is evident that the SM Wilson coefficients get more pronounced deviations in the presence of scenario S1, compared to scenario S2, which also reflects in the overall results of observables such that scenario S2 appears closer to the SM.
\end{itemize}
\subsection{Phenomenological analysis of the Angular coefficients in \texorpdfstring{$B\to\rho(\to\pi\pi)\mu^{+}\mu^{-}$}{} and \texorpdfstring{$B\to a_{1}(\to\rho_{\parallel, \perp}\pi)\mu^{+}\mu^{-}$}{} Decays}
\begin{figure*}[t!]
\begin{tabular}{cc}
\hspace{0.0cm}($\mathbf{a}$)&\hspace{-0.5cm}($\mathbf{b}$)\\
\includegraphics[scale=0.53]{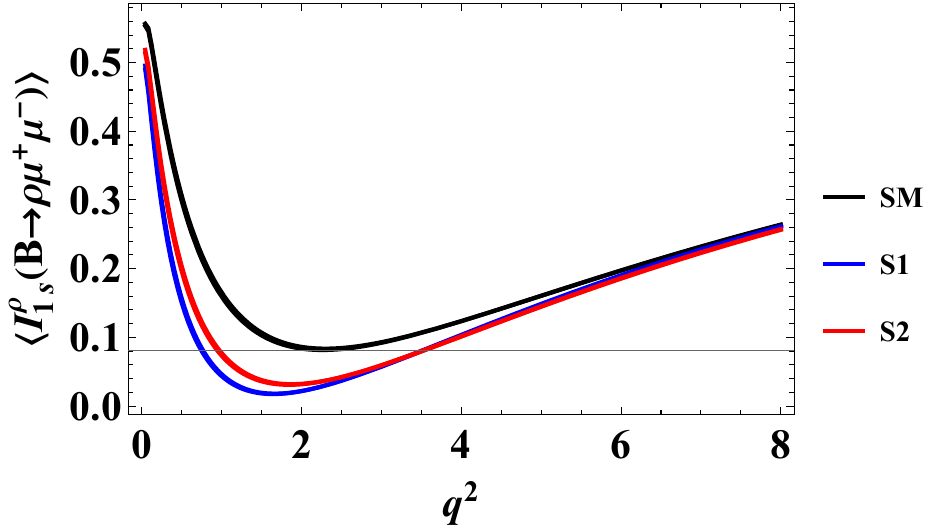}\ \ \
&\hspace{-1.0cm} \ \ \ \includegraphics[scale=0.53]{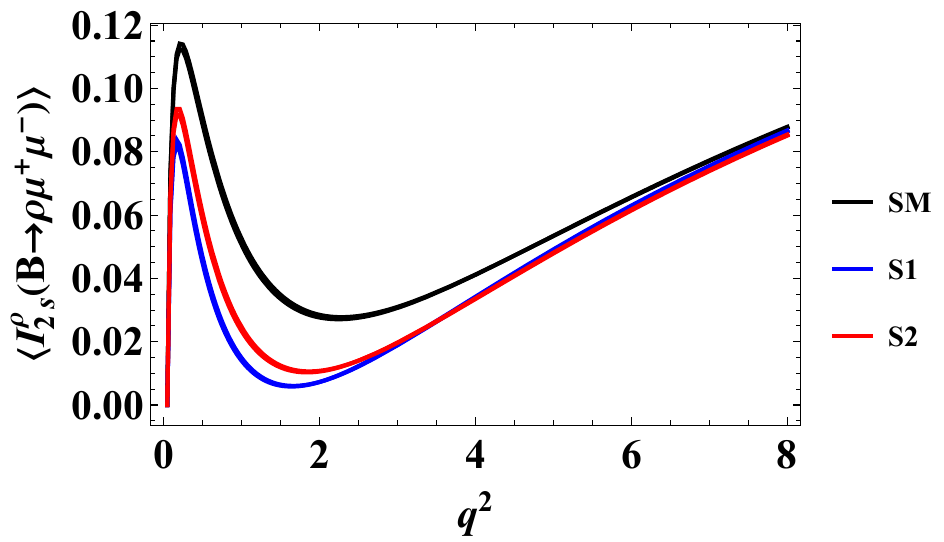}\\\
\hspace{0.0cm}($\mathbf{c}$)&\hspace{-0.5cm}($\mathbf{d}$)\\
\includegraphics[scale=0.53]{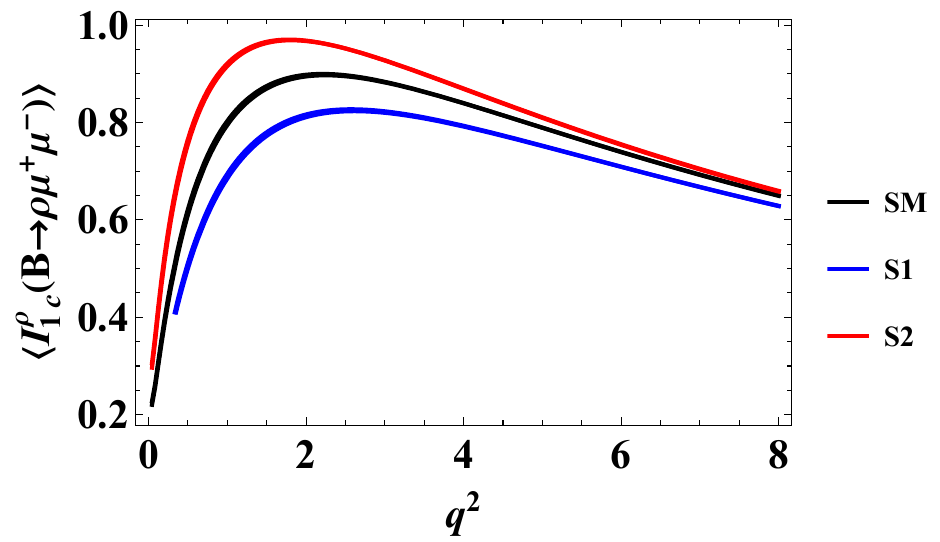}\ \ \
&\hspace{-1.0cm} \ \ \ \includegraphics[scale=0.53]{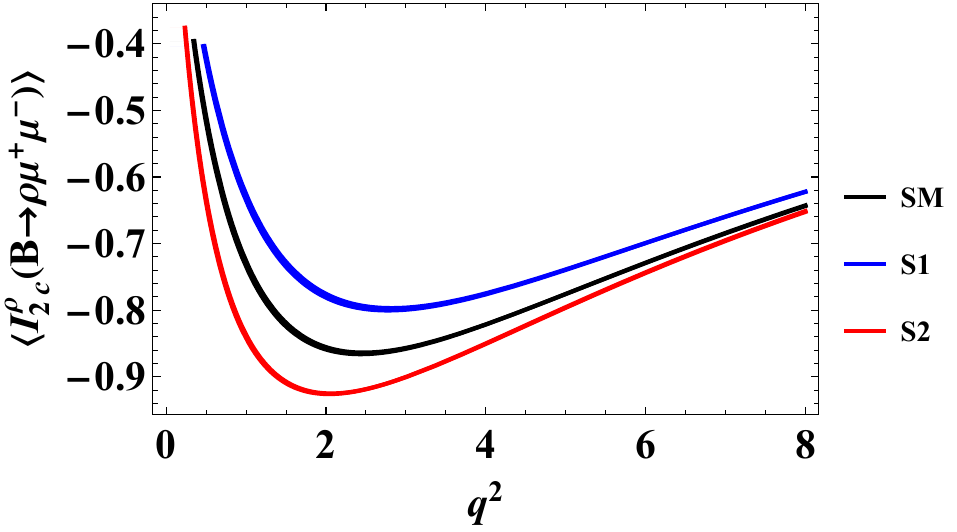}\\\
\hspace{0.0cm}($\mathbf{e}$)&\hspace{-0.5cm}($\mathbf{f}$)\\
\includegraphics[scale=0.54]{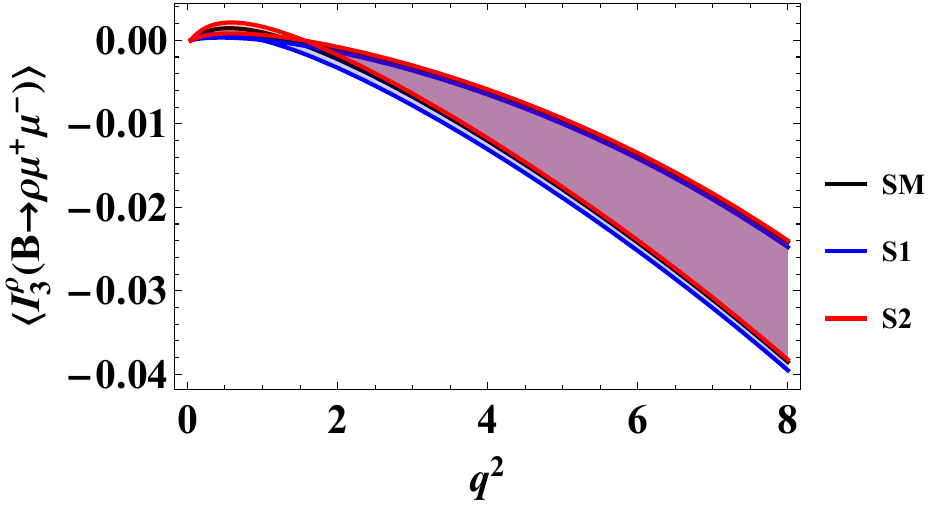}\ \ \
&\hspace{-1.0cm} \ \ \ \includegraphics[scale=0.54]{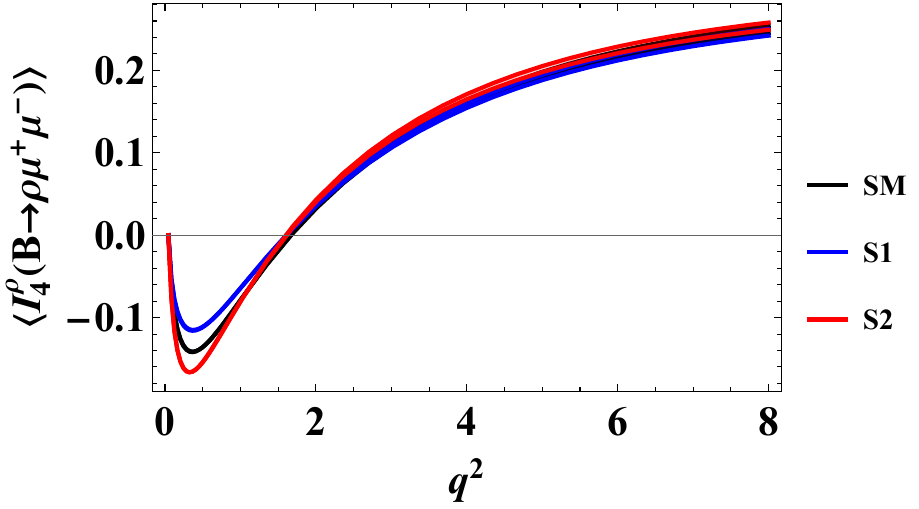}\\\
\hspace{0.0cm}($\mathbf{g}$)&\hspace{-0.5cm}($\mathbf{h}$)\\
\includegraphics[scale=0.53]{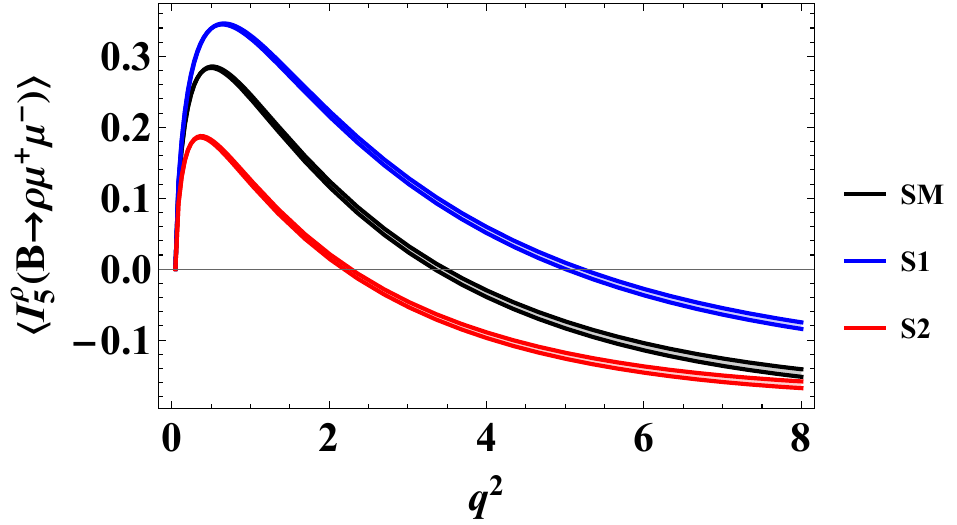}\ \ \
&\hspace{-1.0cm} \ \ \ \includegraphics[scale=0.53]{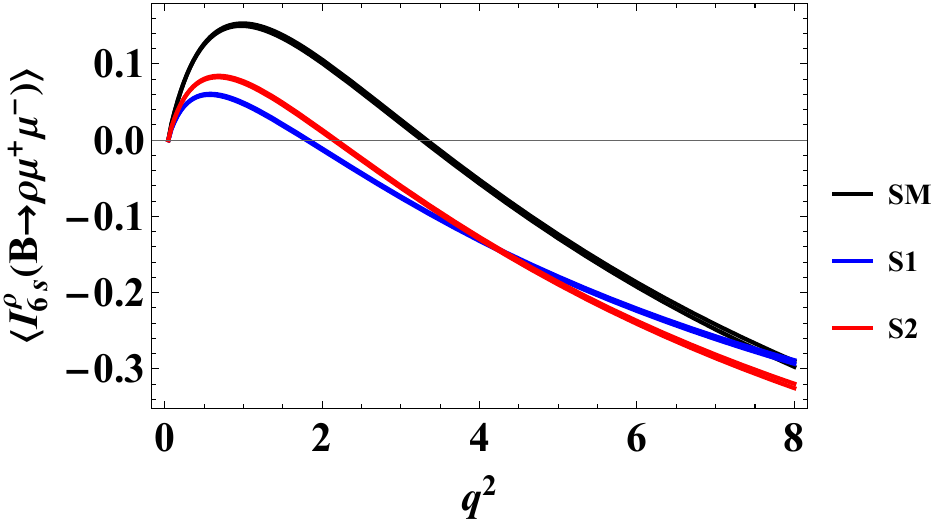}\\\
\end{tabular}
\caption{Angular observables $\langle I^{\rho}_{1s}\rangle, \langle I^{\rho}_{2s}\rangle, \langle I^{\rho}_{1c}\rangle, \langle I^{\rho}_{2c}\rangle, \langle I^{\rho}_{3}\rangle, \langle I^{\rho}_{4}\rangle, \langle I^{\rho}_{5}\rangle$, and $\langle I^{\rho}_{6s}\rangle$ for the decay $B\to\rho(\to\pi\pi)\mu^{+}\mu^{-}$, in the SM and the two scenarios of the family non-universal $Z^{\prime}$ model.}
\label{I1rho}
\end{figure*}
\begin{figure*}[b!]
\begin{tabular}{cc}
\hspace{0.0cm}($\mathbf{a}$)&\hspace{-0.5cm}($\mathbf{b}$)\\
\includegraphics[scale=0.45]{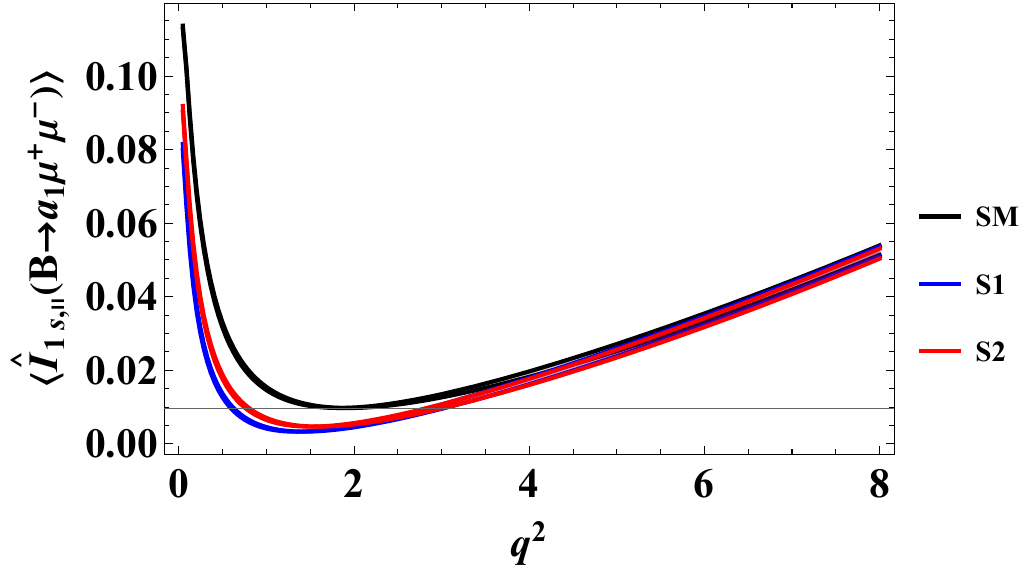}\ \ \
&\hspace{-1.2cm} \ \ \ \includegraphics[scale=0.45]{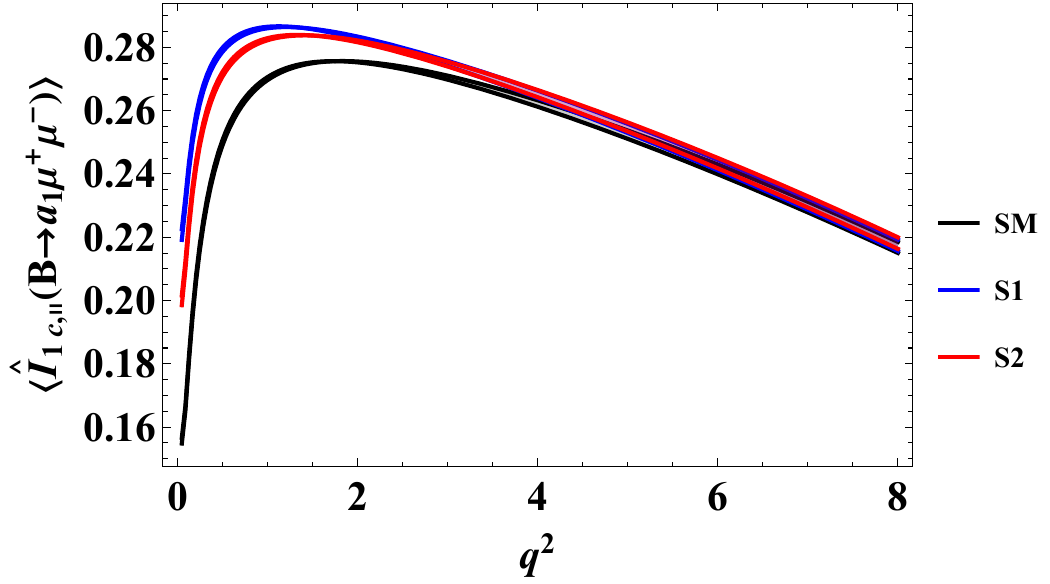}\\\
\hspace{0.0cm}($\mathbf{c}$)&\hspace{-0.5cm}($\mathbf{d}$)\\
\includegraphics[scale=0.45]{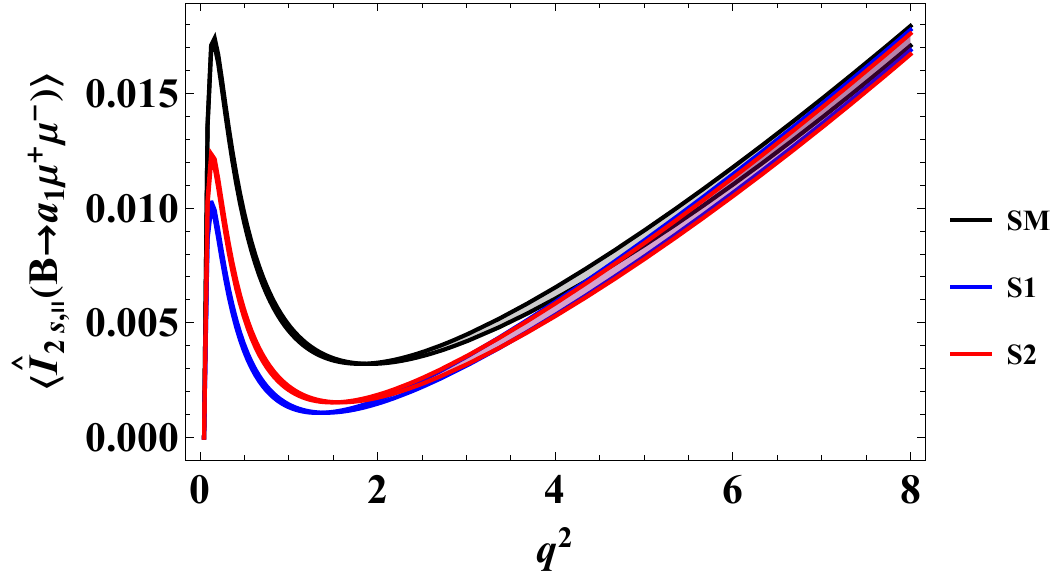}\ \ \
&\hspace{-1.2cm} \ \ \ \includegraphics[scale=0.45]{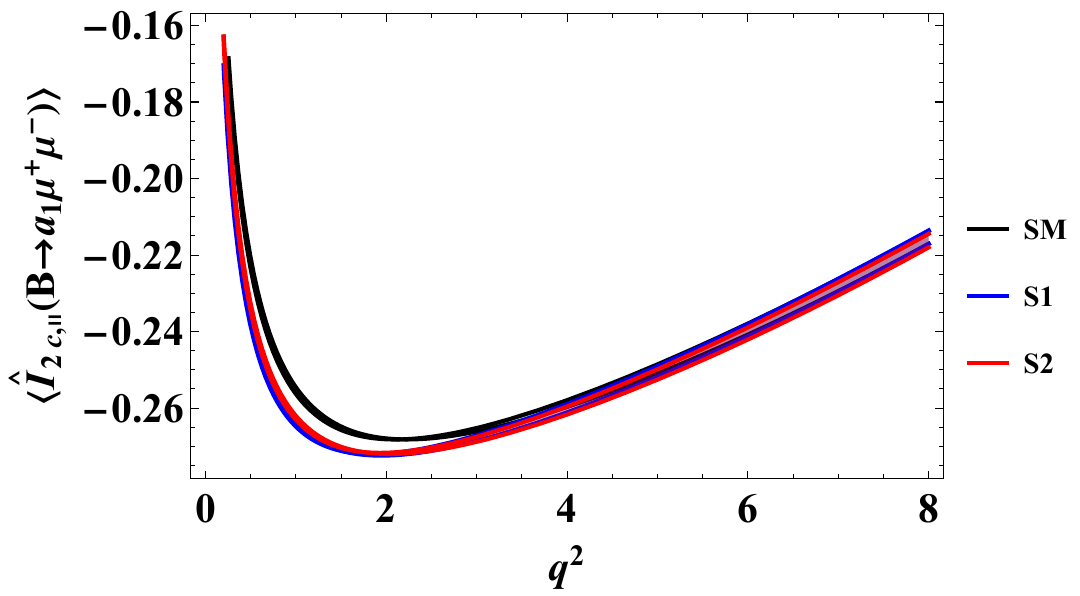}\\\
\hspace{0.0cm}($\mathbf{e}$)&\hspace{-0.5cm}($\mathbf{f}$)\\
\includegraphics[scale=0.45]{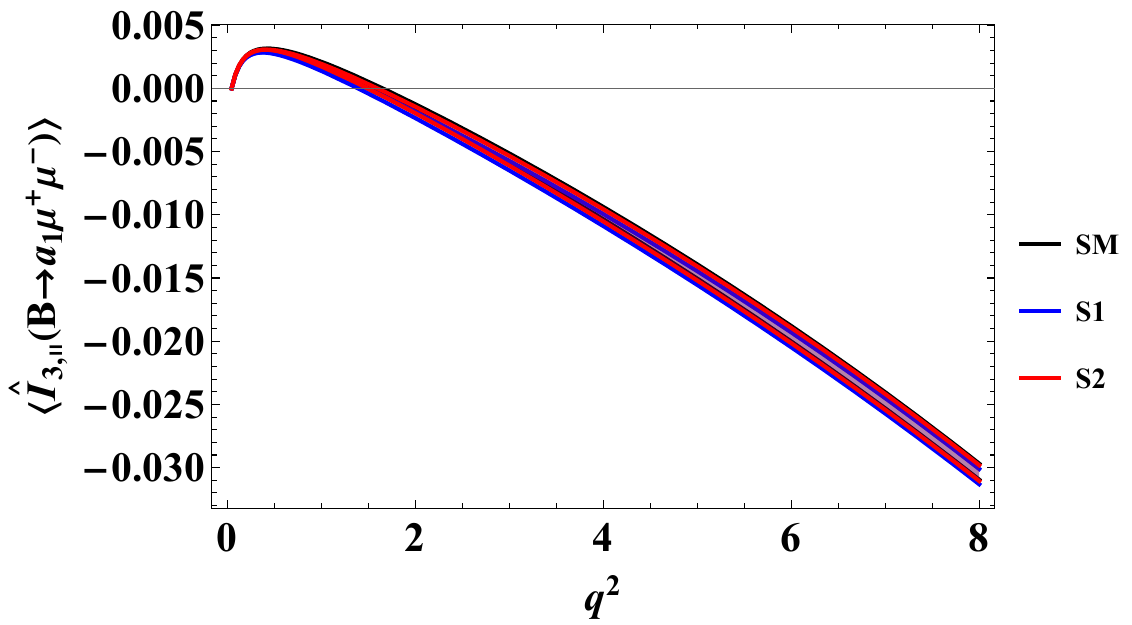}\ \ \
&\hspace{-1.2cm} \ \ \ \includegraphics[scale=0.45]{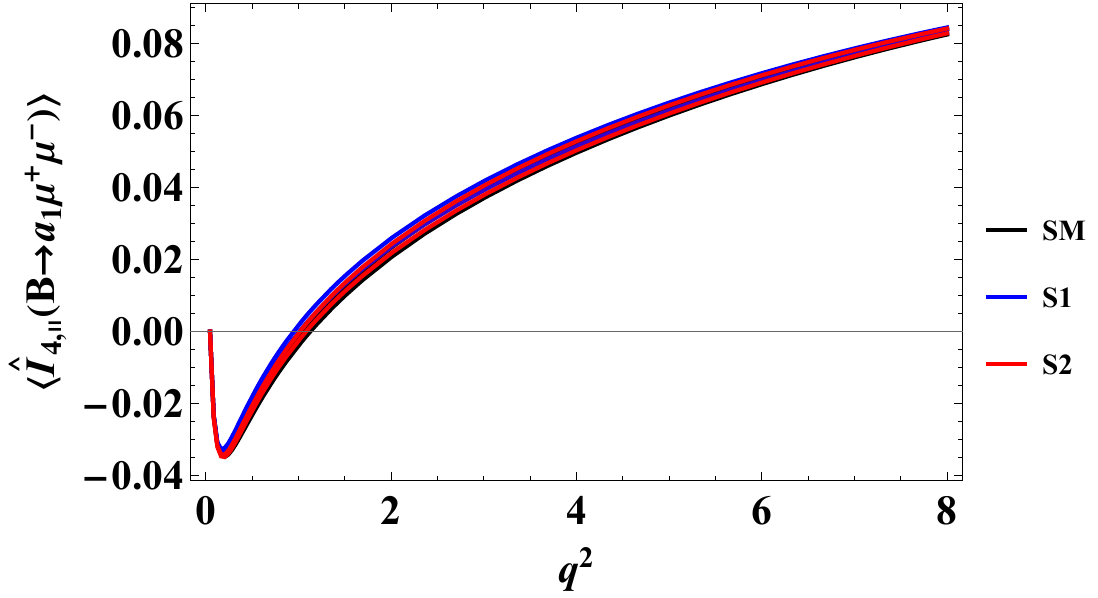}\\\
\hspace{0.0cm}($\mathbf{g}$)&\hspace{-0.5cm}($\mathbf{h}$)\\
\includegraphics[scale=0.45]{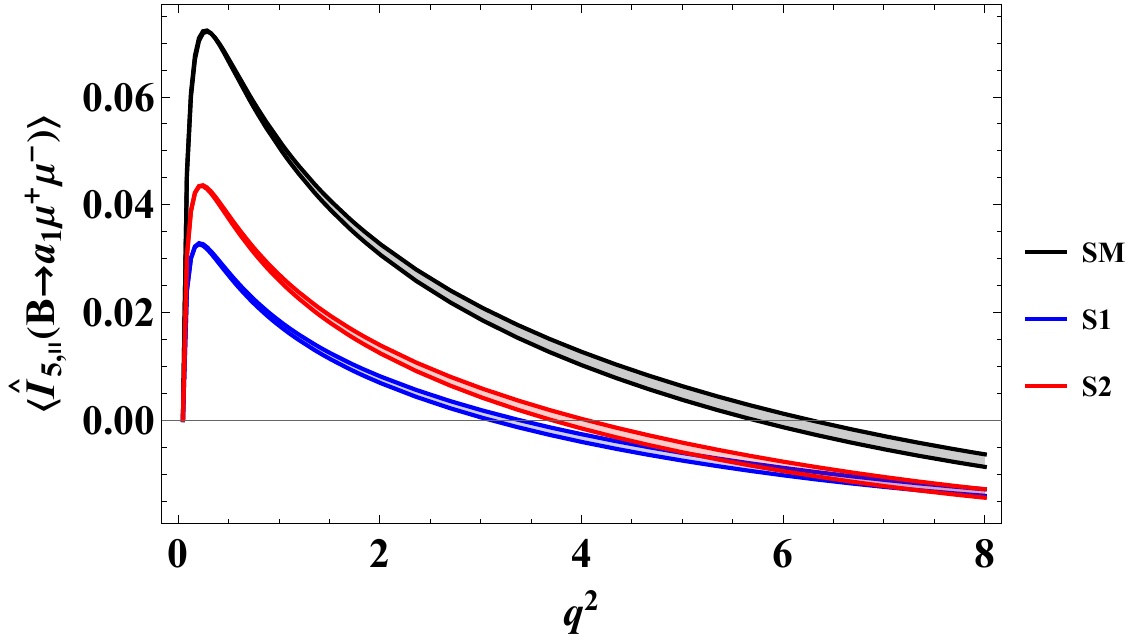}\ \ \
&\hspace{-1.2cm} \ \ \ \includegraphics[scale=0.52]{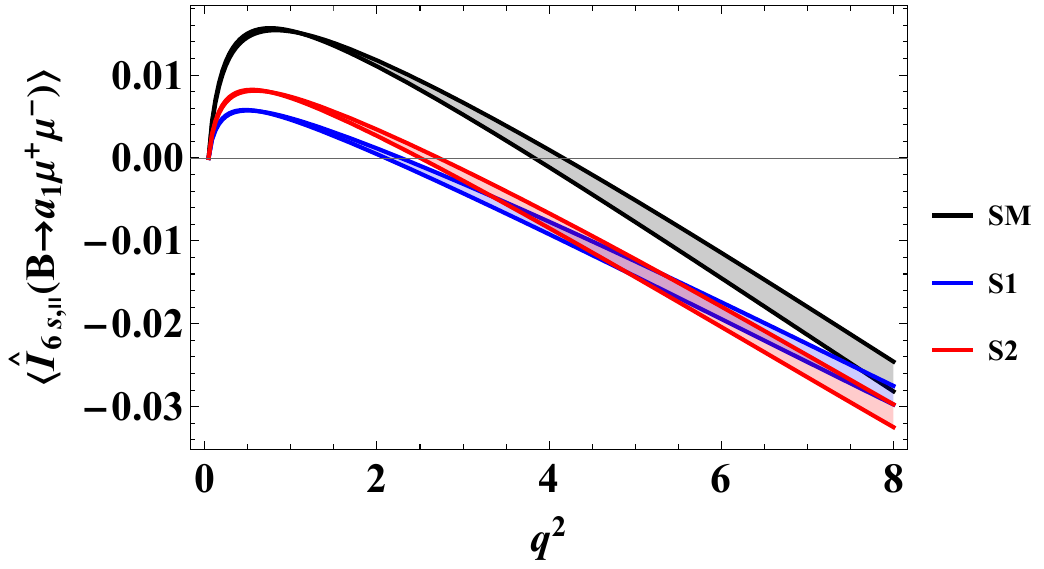}\\\
\end{tabular}
\caption{Angular observables $\langle \widehat{I}^{a_1}_{1s,\parallel}\rangle,\langle \widehat{I}^{a_1}_{1c.\parallel}\rangle,\langle \widehat{I}^{a_1}_{2s,\parallel}\rangle,\langle \widehat{I}^{a_1}_{2c,\parallel}\rangle,\langle \widehat{I}^{a_1}_{3,\parallel}\rangle,\langle \widehat{I}^{a_1}_{4,\parallel}\rangle,\langle \widehat{I}^{a_1}_{5,\parallel}\rangle$, and $\langle \widehat{I}^{a_1}_{6s,\parallel}\rangle$ for the decay $B\to a_{1}(\to\rho_{\parallel}\pi)\mu^{+}\mu^{-}$, in the SM and the two scenarios of the family non-universal $Z^{\prime}$ model.}
\label{Ia1rho}
\end{figure*}

\begin{figure*}[t!]
\begin{tabular}{cc}
\hspace{0.0cm}($\mathbf{a}$)&\hspace{-0.5cm}($\mathbf{b}$)\\
\includegraphics[scale=0.42]{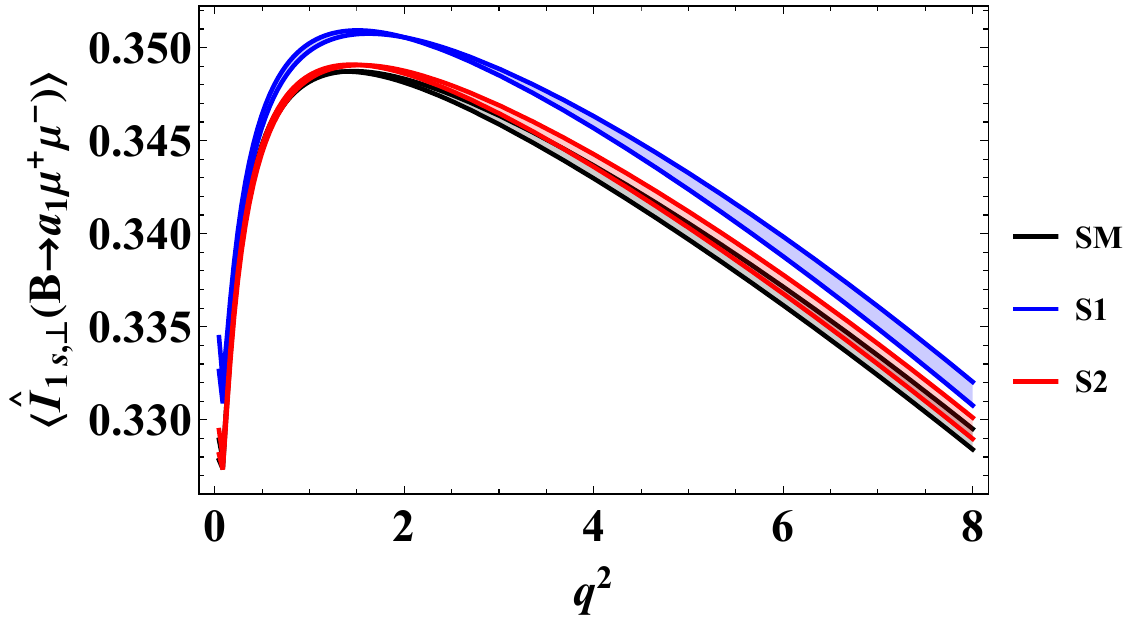}\ \ \
&\hspace{-1.2cm} \ \ \ \includegraphics[scale=0.42]{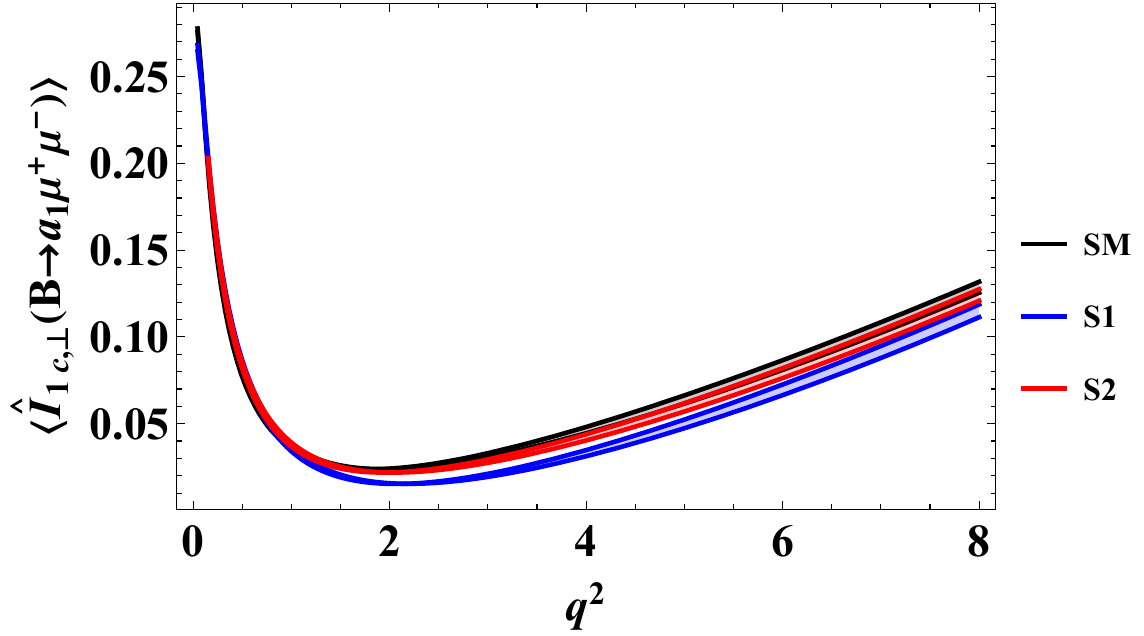}\\\
\hspace{0.0cm}($\mathbf{c}$)&\hspace{-0.5cm}($\mathbf{d}$)\\
\includegraphics[scale=0.42]{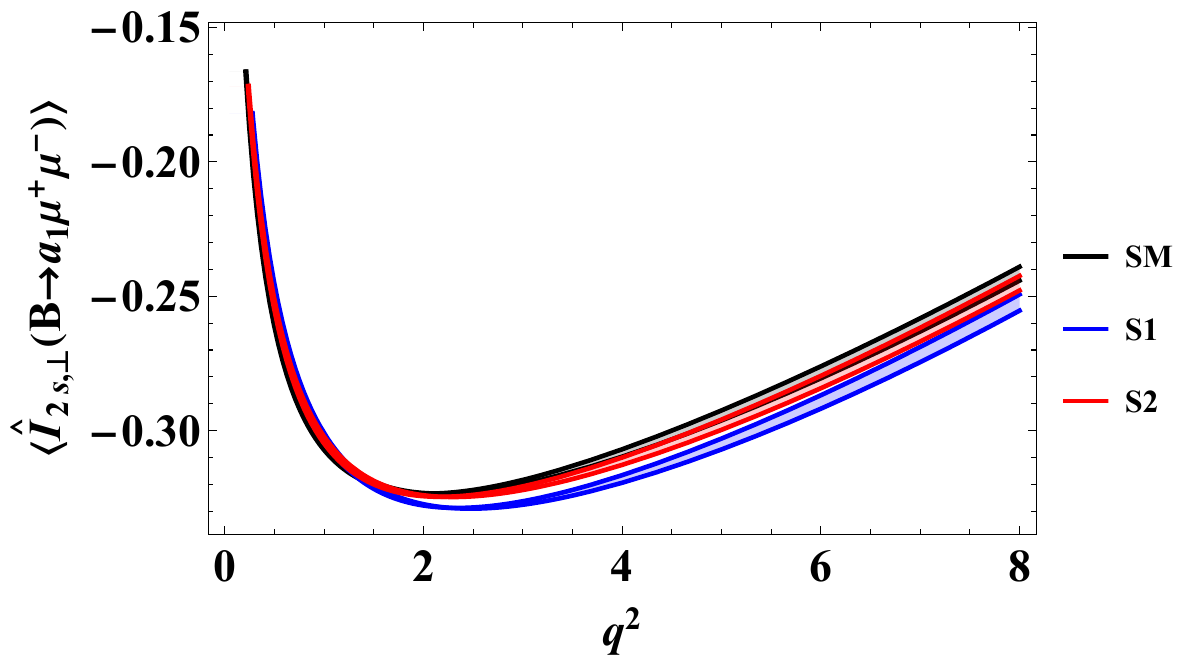}\ \ \
&\hspace{-1.2cm} \ \ \ \includegraphics[scale=0.42]{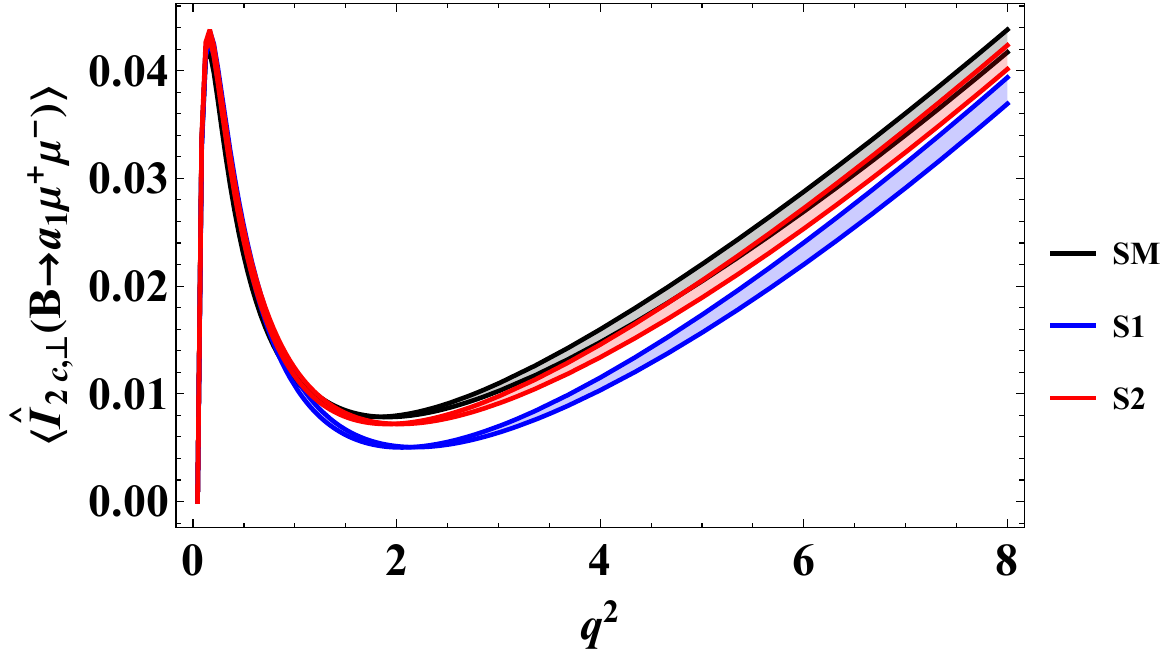}\\\
\hspace{0.0cm}($\mathbf{e}$)&\hspace{-0.5cm}($\mathbf{f}$)\\
\includegraphics[scale=0.42]{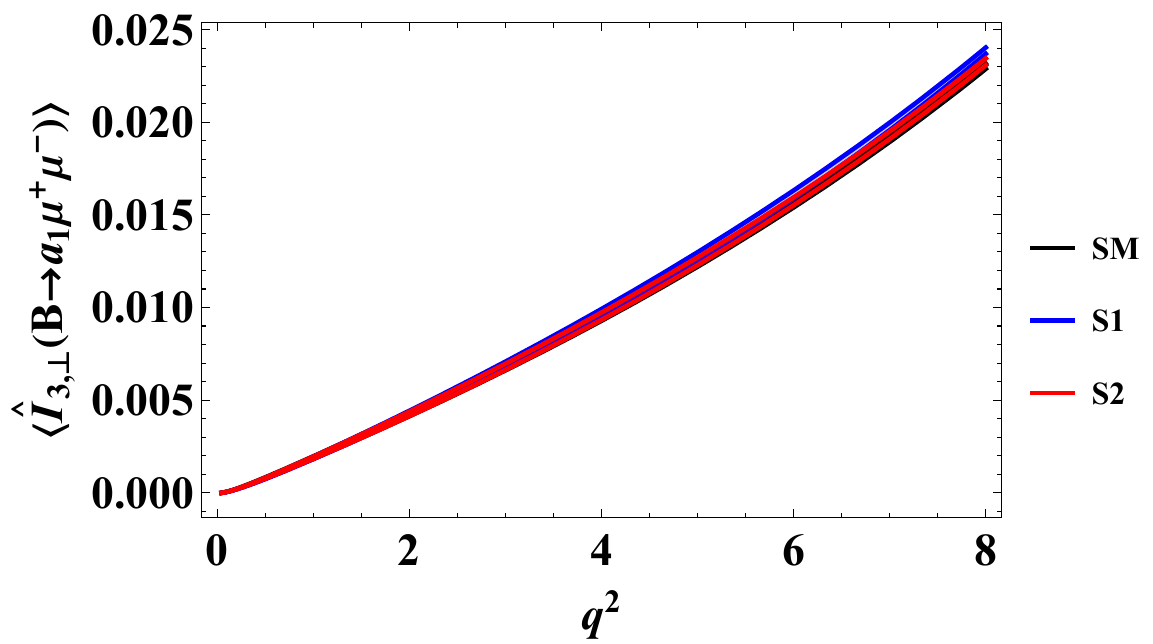}\ \ \
&\hspace{-1.2cm} \ \ \ \includegraphics[scale=0.42]{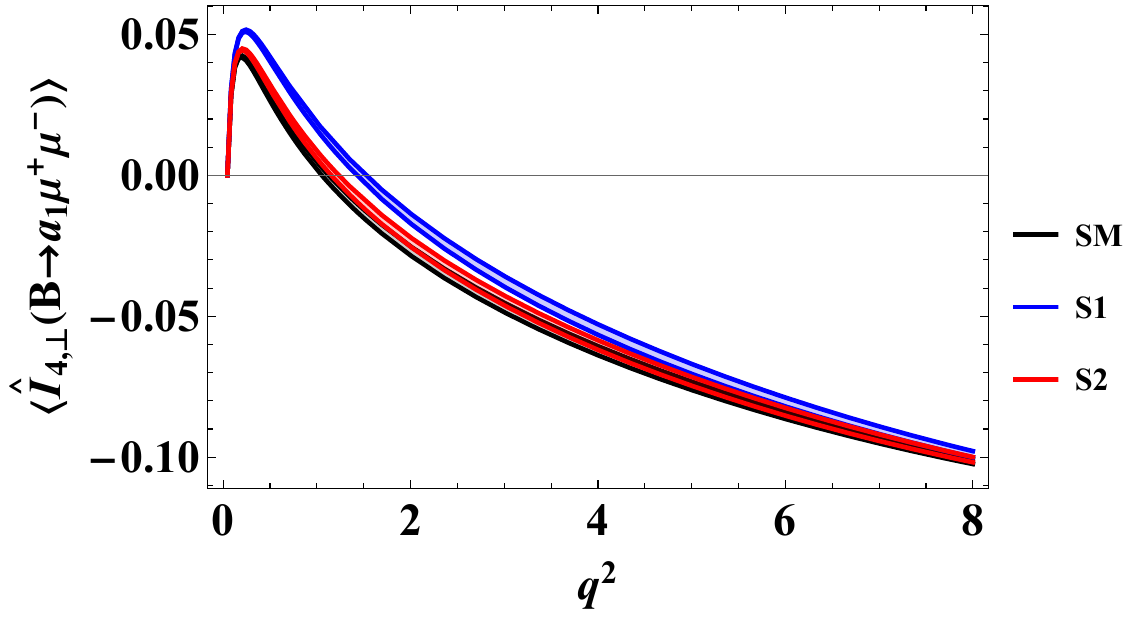}\\\
\hspace{0.0cm}($\mathbf{g}$)&\hspace{-0.5cm}($\mathbf{h}$)\\
\includegraphics[scale=0.42]{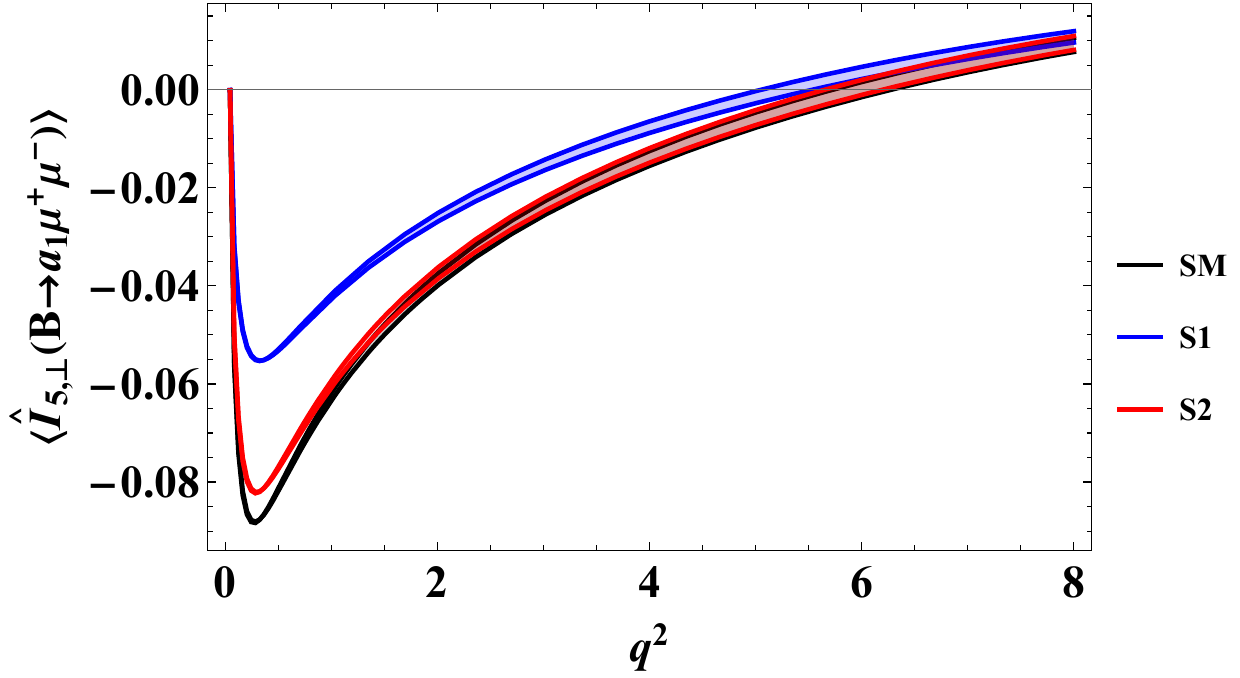}\ \ \
&\hspace{-1.2cm} \ \ \ \includegraphics[scale=0.42]{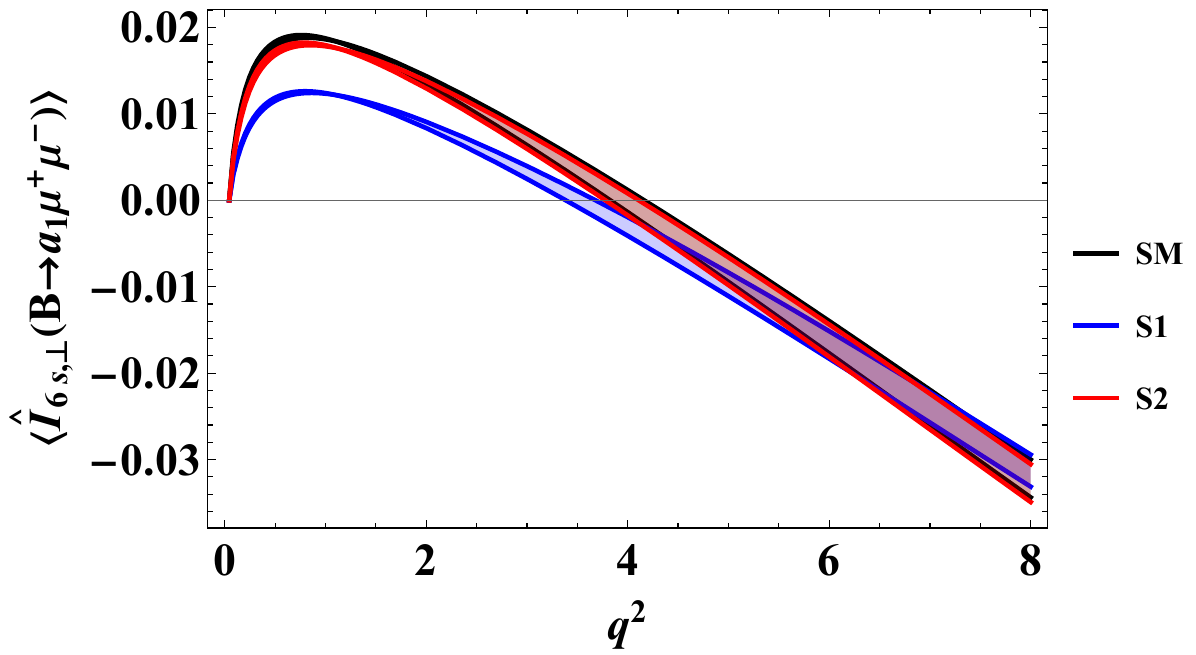}\\\
%\hspace{0.0cm}($\mathbf{e}$)&\hspace{-0.5cm}\\
%\includegraphics[scale=0.38]{I6cperpa1.pdf}\ \ \
\end{tabular}
\caption{Angular observables $\langle \widehat{I}^{a_{1}}_{1s,\perp}\rangle, \langle \widehat{I}^{a_{1}}_{1c,\perp}\rangle, \langle \widehat{I}^{a_{1}}_{2s,\perp}\rangle$, $\langle \widehat{I}^{a_{1}}_{2c,\perp}\rangle$, $\langle \widehat{I}^{a_1}_{3,\perp}\rangle,\langle \widehat{I}^{a_1}_{4,\perp}\rangle,\langle \widehat{I}^{a_1}_{5,\perp}\rangle$, and $\langle \widehat{I}^{a_1}_{6s,\perp}\rangle$ for the decay $B\to a_{1}(\to\rho_{\perp}\pi)\mu^{+}\mu^{-}$, in the SM and the two scenarios of the family non-universal $Z^{\prime}$ model.}
\label{FBA}
\end{figure*}

 
In this section, we present the effects of the family non-universal $Z^{\prime}$ model on the separate normalized angular coefficients such as $\langle I^{\rho}_{n\lambda}\rangle$, in the $B\to\rho(\to\pi\pi)\mu^{+}\mu^{-}$ decay, and $\langle \widehat{I}^{a_{1}}_{n\lambda,\parallel}\rangle$, $\langle \widehat{I}^{a_{1}}_{n\lambda,\perp}\rangle$, in the $B\to a_{1}(\to\rho_{\parallel, \perp})\mu^{+}\mu^{-}$ decays. The predicted numerical values of these observables, in different $q^2$ bins, for the SM as well as for the two different scenarios of family non-universal $Z^{\prime}$ model are given in Tables \ref{table:10}-\ref{table:35}, of appendix \ref{append1}. The listed errors in these tables originate mainly from the uncertainties of the form factors. Furthermore, we also display the results of normalized angular observables as a function of $q^{2}$ in Figs. \ref{I1rho}-\ref{FBA}.
%Numerical values of SM and in family non-universal $Z^{\prime}$ model of angular coefficients along with hadronic uncertainties in different $q^{2}$ bins are presented in Tables \ref{table:10}-\ref{table:36}, of appendix \ref{append1}. Furthermore, we also display the results of normalized angular observables as a function of $q^{2}$ in Figs. \ref{I1rho}-\ref{FBA}.


\begin{itemize}
\item Fig. \ref{I1rho} depicts the normalized angular observables $\langle I^{\rho}_{1s}\rangle, \langle I^{\rho}_{2s}\rangle, \langle I^{\rho}_{1c}\rangle, \langle I^{\rho}_{2c}\rangle, \langle I^{\rho}_{3}\rangle, \langle I^{\rho}_{4}\rangle, \langle I^{\rho}_{5}\rangle$ and $\langle I^{\rho}_{6s}\rangle$ for $B\to\rho(\to\pi\pi)\mu^{+}\mu^{-}$ decay in the SM and the two scenarios S1, S2 of family non universal $Z^{\prime}$ model. Figs. \ref{I1rho}(a)-\ref{I1rho}(d) show clear discrimination between the SM predictions and the two scenarios S1 and S2 of the family non-universal $Z^{\prime}$ model for the angular observables $\langle I^{\rho}_{1s}\rangle,\langle I^{\rho}_{2s}\rangle,\langle I^{\rho}_{1c}\rangle$ and $\langle I^{\rho}_{2c}\rangle$ in the range of $q^{2}=(0.1-7)$ $\text{GeV}^{2}$. Furthermore, the angular observables $\langle I^{\rho}_{3}\rangle, \langle I^{\rho}_{4}\rangle, \langle I^{\rho}_{5}\rangle$ and $\langle I^{\rho}_{6s}\rangle$ are shown in Fig. \ref{I1rho}(e)-\ref{I1rho}(h). For the observables $\langle  I^{\rho}_{3}\rangle$ and $\langle  I^{\rho}_{4}\rangle$ shown in Fig. \ref{I1rho}(e) and \ref{I1rho}(f), the SM predictions and effects of the family non-universal $Z^{\prime}$ model overlap with each other and hence no discrimination can be observed for these observables, except for some region of $q^2$ near $0.5$ $\text{GeV}^{2}$ for $\langle  I^{\rho}_{4}\rangle$. For the observables $\langle  I^{\rho}_{5}\rangle$ and $\langle  I^{\rho}_{6s}\rangle$, the discrimination between the SM and two scenarios of the family non-universal $Z^{\prime}$ model can be found for almost the whole kinematical range of $q^2$. Further, the two scenarios remain distinguishable from each other in almost the whole $q^2$ range.  
\end{itemize} 
\begin{itemize}
    \item Fig. \ref{Ia1rho} presents the normalized angular observables $\langle \widehat{I}^{a_1}_{1s,\parallel}\rangle,\langle \widehat{I}^{a_1}_{1c.\parallel}\rangle,\langle \widehat{I}^{a_1}_{2s,\parallel}\rangle,\langle \widehat{I}^{a_1}_{2c,\parallel}\rangle,\langle \widehat{I}^{a_1}_{3,\parallel}\rangle,\langle \widehat{I}^{a_1}_{4,\parallel}\rangle$, $\langle \widehat{I}^{a_1}_{5,\parallel}\rangle$, and $\langle \widehat{I}^{a_1}_{6s,\parallel}\rangle$ for the $B\to a_{1}(\to\rho_{\parallel}\pi)\mu^{+}\mu^{-}$ decay in the SM as well as in S1 and S2 of the family non universal $Z^{\prime}$ model. Figs. \ref{Ia1rho}(a)-\ref{Ia1rho}(d) show some discrimination between the SM predictions and the two scenarios S1 and S2 of the family non-universal $Z^{\prime}$ model for the angular observables $\langle \widehat{I}^{a_1}_{1s,\parallel}\rangle,\langle \widehat{I}^{a_1}_{1c.\parallel}\rangle,\langle \widehat{I}^{a_1}_{2s,\parallel}\rangle,\langle \widehat{I}^{a_1}_{2c,\parallel}\rangle$ in the range of $q^{2}=(0.1-3)$ $\text{GeV}^{2}$, whereas the two scenarios are not much distinct from each other. Furthermore, the angular observables $\langle \widehat{I}^{a_1}_{3,\parallel}\rangle,\langle \widehat{I}^{a_1}_{4,\parallel}\rangle$, $\langle \widehat{I}^{a_1}_{5,\parallel}\rangle$, and $\langle \widehat{I}^{a_1}_{6s,\parallel}\rangle$ are shown in Fig. \ref{Ia1rho}(e)-\ref{Ia1rho}(h). For the observables $\langle \widehat{I}^{a_1}_{3,\parallel}\rangle$ and $\langle \widehat{I}^{a_1}_{4,\parallel}\rangle$, no segregation is observed among SM and S1, S2 scenarios of the family non-universal $Z^{\prime}$ model as they are found to be largely overlapping with each other as shown in Fig. \ref{Ia1rho}(e) and \ref{Ia1rho}(f). 
    For the observables $\langle \widehat{I}^{a_1}_{5,\parallel}\rangle$ and $\langle \widehat{I}^{a_1}_{6s,\parallel}\rangle$, the discrimination between the SM and two scenarios of the family non-universal $Z^{\prime}$ model can be found for almost the whole kinematical range of $q^2$, whereas the two scenarios remain distinguishable from each other in the range of $q^{2}=(0.1-4)$ $\text{GeV}^{2}$ and $q^{2}=(0.1-2.5)$ $\text{GeV}^{2}$, for $\langle \widehat{I}^{a_1}_{5,\parallel}\rangle$ and $\langle \widehat{I}^{a_1}_{6s,\parallel}\rangle$, respectively, as shown in Fig. \ref{Ia1rho}(g) and \ref{Ia1rho}(h).  
    \item Fig. \ref{FBA} presents the normalized angular observables $\langle \widehat{I}^{a_1}_{1s,\perp}\rangle,\langle \widehat{I}^{a_1}_{1c.\perp}\rangle,\langle \widehat{I}^{a_1}_{2s,\perp}\rangle,\langle \widehat{I}^{a_1}_{2c,\perp}\rangle,\langle \widehat{I}^{a_1}_{3,\perp}\rangle$, $\langle \widehat{I}^{a_1}_{4,\perp}\rangle$, $\langle \widehat{I}^{a_1}_{5,\perp}\rangle$, and $\langle \widehat{I}^{a_1}_{6s,\perp}\rangle$ for the $B\to a_{1}(\to\rho_{\perp}\pi)\mu^{+}\mu^{-}$ decay in the SM as well as in S1 and S2 of the family non universal $Z^{\prime}$ model. Although the angular observable $\langle \widehat{I}^{a_1}_{6c,\perp}\rangle$ is non-zero, we do not present its results as its $q^2$ behaviour is similar to $\langle \widehat{I}^{a_1}_{6s,\perp}\rangle$, as coming from Eqs. (\ref{6sperp}), (\ref{6cperp}). In all the angular observables, scenario S1 of the family non-universal $Z^{\prime}$ model overlaps with the SM error band, whereas scenario S2 
    remains distinct from the SM and the scenario S1 in different ranges of the $q^{2}$ for different angular observables as shown in Fig. \ref{FBA}(a)-\ref{FBA}(h).
    %overlapping with two scenarios of the family non-universal $Z^{\prime}$ model is slightly visible in the range $q^{2}=(2-8)$ $\text{GeV}^{2}$. On the other hand, in the angular observables $\langle \widehat{I}^{a_1}_{3,\perp}\rangle,\langle \widehat{I}^{a_1}_{5.\perp}\rangle$ and $\langle \widehat{I}^{a_1}_{6s,\perp}\rangle$, SM predictions completely overlap with scenarios S1 and S2 of the family non-universal $Z^{\prime}$ model as shown in Fig. \ref{FBA}(e), Fig. \ref{FBA}(g) and Fig. \ref{FBA}(h). %for the whole range of $q^{2}$.
\end{itemize}

\section{Conclusions}\label{Conc}
Investigating B meson decays allows us to test the SM parameters along with exploring New Physics. Several exclusive semileptonic decays that involve flavor-changing neutral current transitions and flavor-changing charged current transitions exhibit notable deviations from SM predictions. Semileptonic decays involving $b\rightarrow s$ current have been studied during recent years and showed deviations from the SM predictions. In this work, the FCNC processes governed by $b\rightarrow dl^{+}l^{-}$ transition have been studied in the family non-universal $Z^{\prime}$ model. The four-fold angular decay distributions of $B\to\rho(\to\pi\pi)\mu^{+}\mu^{-}$, and $B\to a_{1}(\to\rho_{\parallel, \perp}\pi)\mu^{+}\mu^{-}$ decays have been derived using the helicity formalism. For both decays, various physical observables have been extracted and studied in the SM and the two scenarios of the family non-universal $Z^{\prime}$ model.
%the differential branching ratio for $B\to\rho(\to\pi\pi)\mu^{+}\mu^{-}$ clearly indicates the difference between SM and family non-universal $Z^{\prime}$ model scenarios whereas for $B\to a_{1}(\to\rho_{\parallel, \perp}\pi)\mu^{+}\mu^{-}$ shows no dividing line between SM and family non-universal $Z^{\prime}$ model scenarios.

To conclude the analysis, a noticeable difference has been observed between SM and the NP predicted values of the studied physical observables and in the majority of the normalized angular coefficients for the $B\to\rho(\to\pi\pi)\mu^{+}\mu^{-}$ decay, while the overall effect of NP is less distinct in most of the observables for the case of $B\to a_{1}(\to\rho_{\parallel, \perp}\pi)\mu^{+}\mu^{-}$ decay. For instance, in case of the differential branching ratio the SM and the $Z^{\prime}$ model scenarios are distinguishable for $B\to\rho(\to\pi\pi)\mu^{+}\mu^{-}$ decay while no clear distinction has been observed for $B\rightarrow a_{1}\mu^{+}\mu^{-}$ decay. In case of the forward-backward asymmetry and the longitudinal polarization fraction, the pattern of deviations reported is similar for both decays, however, the distinction among the two scenarios appears for larger range of $q^{2}$ in case of $B\to\rho(\to\pi\pi)\mu^{+}\mu^{-}$ decay in comparison to $B\rightarrow a_{1}\mu^{+}\mu^{-}$ decay. Similarly, most of the normalized angular observables for the $B\to\rho(\to\pi\pi)\mu^{+}\mu^{-}$ decay show notable distinction between the SM and the scenarios (S1, S2) predictions of the family non-universal $Z^{\prime}$ model, whereas, except for some, most of the normalized angular observables for the $B\to a_{1}(\to\rho_{\parallel, \perp}\pi)\mu^{+}\mu^{-}$ decay, show little or no distinction between the SM and the NP scenarios in the family non-universal $Z^{\prime}$ model. Rare $b\to d \ell \ell$ decays can be studied at high luminosity flavor facilities, such as LHCb \cite{Cerri:2018ypt}, and Belle II \cite{Belle-II:2018jsg}. For the present study, measurements of various normalized angluar observables of the order of $1\%$ relative to the branching ratio of order $10^{-9}$ at $3\sigma$ level require approximately $10^{13}$ $B\bar B$ pairs, and for the integrated luminosity goal of $300$ fb$^{-1}$ in HL-LHC $\sim 10^{14}$ $b\bar b$ pairs are expected to be produced \cite{Cerri:2018ypt, DiCanto:2022icc}. Therefore, the precise measurements of these observables in experiments conducted at LHCb and future collider facilities appear to be possible and it will provide valuable supplementary data needed to elucidate the underlying characteristics of NP within $b\rightarrow dl^{+}l^{-}$ decays.
\section*{Acknowledgments}
This work is supported by the Higher Education Commission of Pakistan through Grant no. NRPU/20-15142.
\appendix
\section{ Wilson Coefficients Expressions in the SM}\label{append}
The explicit expressions used for the Wilson coefficients are given as follows \cite{Bobeth:1999mk,Beneke:2001at,Asatrian:2001de,Asatryan:2001zw,Greub:2008cy,Du:2015tda},
\begin{eqnarray}
C_{7}^{\text{eff}}(q^2)&=&C_{7}-\frac{1}{3}\left(C_{3}+\frac{4}{3}C_{4}+20C_{5}+\frac{80}{3}C_{6}\right)
-\frac{\alpha_{s}}{4\pi}\left[(C_{1}-6C_{2})F^{(7)}_{1,c}(q^2)+C_{8}F^{(7)}_{8}(q^2)\right]\notag\\&-&\frac{\alpha_{s}}{4\pi}\lambda^{(q)}_{u}\left(C_{1}-6C_{2}\right)\left(F^{7}_{1,c}-F^{7}_{1,u}\right),\notag\\
C_{9}^{\text{eff}}(q^2)&=&C_{9}+\frac{4}{3}\left(C_{3}+\frac{16}{3}C_{5}+\frac{16}{9}C_{6}\right)
-h(0, q^2)\left(\frac{1}{2}C_{3}+\frac{2}{3}C_{4}+8C_{5}+\frac{32}{3}C_{6}\right)\notag\\
&-&h(m_{b}^{\text{pole}}, q^2)\big(\frac{7}{2}C_{3}+\frac{2}{3}C_{4}+38C_{5}+\frac{32}{3}C_{6}\big)+h(m_{c}^{\text{pole}}, q^2)
\big(\frac{4}{3}C_{1}+C_{2}+6C_{3}+60C_{5}\big)\notag\\
&+&\lambda^{(q)}_{u}\left[h(m_{c},q^{2})-h(0,q^{2})\right]\left(\frac{4}{3}C_{1}+C_{2}\right)
-\frac{\alpha_{s}}{4\pi}\left[C_{1}F^{(9)}_{1,c}(q^2)+C_{2}F^{(9)}_{2,c}(q^2)+C_{8}F^{(9)}_{8}(q^2)\right]\notag\\&-&\frac{\alpha_{s}}{4\pi}\lambda^{(q)}_{u}\left[C_{1}(F^{(9)}_{1,c}-F^{(9)}_{1,u})+C_{2}(F^{(9)}_{2,c}-F^{(9)}_{2,u})\right],\label{WC3}
\end{eqnarray}

where the functions $h(m_{q}^{\text{pole}}, q^2)$ with $q=c, b$, and functions $F^{(7,9)}_{8}(q^2)$ are
defined in \cite{Beneke:2001at}, while the functions $F^{(7,9)}_{1,c}(q^2)$, $F^{(7,9)}_{2,c}(q^2)$ are
given in \cite{Asatryan:2001zw} for low $q^{2}$ and in \cite{Greub:2008cy} for high $q^{2}$. The quark masses appearing in all of these functions are defined in the pole scheme.

\section{Binned Predictions of Physical Observables}\label{append1}
In this appendix, we give the SM as well as the family non-universal $Z^{\prime}$ model predictions of physical observables in different $q^2$ bins.

\begin{table}[htt!]
\begin{center}
%\captionsetup{margin=0.5cm}
\caption{Predictions of observables in the decay $B\to\rho(\to \pi\pi)\mu^{+}\mu^{-}$, such as differential branching ratios, $\frac{d\mathcal{B}}{dq^{2}} (B\to\rho\mu^{+}\mu^{-})$, $\frac{d\mathcal{B}}{dq^{2}} (B\to\rho(\to \pi\pi)\mu^{+}\mu^{-})$, lepton forward-backward asymmetry $A_{\text{FB}}^{\rho}$, longitudinal helicity fraction $f_{L}^{\rho}$, and in the decay $B\to a_{1}(\to\rho\pi)\mu^{+}\mu^{-}$, such as differential branching ratios $\frac{d\mathcal{B}}{dq^{2}} (B\to a_{1}\mu^{+}\mu^{-})$, $\frac{d\mathcal{B}}{dq^{2}}(B\to a_{1}(\to\rho_{\parallel}\pi)\mu^{+}\mu^{-})$, $\frac{d\mathcal{B}}{dq^{2}} (B\to a_{1}(\to \rho_{\perp}\pi)\mu^{+}\mu^{-})$, lepton forward-backward asymmetry   $A_{\text{FB}}^{a_{1}}$, and longitudinal helicity fraction $f_{L}^{a_{1}}$, in $q^2 = 0.1 - 1.0$ GeV$^{2}$ bin, for the SM as well as the NP scenarios (S1, S2) of $Z^{\prime}$ model listed in Table \ref{TZP1}. The errors presented mainly come from the uncertainties of the form factors.}
\label{table:1}
\begin{tabular}{ |p{6cm}||p{3cm}|p{3cm}|p{3cm}|  }
 \hline
 \multicolumn{4}{|c|}{\textbf{$q^{2}= 0.1 - 1.0$} GeV$^{2}$} \\
 \hline
 \textbf{Observables}& \textbf{SM} &\textbf{$Z^{\prime}$ Scenario 1}&\textbf{$Z^{\prime}$ Scenario 2}\\
 \hline
 $\frac{d\mathcal{B}}{dq^{2}} (B\to\rho\mu^{+}\mu^{-}) \times 10^{+9}$&$1.889^{+0.339}_{-0.339}$& $2.410^{+0.497}_{-0.497}$&  $2.176^{+0.430}_{-0.430}$\\
 \hline
 $\frac{d\mathcal{B}}{dq^{2}} (B\to\rho(\to \pi\pi)\mu^{+}\mu^{-})\times 10^{+9}$&$1.889^{+0.339}_{-0.339}$& $2.410^{+0.497}_{-0.497}$&  $2.176^{+0.430}_{-0.430}$\\
 \hline
$A_{\text{FB}}^{\rho}$& $0.077^{+0.002}_{-0.002}$&  $0.037^{+0.001}_{-0.001}$&   $0.049^{+0.001}_{-0.001}$\\
 \hline 
 $f_{L}^{\rho}$ &$0.510^{+0.005}_{-0.005}$&$0.717^{+0.005}_{-0.005}$& $0.652^{+0.006}_{-0.006}$\\
 \hline
 $\frac{d\mathcal{B}}{dq^{2}} (B\to a_{1}\mu^{+}\mu^{-})\times 10^{+9}$&$11.1295^{+7.3804}_{-7.3804}$& $19.208^{+12.684}_{-12.684}$&  $15.772^{+1.042}_{-1.042}$\\
 \hline
 $\frac{d\mathcal{B}}{dq^{2}} (B\to a_{1}(\to \rho_{\parallel}\pi)\mu^{+}\mu^{-})\times 10^{+9}$&$3.1905^{+2.1157}_{-2.1157}$& $5.506^{+3.636}_{-3.636}$&  $4.521^{+2.989}_{-2989}$\\
 \hline
 $\frac{d\mathcal{B}}{dq^{2}} (B\to a_{1}(\to \rho_{\perp}\pi)\mu^{+}\mu^{-})\times 10^{+9}$&$7.7907^{+5.1663}_{-5.1663}$& $13.446^{+8.879}_{-8.879}$&  $11.040^{+7.3001}_{-7.3001}$\\
 \hline
$A_{\text{FB}}^{a_{1}}$& $0.0347^{+0.0005}_{-0.0005}$&  $0.0135^{+0.00005}_{-0.00005}$&   $0.0191^{+0.0002}_{-0.0002}$\\
 \hline 
 $f_{L}^{a_{1}}$ &$0.8022^{+0.0036}_{-0.0036}$&$0.7952^{+0.0034}_{-0.0034}$& $0.7910^{+0.0038}_{-0.0038}$\\
 \hline
\end{tabular}
\end{center}
\end{table}

\begin{table}[ht!]
\begin{center}
\caption{Same as in Table \ref{table:1}, but for $q^2 = 1.0 - 2.0$ GeV$^{2}$ bin.}\label{table:2}
\begin{tabular}{ |p{6cm}||p{3cm}|p{3cm}|p{3cm}|  }
 \hline
 \multicolumn{4}{|c|}{\textbf{$q^{2}= 1.0 - 2.0$} GeV$^{2}$} \\
 \hline
 \textbf{Observables}& \textbf{SM} &\textbf{$Z^{\prime}$ Scenario 1}&\textbf{$Z^{\prime}$ Scenario 2}\\
 \hline
 $\frac{d\mathcal{B}}{dq^{2}} (B\to\rho\mu^{+}\mu^{-}) \times 10^{+9}$&$0.856^{+0.166}_{-0.166}$& $1.462^{+0.279}_{-0.279}$&  $1.188^{+0.275}_{-0.275}$\\
 \hline
 $\frac{d\mathcal{B}}{dq^{2}} (B\to\rho(\to \pi\pi)\mu^{+}\mu^{-})\times 10^{+9}$&$0.856^{+0.166}_{-0.166}$& $1.462^{+0.279}_{-0.279}$&  $1.188^{+0.275}_{-0.275}$\\
 \hline
$A_{\text{FB}}^{\rho}$& $0.100^{+0.002}_{-0.002}$&  $0.015^{+0.001}_{-0.001}$&   $0.035^{+0.001}_{-0.001}$\\
 \hline 
 $f_{L}^{\rho}$ &$0.840^{+0.001}_{-0.001}$&$0.968^{+0.001}_{-0.001}$& $0.943^{+0.001}_{-0.001}$\\
 \hline
 $\frac{d\mathcal{B}}{dq^{2}} (B\to a_{1}\mu^{+}\mu^{-})\times 10^{+9}$&$8.1566^{+5.3824}_{-5.3824}$& $15.5392^{+10.2335}_{-10.2335}$&  $12.330^{+8.125}_{-8.125}$\\
 \hline
 $\frac{d\mathcal{B}}{dq^{2}} (B\to a_{1}(\to \rho_{\parallel}\pi)\mu^{+}\mu^{-})\times 10^{+9}$&$2.3382^{+1.5430}_{-1.5430}$& $4.4546^{+2.9336}_{-2.9336}$&  $3.534^{+2.329}_{-2.329}$\\
 \hline
 $\frac{d\mathcal{B}}{dq^{2}} (B\to a_{1}(\to \rho_{\perp}\pi)\mu^{+}\mu^{-})\times 10^{+9}$&$5.7096^{+3.7677}_{-3.7677}$& $10.8774^{+7.1634}_{-7.1634}$&  $8.629^{+5.687}_{-5.687}$\\
 \hline
$A_{\text{FB}}^{a_{1}}$& $0.0357^{+0.0004}_{-0.0004}$&  $0.0075^{+0.0006}_{-0.0006}$&   $0.0140^{+0.0006}_{-0.0006}$\\
 \hline 
 $f_{L}^{a_{1}}$ &$0.9479^{+0.0008}_{-0.0008}$&$0.9533^{+0.0062}_{-0.0062}$& $0.9492^{+0.0011}_{-0.0011}$\\
 \hline
\end{tabular}
\end{center}
\end{table}

\begin{table}[ht!]
\begin{center}
\caption{Same as in Table \ref{table:1}, but for $q^2 = 2.0 - 3.0$ GeV$^{2}$ bin.}\label{table:3}
\begin{tabular}{ |p{6cm}||p{3cm}|p{3cm}|p{3cm}|  }
 \hline
 \multicolumn{4}{|c|}{\textbf{$q^{2}= 2.0 - 3.0$} GeV$^{2}$} \\
 \hline
 \textbf{Observables}& \textbf{SM} &\textbf{$Z^{\prime}$ Scenario 1}&\textbf{$Z^{\prime}$ Scenario 2}\\
 \hline
 $\frac{d\mathcal{B}}{dq^{2}} (B\to\rho\mu^{+}\mu^{-}) \times 10^{+9}$&$0.789^{+0.146}_{-0.146}$& $1.443^{+0.266}_{-0.266}$&  $1.147^{+0.212}_{-0.212}$\\
 \hline
 $\frac{d\mathcal{B}}{dq^{2}} (B\to\rho(\to \pi\pi)\mu^{+}\mu^{-})\times 10^{+9}$&$0.789^{+0.146}_{-0.146}$& $1.443^{+0.266}_{-0.266}$&  $1.147^{+0.212}_{-0.212}$\\
 \hline
$A_{\text{FB}}^{\rho}$& $0.048^{+0.002}_{-0.002}$&  $0.033^{+0.001}_{-0.001}$&   $0.019^{+0.001}_{-0.001}$\\
 \hline 
 $f_{L}^{\rho}$ &$0.880^{+0.001}_{-0.001}$&$0.949^{+0.000}_{-0.000}$& $0.942^{+0.000}_{-0.000}$\\
 \hline
 $\frac{d\mathcal{B}}{dq^{2}} (B\to a_{1}\mu^{+}\mu^{-})\times 10^{+9}$&$7.5818^{+4.9957}_{-4.9957}$& $14.6876^{+9.6632}_{-9.6632}$&  $11.570^{+7.618}_{-7.618}$\\
 \hline
 $\frac{d\mathcal{B}}{dq^{2}} (B\to a_{1}(\to \rho_{\parallel}\pi)\mu^{+}\mu^{-})\times 10^{+9}$&$2.1735^{+1.4321}_{-1.4321}$& $4.2105^{+2.7701}_{-2.7701}$&  $3.317^{+2.1837}_{-2.1837}$\\
 \hline
 $\frac{d\mathcal{B}}{dq^{2}} (B\to a_{1}(\to \rho_{\perp}\pi)\mu^{+}\mu^{-})\times 10^{+9}$&$5.3073^{+3.4970}_{-3.4970}$& $10.2813^{+6.7642}_{-6.7642}$&  $8.0399^{+5.3320}_{-5.3320}$\\
 \hline
$A_{\text{FB}}^{a_{1}}$& $0.0230^{+0.0014}_{-0.0014}$&  $-0.0036^{+0.0012}_{-0.0012}$&   $0.0015^{+0.0014}_{-0.0014}$\\
 \hline 
 $f_{L}^{a_{1}}$ &$0.9484^{+0.0012}_{-0.0012}$&$0.9584^{+0.0088}_{-0.0088}$& $0.9539^{+0.0010}_{-0.0010}$\\
 \hline
\end{tabular}
\end{center}
\end{table}

\begin{table}[ht!]
\begin{center}
\caption{Same as in Table \ref{table:1}, but for $q^2 = 3.0 - 4.0$ GeV$^{2}$ bin.}\label{table:4}
\begin{tabular}{ |p{6cm}||p{3cm}|p{3cm}|p{3cm}|  }
 \hline
 \multicolumn{4}{|c|}{\textbf{$q^{2}= 3.0 - 4.0$} GeV$^{2}$} \\
 \hline
 \textbf{Observables}& \textbf{SM} &\textbf{$Z^{\prime}$ Scenario 1}&\textbf{$Z^{\prime}$ Scenario 2}\\
 \hline
 $\frac{d\mathcal{B}}{dq^{2}} (B\to\rho\mu^{+}\mu^{-}) \times 10^{+9}$&$0.797^{+0.143}_{-0.143}$& $1.503^{+0.269}_{-0.269}$&  $1.212^{+0.212}_{-0.212}$\\
 \hline
 $\frac{d\mathcal{B}}{dq^{2}} (B\to\rho(\to \pi\pi)\mu^{+}\mu^{-})\times 10^{+9}$&$0.797^{+0.143}_{-0.143}$& $1.503^{+0.269}_{-0.269}$&  $1.212^{+0.212}_{-0.212}$\\
 \hline
$A_{\text{FB}}^{\rho}$& $-0.012^{+0.002}_{-0.002}$&  $-0.078^{+0.001}_{-0.001}$&   $-0.072^{+0.002}_{-0.002}$\\
 \hline 
 $f_{L}^{\rho}$ &$0.8501^{+0.0003}_{-0.0003}$&$0.893^{+0.0005}_{-0.0005}$& $0.8931^{+0.0004}_{-0.0004}$\\
 \hline
 $\frac{d\mathcal{B}}{dq^{2}} (B\to a_{1}\mu^{+}\mu^{-})\times 10^{+9}$&$7.2480^{+4.7714}_{-4.7714}$& $14.1509^{+9.3042}_{-9.3042}$&  $11.600^{+7.618}_{-7.618}$\\
 \hline
 $\frac{d\mathcal{B}}{dq^{2}} (B\to a_{1}(\to \rho_{\parallel}\pi)\mu^{+}\mu^{-})\times 10^{+9}$&$2.0778^{+1.3678}_{-1.3678}$& $4.0566^{+2.6672}_{-2.6672}$&  $3.184^{+2.090}_{-2.090}$\\
 \hline
 $\frac{d\mathcal{B}}{dq^{2}} (B\to a_{1}(\to \rho_{\perp}\pi)\mu^{+}\mu^{-})\times 10^{+9}$&$5.0736^{+3.3400}_{-3.3400}$& $9.9056^{+6.5129}_{-6.5129}$&  $7.775^{+5.115}_{-5.115}$\\
 \hline
$A_{\text{FB}}^{a_{1}}$& $0.0079^{+0.0023}_{-0.0023}$&  $-0.0157^{+0.0017}_{-0.0017}$&   $0.0125^{+0.0020}_{-0.0020}$\\
 \hline 
 $f_{L}^{a_{1}}$ &$0.9264^{+0.0028}_{-0.0028}$&$0.9384^{+0.0093}_{-0.0093}$& $0.9337^{+0.0027}_{-0.0027}$\\
 \hline
\end{tabular}
\end{center}
\end{table}

\begin{table}[ht!]
\begin{center}
\caption{Same as in Table \ref{table:1}, but for $q^2 = 4.0 - 5.0$ GeV$^{2}$ bin.}\label{table:5}
\begin{tabular}{ |p{6cm}||p{3cm}|p{3cm}|p{3cm}|  }
 \hline
 \multicolumn{4}{|c|}{\textbf{$q^{2}= 4.0 - 5.0$} GeV$^{2}$} \\
 \hline
 \textbf{Observables}& \textbf{SM} &\textbf{$Z^{\prime}$ Scenario 1}&\textbf{$Z^{\prime}$ Scenario 2}\\
 \hline
 $\frac{d\mathcal{B}}{dq^{2}} (B\to\rho\mu^{+}\mu^{-}) \times 10^{+9}$&$0.828^{+0.144}_{-0.144}$& $1.587^{+0.276}_{-0.276}$&  $1.242^{+0.216}_{-0.216}$\\
 \hline
 $\frac{d\mathcal{B}}{dq^{2}} (B\to\rho(\to \pi\pi)\mu^{+}\mu^{-})\times 10^{+9}$&$0.828^{+0.144}_{-0.144}$& $1.587^{+0.276}_{-0.276}$&  $1.242^{+0.216}_{-0.216}$\\
 \hline
$A_{\text{FB}}^{\rho}$& $-0.068^{+0.002}_{-0.002}$&  $-0.117^{+0.002}_{-0.002}$&   $-0.119^{+0.001}_{-0.001}$\\
 \hline 
 $f_{L}^{\rho}$ &$0.8101^{+0.0002}_{-0.0002}$&$0.8315^{+0.0006}_{-0.0006}$& $0.8352^{+0.0005}_{-0.0005}$\\
 \hline
 $\frac{d\mathcal{B}}{dq^{2}} (B\to a_{1}\mu^{+}\mu^{-})\times 10^{+9}$&$6.9869^{+4.5967}_{-4.5967}$& $13.7062^{+9.0082}_{-9.0082}$&  $11.110^{+7.307}_{-7.307}$\\
 \hline
 $\frac{d\mathcal{B}}{dq^{2}} (B\to a_{1}(\to \rho_{\parallel}\pi)\mu^{+}\mu^{-})\times 10^{+9}$&$2.0029^{+1.3177}_{-1.3177}$& $3.9291^{+2.5823}_{-2.5823}$&  $3.076^{+2.023}_{-2.023}$\\
 \hline
 $\frac{d\mathcal{B}}{dq^{2}} (B\to a_{1}(\to \rho_{\perp}\pi)\mu^{+}\mu^{-})\times 10^{+9}$&$4.8908^{+3.2177}_{-3.2177}$& $9.5944^{+6.3057}_{-6.3057}$&  $7.512^{+4.939}_{-4.939}$\\
 \hline
$A_{\text{FB}}^{a_{1}}$& $-0.0083^{+0.0031}_{-0.0031}$&  $-0.0284^{+0.0022}_{-0.0022}$&   $-0.0271^{+0.0026}_{-0.0026}$\\
 \hline 
 $f_{L}^{a_{1}}$ &$0.8957^{+0.0040}_{-0.0040}$&$0.9087^{+0.0090}_{-0.0090}$& $0.9040^{+0.0040}_{-0.0040}$\\
 \hline
\end{tabular}
\end{center}
\end{table}

\begin{table}[ht!]
\begin{center}
\caption{Same as in Table \ref{table:1}, but for $q^2 = 5.0 - 6.0$ GeV$^{2}$ bin.}\label{table:6}
\begin{tabular}{ |p{6cm}||p{3cm}|p{3cm}|p{3cm}|  }
 \hline
 \multicolumn{4}{|c|}{\textbf{$q^{2}= 5.0 - 6.0$} GeV$^{2}$} \\
 \hline
 \textbf{Observables}& \textbf{SM} &\textbf{$Z^{\prime}$ Scenario 1}&\textbf{$Z^{\prime}$ Scenario 2}\\
 \hline
 $\frac{d\mathcal{B}}{dq^{2}} (B\to\rho\mu^{+}\mu^{-}) \times 10^{+9}$&$0.868^{+0.147}_{-0.147}$& $1.682^{+0.285}_{-0.285}$&  $1.312^{+0.220}_{-0.220}$\\
 \hline
 $\frac{d\mathcal{B}}{dq^{2}} (B\to\rho(\to \pi\pi)\mu^{+}\mu^{-})\times 10^{+9}$&$0.868^{+0.147}_{-0.147}$& $1.682^{+0.285}_{-0.285}$&  $1.312^{+0.220}_{-0.220}$\\
 \hline
$A_{\text{FB}}^{\rho}$& $-0.119^{+0.002}_{-0.002}$&  $-0.152^{+0.001}_{-0.001}$&   $-0.161^{+0.001}_{-0.001}$\\
 \hline 
 $f_{L}^{\rho}$ &$0.7601^{+0.0002}_{-0.0002}$&$0.7735^{+0.0006}_{-0.0006}$& $0.7786^{+0.0006}_{-0.0006}$\\
 \hline
 $\frac{d\mathcal{B}}{dq^{2}} (B\to a_{1}\mu^{+}\mu^{-})\times 10^{+9}$&$6.7521^{+4.4407}_{-4.4407}$& $13.2899^{+8.7329}_{-8.7329}$&  $10.700^{+7.056}_{-7.056}$\\
 \hline
 $\frac{d\mathcal{B}}{dq^{2}} (B\to a_{1}(\to \rho_{\parallel}\pi)\mu^{+}\mu^{-})\times 10^{+9}$&$1.9356^{+1.2730}_{-1.2730}$& $3.8098^{+2.5034}_{-2.5034}$&  $2.977^{+1.893}_{-1.893}$\\
 \hline
 $\frac{d\mathcal{B}}{dq^{2}} (B\to a_{1}(\to \rho_{\perp}\pi)\mu^{+}\mu^{-})\times 10^{+9}$&$4.7265^{+3.1085}_{-3.1085}$& $9.3029^{+6.1130}_{-6.1130}$&  $7.269^{+4.778}_{-4.778}$\\
 \hline
$A_{\text{FB}}^{a_{1}}$& $-0.0252^{+0.0037}_{-0.0037}$&  $-0.0415^{+0.0025}_{-0.0025}$&   $-0.0423^{+0.0030}_{-0.0030}$\\
 \hline 
 $f_{L}^{a_{1}}$ &$0.8599^{+0.0049}_{-0.0049}$&$0.8732^{+0.0084}_{-0.0084}$& $0.8687^{+0.0050}_{-0.0050}$\\
 \hline
\end{tabular}
\end{center}
\end{table}

\begin{table}[ht!]
\begin{center}
\caption{Same as in Table \ref{table:1}, but for $q^2 = 6.0 - 7.0$ GeV$^{2}$ bin.}\label{table:7}
\begin{tabular}{ |p{6cm}||p{3cm}|p{3cm}|p{3cm}|  }
 \hline
 \multicolumn{4}{|c|}{\textbf{$q^{2}= 6.0 - 7.0$} GeV$^{2}$} \\
 \hline
 \textbf{Observables}& \textbf{SM} &\textbf{$Z^{\prime}$ Scenario 1}&\textbf{$Z^{\prime}$ Scenario 2}\\
 \hline
 $\frac{d\mathcal{B}}{dq^{2}} (B\to\rho\mu^{+}\mu^{-}) \times 10^{+9}$&$0.912^{+0.150}_{-0.150}$& $1.781^{+0.294}_{-0.294}$&  $1.386^{+0.285}_{-0.285}$\\
 \hline
 $\frac{d\mathcal{B}}{dq^{2}} (B\to\rho(\to \pi\pi)\mu^{+}\mu^{-})\times 10^{+9}$&$0.912^{+0.150}_{-0.150}$& $1.781^{+0.294}_{-0.294}$&  $1.386^{+0.285}_{-0.285}$\\
 \hline
$A_{\text{FB}}^{\rho}$& $-0.164^{+0.002}_{-0.002}$&  $-0.181^{+0.001}_{-0.001}$&   $-0.197^{+0.002}_{-0.002}$\\
 \hline 
 $f_{L}^{\rho}$ &$0.7101^{+0.0003}_{-0.0003}$&$0.7207^{+0.0006}_{-0.0006}$& $0.7263^{+0.0006}_{-0.0006}$\\
 \hline
 $\frac{d\mathcal{B}}{dq^{2}} (B\to a_{1}\mu^{+}\mu^{-})\times 10^{+9}$&$6.5244^{+4.2905}_{-4.2905}$& $12.8742^{+8.4599}_{-8.4599}$&  $10.390^{+6.827}_{-6.827}$\\
 \hline
 $\frac{d\mathcal{B}}{dq^{2}} (B\to a_{1}(\to \rho_{\parallel}\pi)\mu^{+}\mu^{-})\times 10^{+9}$&$1.8703^{+1.2300}_{-1.2300}$& $3.6906^{+2.4252}_{-2.4252}$&  $2.879^{+1.893}_{-1.893}$\\
 \hline
 $\frac{d\mathcal{B}}{dq^{2}} (B\to a_{1}(\to \rho_{\perp}\pi)\mu^{+}\mu^{-})\times 10^{+9}$&$4.5671^{+3.0034}_{-3.0034}$& $9.0120^{+5.9220}_{-5.9220}$&  $7.030^{+4.621}_{-4.621}$\\
 \hline
$A_{\text{FB}}^{a_{1}}$& $-0.0425^{+0.0042}_{-0.0042}$&  $-0.0548^{+0.0028}_{-0.0028}$&   $-0.0578^{+0.0033}_{-0.0033}$\\
 \hline 
 $f_{L}^{a_{1}}$ &$0.8204^{+0.0056}_{-0.0056}$&$0.8336^{+0.0076}_{-0.0076}$& $0.8292^{+0.0058}_{-0.0058}$\\
 \hline
\end{tabular}
\end{center}
\end{table}

\begin{table}[ht!]
\begin{center}
\caption{Same as in Table \ref{table:1}, but for $q^2 = 7.0 - 8.0$ GeV$^{2}$ bin.}\label{table:8}
\begin{tabular}{ |p{6cm}||p{3cm}|p{3cm}|p{3cm}|  }
 \hline
 \multicolumn{4}{|c|}{\textbf{$q^{2}= 7.0 - 8.0$} GeV$^{2}$} \\
 \hline
 \textbf{Observables}& \textbf{SM} &\textbf{$Z^{\prime}$ Scenario 1}&\textbf{$Z^{\prime}$ Scenario 2}\\
 \hline
 $\frac{d\mathcal{B}}{dq^{2}} (B\to\rho\mu^{+}\mu^{-}) \times 10^{+9} $&$0.959^{+0.150}_{-0.150}$& $1.882^{+0.303}_{-0.303}$&  $1.463^{+0.235}_{-0.235}$\\
 \hline
 $\frac{d\mathcal{B}}{dq^{2}} (B\to\rho(\to \pi\pi)\mu^{+}\mu^{-})\times 10^{+9} $&$0.959^{+0.150}_{-0.150}$&$1.882^{+0.303}_{-0.303}$&  $1.463^{+0.235}_{-0.235}$\\
 \hline
$A_{\text{FB}}^{\rho}$& $-0.202^{+0.003}_{-0.003}$&  $-0.207^{+0.002}_{-0.002}$&   $-0.228^{+0.002}_{-0.002}$\\
 \hline 
 $f_{L}^{\rho}$ &$0.6602^{+0.0004}_{-0.0004}$&$0.6733^{+0.0006}_{-0.0006}$& $0.6790^{+0.0006}_{-0.0006}$\\
 \hline
 $\frac{d\mathcal{B}}{dq^{2}} (B\to a_{1}\mu^{+}\mu^{-})\times 10^{+9}$&$6.2930^{+4.1389}_{-4.1389}$& $12.4428^{+8.1785}_{-8.1785}$&  $10.004^{+6.622}_{-6.622}$\\
 \hline
 $\frac{d\mathcal{B}}{dq^{2}} (B\to a_{1}(\to \rho_{\parallel}\pi)\mu^{+}\mu^{-})\times 10^{+9}$&$1.8040^{+1.1865}_{-1.1865}$& $3.5669^{+2.3445}_{-2.3445}$&  $2.778^{+1.827}_{-1.827}$\\
 \hline
 $\frac{d\mathcal{B}}{dq^{2}} (B\to a_{1}(\to \rho_{\perp}\pi)\mu^{+}\mu^{-})\times 10^{+9}$&$4.4051^{+2.8972}_{-2.8972}$& $8.7100^{+5.7249}_{-5.7249}$&  $6.786^{+4.461}_{-4.461}$\\
 \hline
$A_{\text{FB}}^{a_{1}}$& $-0.0600^{+0.0046}_{-0.0046}$&  $-0.0681^{+0.0029}_{-0.0029}$&   $-0.0734^{+0.0035}_{-0.0035}$\\
 \hline 
 $f_{L}^{a_{1}}$ &$0.7778^{+0.0059}_{-0.0059}$&$0.7905^{+0.0068}_{-0.0068}$& $0.7864^{+0.0029}_{-0.0029}$\\
 \hline
\end{tabular}
\end{center}
\end{table}

\begin{table}[ht!]
\begin{center}
\caption{Predictions of averaged values of angular observables for the $B\to\rho(\to \pi\pi)\mu^{+}\mu^{-}$ decay, in $q^2 = 0.1 - 1.0$ GeV$^{2}$ bin, for the SM as well as the NP scenarios (S1, S2) of $Z^{\prime}$ model listed in Table \ref{TZP1}. The errors presented mainly come from the uncertainties of the form factors.}\label{table:10}
\begin{tabular}{ |p{5cm}||p{3cm}|p{3cm}|p{3cm}|  }
 \hline
 \multicolumn{4}{|c|}{\textbf{$q^{2}= 0.1 - 1.0$} GeV$^{2}$} \\
 \hline
 \textbf{Angular Observables}& \textbf{SM} &\textbf{$Z^{\prime}$ Scenario 1}&\textbf{$Z^{\prime}$ Scenario 2}\\
 \hline
 $\langle I^{\rho}_{1s}\rangle$   & $0.3524^{+0.0043}_{-0.0043}$   & $0.2056^{+0.0039}_{-0.0039}$&   $0.2533^{+0.0042}_{-0.0042}$\\
 \hline
 $\langle I^{\rho}_{1c}\rangle$&$0.5455^{+0.0060}_{-0.0060}$& $0.7569^{+0.0056}_{-0.0056}$&  $0.6882^{+0.0060}_{-0.0060}$\\
 \hline
$\langle I^{\rho}_{2s}\rangle$&$0.0908^{+0.0011}_{-0.0011}$&$0.0517^{+0.0009}_{-0.0009}$& $0.0641^{+0.0011}_{-0.0011}$\\
 \hline
$\langle I^{\rho}_{2c}\rangle$&$-0.4308^{+0.0055}_{-0.0055}$& $-0.4035^{+0.0054}_{-0.0054}$& $-0.4134^{+0.0055}_{-0.0055}$\\
 \hline
  $\langle I^{\rho}_{3}\rangle$  &$0.0008^{+0.0003}_{-0.0003}$& $0.0012^{+0.0005}_{-0.0005}$& $0.0012^{+0.0005}_{-0.0005}$\\
 \hline
 $\langle I^{\rho}_{4}\rangle$&$-0.1193^{+0.0003}_{-0.0003}$& $-0.1354^{+0.0003}_{-0.0003}$& $-0.1371^{+0.0003}_{-0.0003}$\\
 \hline
$\langle I^{\rho}_{5}\rangle$&$0.0024^{+0.0002}_{-0.0002}$& $0.0013^{+0.00005}_{-0.00005}$& $0.0021^{+0.0001}_{-0.0001}$\\
 \hline
$\langle I^{\rho}_{6s}\rangle$&$0.1028^{+0.0003}_{-0.0003}$&$0.0639^{+0.0002}_{-0.0002}$&  $0.0918^{+0.0002}_{-0.0002}$\\
 \hline 
 $\langle I^{\rho}_{6c}\rangle$   &$0$& $0$&  $0$\\
 \hline
\end{tabular}
\end{center}
\end{table}

\begin{table}[ht!]
\begin{center}
\caption{Same as in Table \ref{table:10}, but for $q^2 = 1.0 - 2.0$ GeV$^{2}$ bin.}\label{table:11}
\begin{tabular}{ |p{5cm}||p{3cm}|p{3cm}|p{3cm}|  }
 \hline
 \multicolumn{4}{|c|}{\textbf{$q^{2}= 1.0 - 2.0$} GeV$^{2}$} \\
 \hline
 \textbf{Angular Observables}& \textbf{SM} &\textbf{$Z^{\prime}$ Scenario 1}&\textbf{$Z^{\prime}$ Scenario 2}\\
 \hline
 $\langle I^{\rho}_{1s}\rangle$  & $0.1122^{+0.0022}_{-0.0022}$ &  $0.0238^{+0.0006}_{-0.0006}$ &   $0.0434^{+0.0011}_{-0.0011}$\\
 \hline
 $\langle I^{\rho}_{1c}\rangle$ & $0.8629^{+0.0029}_{-0.0029}$ &$0.9830^{+0.0008}_{-0.0008}$&  $0.9572^{+0.0014}_{-0.0014}$ \\
 \hline
$\langle I^{\rho}_{2s}\rangle$&  $0.0363^{+0.0007}_{-0.0007}$& $0.0078^{+0.0002}_{-0.0002}$&  $0.0139^{+0.0003}_{-0.0003}$\\
 \hline
$\langle I^{\rho}_{2c}\rangle$&   $-0.8108^{+0.0031}_{-0.0031}$& $-0.8350^{+0.0033}_{-0.0033}$&$-0.8125^{+0.0032}_{-0.0032}$\\
 \hline
  $\langle I^{\rho}_{3}\rangle$   &$-0.0003^{+0.0001}_{-0.0001}$& $0.0013^{+0.0005}_{-0.0005}$& $0.0002^{+0.0001}_{-0.0001}$\\
 \hline
 $\langle I^{\rho}_{4}\rangle$ &$-0.0213^{+0.0003}_{-0.0003}$& $-0.0030^{+0.0006}_{-0.0006}$&  $-0.0150^{+0.0004}_{-0.0004}$\\
 \hline
$\langle I^{\rho}_{5}\rangle$&$-0.0124^{+0.0004}_{-0.0004}$&$-0.0037^{+0.0011}_{-0.0011}$& $-0.0090^{+0.0003}_{-0.0003}$\\
 \hline
$\langle I^{\rho}_{6s}\rangle$&$0.1335^{+0.0022}_{-0.0022}$& $0.0863^{+0.0016}_{-0.0016}$& $0.1257^{+0.0020}_{-0.0020}$\\
 \hline 
 $\langle I^{\rho}_{6c}\rangle$  &$0$& $0$&  $0$\\
 \hline
\end{tabular}
\end{center}
\end{table}

\begin{table}[ht!]
\begin{center}
\caption{Same as in Table \ref{table:10}, but for $q^2 = 2.0 - 3.0$ GeV$^{2}$ bin.}\label{table:12}
\begin{tabular}{|p{5cm}||p{3cm}|p{3cm}|p{3cm}|  }
 \hline
 \multicolumn{4}{|c|}{\textbf{$q^{2}= 2.0 - 3.0$} GeV$^{2}$} \\
 \hline
 \textbf{Angular Observables}& \textbf{SM} &\textbf{$Z^{\prime}$ Scenario 1}&\textbf{$Z^{\prime}$ Scenario 2}\\
 \hline
 $\langle I^{\rho}_{1s}\rangle$   &$0.0854^{+0.0008}_{-0.0008}$&$0.0380^{+0.0002}_{-0.0002}$&  $0.0334^{+0.0003}_{-0.0003}$\\
 \hline
 $\langle I^{\rho}_{1c}\rangle$& $0.8940^{+0.0011}_{-0.0011}$ &  $0.9579^{+0.0003}_{-0.0003}$ &   $0.9506^{+0.0001}_{-0.0001}$ \\
 \hline
$\langle I^{\rho}_{2s}\rangle$&$0.0284^{+0.0003}_{-0.0003}$&$0.0125^{+0.0001}_{-0.0001}$&  $0.0144^{+0.0001}_{-0.0001}$\\
 \hline
$\langle I^{\rho}_{2c}\rangle$&$-0.8623^{+0.0012}_{-0.0012}$&$-0.9200^{+0.0010}_{-0.0010}$&  $-0.8775^{+0.0012}_{-0.0012}$\\
 \hline
  $\langle I^{\rho}_{3}\rangle$  &$-0.0032^{+0.0012}_{-0.0012}$&$-0.0032^{+0.0012}_{-0.0012}$&  $-0.0029^{+0.0011}_{-0.0011}$\\
 \hline
 $\langle I^{\rho}_{4}\rangle$ &$0.0737^{+0.0015}_{-0.0015}$&$0.0941^{+0.0020}_{-0.0020}$&  $0.0830^{+0.0018}_{-0.0018}$\\
 \hline
$\langle I^{\rho}_{5}\rangle$&$-0.0227^{+0.0006}_{-0.0006}$&$-0.0071^{+0.0001}_{-0.0001}$& $-0.0158^{+0.0004}_{-0.0004}$\\
 \hline
$\langle I^{\rho}_{6s}\rangle$&$0.0644^{+0.0025}_{-0.0025}$&$0.0290^{+0.0019}_{-0.0019}$& $0.0612^{+0.0024}_{-0.0024}$\\
 \hline 
 $\langle I^{\rho}_{6c}\rangle$   &$0$& $0$&  $0$\\
 \hline
\end{tabular}
\end{center}
\end{table}

\begin{table}[ht!]
\begin{center}
\caption{Same as in Table \ref{table:10}, but for $q^2 = 3.0 - 4.0$ GeV$^{2}$ bin.}\label{table:13}
\begin{tabular}{ |p{5cm}||p{3cm}|p{3cm}|p{3cm}|  }
 \hline
 \multicolumn{4}{|c|}{\textbf{$q^{2}= 3.0 - 4.0$} GeV$^{2}$} \\
 \hline
 \textbf{Angular Observables}& \textbf{SM} &\textbf{$Z^{\prime}$ Scenario 1}&\textbf{$Z^{\prime}$ Scenario 2}\\
 \hline
 $\langle I^{\rho}_{1s}\rangle$   &$0.1072^{+0.0002}_{-0.0002}$& $0.0804^{+0.0004}_{-0.0004}$& $0.0801^{+0.0003}_{-0.0003}$\\
 \hline
 $\langle I^{\rho}_{1c}\rangle$& $0.8624^{+0.0003}_{-0.0003}$ & $0.8984^{+0.0005}_{-0.0005}$ &  $0.8988^{+0.0004}_{-0.0004}$ \\
 \hline
$\langle I^{\rho}_{2s}\rangle$&$0.0357^{+0.00009}_{-0.00009}$&$0.0265^{+0.00001}_{-0.00001}$&  $0.0266^{+0.00009}_{-0.00009}$\\
 \hline
$\langle I^{\rho}_{2c}\rangle$&$-0.8408^{+0.0004}_{-0.0004}$& $-0.9098^{+0.00008}_{-0.00008}$& $-0.8611^{+0.0003}_{-0.0003}$\\
 \hline
  $\langle I^{\rho}_{3}\rangle$  &$-0.0070^{+0.0024}_{-0.0024}$& $-0.0070^{+0.0021}_{-0.0021}$&  $-0.0067^{+0.0023}_{-0.0023}$\\
 \hline
 $\langle I^{\rho}_{4}\rangle$ &$0.1364^{+0.0025}_{-0.0025}$& $0.1535^{+0.0029}_{-0.0029}$&$0.1452^{+0.0027}_{-0.0027}$\\
 \hline
$\langle I^{\rho}_{5}\rangle$&$-0.0288^{+0.0007}_{-0.0007}$&$-0.0092^{+0.0002}_{-0.0002}$& $-0.0197^{+0.0004}_{-0.0004}$\\
 \hline
$\langle I^{\rho}_{6s}\rangle$&$-0.0159^{+0.0025}_{-0.0025}$& $-0.0405^{+0.0020}_{-0.0020}$& $-0.0167^{+0.0025}_{-0.0025}$\\
 \hline 
 $\langle I^{\rho}_{6c}\rangle$ &$0$& $0$&  $0$\\
 \hline
\end{tabular}
\end{center}
\end{table}

\begin{table}[ht!]
\begin{center}
\caption{Same as in Table \ref{table:10}, but for $q^2 = 4.0 - 5.0$ GeV$^{2}$ bin.}\label{table:14}
\begin{tabular}{ |p{5cm}||p{3cm}|p{3cm}|p{3cm}|  }
 \hline
 \multicolumn{4}{|c|}{\textbf{$q^{2}= 4.0 - 5.0$} GeV$^{2}$} \\
 \hline
 \textbf{Angular Observables}& \textbf{SM} &\textbf{$Z^{\prime}$ Scenario 1}&\textbf{$Z^{\prime}$ Scenario 2}\\
 \hline
 $\langle I^{\rho}_{1s}\rangle$ &$0.1421^{+0.00002}_{-0.00002}$& $0.1260^{+0.0004}_{-0.0004}$& $0.1235^{+0.0004}_{-0.0004}$\\
 \hline
 $\langle I^{\rho}_{1c}\rangle$& $0.8145^{+0.00003}_{-0.00003}$ &  $0.8356^{+0.0005}_{-0.0005}$ &  $0.8393^{+0.0005}_{-0.0005}$ \\
 \hline
$\langle I^{\rho}_{2s}\rangle$&$0.0473^{+0.000005}_{-0.000005}$&$0.0417^{+0.0001}_{-0.0001}$& $0.0409^{+0.00001}_{-0.00001}$\\
 \hline
$\langle I^{\rho}_{2c}\rangle$&$-0.7987^{+0.00001}_{-0.00001}$& $-0.8690^{+0.0005}_{-0.0005}$&  $-0.8202^{+0.0001}_{-0.0001}$\\
 \hline
  $\langle I^{\rho}_{3}\rangle$&$-0.0113^{+0.0036}_{-0.0036}$& $-0.0174^{+0.0030}_{-0.0030}$&  $-0.0111^{+0.0035}_{-0.0035}$\\
 \hline
 $\langle I^{\rho}_{4}\rangle$&$0.1786^{+0.0032}_{-0.0032}$& $0.1922^{+0.0035}_{-0.0035}$& $0.1861^{+0.0035}_{-0.0035}$\\
 \hline
$\langle I^{\rho}_{5}\rangle$&$-0.0323^{+0.0007}_{-0.0007}$&$-0.0106^{+0.0002}_{-0.0002}$&$-0.0219^{+0.0005}_{-0.0005}$\\
 \hline
$\langle I^{\rho}_{6s}\rangle$&$-0.0914^{+0.0026}_{-0.0026}$& $-0.1070^{+0.0020}_{-0.0020}$&$-0.0910^{+0.0025}_{-0.0025}$\\
 \hline 
 $\langle I^{\rho}_{6c}\rangle$   &$0$& $0$&  $0$\\
 \hline
\end{tabular}
\end{center}
\end{table}

\begin{table}[ht!]
\begin{center}
\caption{Same as in Table \ref{table:10}, but for $q^2 = 5.0 - 6.0$ GeV$^{2}$ bin.}\label{table:15}
\begin{tabular}{ |p{5cm}||p{3cm}|p{3cm}|p{3cm}|  }
 \hline
 \multicolumn{4}{|c|}{\textbf{$q^{2}= 5.0 - 6.0$} GeV$^{2}$} \\
 \hline
 \textbf{Angular Observables}& \textbf{SM} &\textbf{$Z^{\prime}$ Scenario 1}&\textbf{$Z^{\prime}$ Scenario 2}\\
 \hline
 $\langle I^{\rho}_{1s}\rangle$&$0.1792^{+0.0002}_{-0.0002}$& $0.1667^{+0.0004}_{-0.0004}$&  $0.1659^{+0.0004}_{-0.0004}$\\
 \hline
 $\langle I^{\rho}_{1c}\rangle$ & $0.7640^{+0.0002}_{-0.0002}$ &  $0.7767^{+0.0006}_{-0.0006}$ &   $0.7818^{+0.0006}_{-0.0006}$ \\
 \hline
$\langle I^{\rho}_{2s}\rangle$&$0.0596^{+0.0001}_{-0.0001}$&$0.0561^{+0.0001}_{-0.0001}$&  $0.0550^{+0.0001}_{-0.0001}$\\
 \hline
$\langle I^{\rho}_{2c}\rangle$&$-0.7520^{+0.0002}_{-0.0002}$& $-0.8191^{+0.0007}_{-0.0007}$&  $-0.7730^{+0.0004}_{-0.0004}$\\
 \hline
  $\langle I^{\rho}_{3}\rangle$ &$-0.0163^{+0.0048}_{-0.0048}$&$-0.0164^{+0.0048}_{-0.0049}$&  $-0.0161^{+0.0047}_{-0.0047}$\\
 \hline
 $\langle I^{\rho}_{4}\rangle $&$0.2078^{+0.0036}_{-0.0036}$& $0.2185^{+0.0039}_{-0.0039}$&  $0.2140^{+0.0031}_{-0.0031}$\\
 \hline
$\langle I^{\rho}_{5}\rangle$&$-0.0343^{+0.0008}_{-0.0008}$& $-0.0116^{+0.0002}_{-0.0002}$&  $-0.0231^{+0.0005}_{-0.0005}$\\
 \hline
$\langle I^{\rho}_{6s}\rangle$&$-0.1589^{+0.0028}_{-0.0028}$&$-0.1668^{+0.0022}_{-0.0022}$& $-0.1578^{+0.0027}_{-0.0027}$\\
 \hline 
 $\langle I^{\rho}_{6c}\rangle$& $0$& $0$&  $0$\\
 \hline
\end{tabular}
\end{center}
\end{table}

\begin{table}[ht!]
\begin{center}
\caption{Same as in Table \ref{table:10}, but for $q^2 = 6.0 - 7.0$ GeV$^{2}$ bin.}\label{table:16}
\begin{tabular}{ |p{5cm}||p{3cm}|p{3cm}|p{3cm}|  }
 \hline
 \multicolumn{4}{|c|}{\textbf{$q^{2}= 6.0 - 7.0$} GeV$^{2}$} \\
 \hline
 \textbf{Angular Observables}& \textbf{SM} &\textbf{$Z^{\prime}$ Scenario 1}&\textbf{$Z^{\prime}$ Scenario 2}\\
 \hline
 $\langle I^{\rho}_{1s}\rangle$&$0.2151^{+0.0003}_{-0.0003}$& $0.2093^{+0.0005}_{-0.0005}$&  $0.2051^{+0.0005}_{-0.0005}$\\
 \hline
 $\langle I^{\rho}_{1c}\rangle$& $0.7156^{+0.0004}_{-0.0004}$ &  $0.7232^{+0.0006}_{-0.0006}$ &   $0.7288^{+0.00056}_{-0.0006}$ \\
 \hline
$\langle I^{\rho}_{2s}\rangle $&$0.0715^{+0.0001}_{-0.0001}$& $0.0693^{+0.0001}_{-0.0001}$& $0.0680^{+0.0002}_{-0.0002}$\\
 \hline
$\langle I^{\rho}_{2c}\rangle$&$-0.7062^{+0.0004}_{-0.0004}$& $-0.7685^{+0.0009}_{-0.0009}$&  $-0.7259^{+0.0005}_{-0.0005}$\\
 \hline
  $\langle I^{\rho}_{3}\rangle$ &$-0.0218^{+0.0058}_{-0.0058}$&$-0.0220^{+0.0058}_{-0.0058}$&  $-0.0216^{+0.0057}_{-0.0057}$\\
 \hline
 $\langle I^{\rho}_{4}\rangle$& $0.2286^{+0.0040}_{-0.0040}$&  $0.2371^{+0.0042}_{-0.0042}$&   $0.2338^{+0.0041}_{-0.0041}$\\
 \hline
$\langle I^{\rho}_{5}\rangle$&$-0.0354^{+0.0008}_{-0.0008}$& $-0.0122^{+0.0003}_{-0.0003}$&  $-0.0239^{+0.0006}_{-0.0006}$\\
 \hline
$\langle I^{\rho}_{6s}\rangle $&$-0.2182^{+0.0031}_{-0.0031}$&$-0.2192^{+0.0025}_{-0.0025}$& $-0.2166^{+0.0030}_{-0.0030}$\\
 \hline 
 $\langle I^{\rho}_{6c}\rangle$   &$0$& $0$&  $0$\\
 \hline
\end{tabular}
\end{center}
\end{table}

\begin{table}[ht!]
\begin{center}
\caption{Same as in Table \ref{table:10}, but for $q^2 = 7.0 - 8.0$ GeV$^{2}$ bin.}\label{table:17}
\begin{tabular}{ |p{5cm}||p{3cm}|p{3cm}|p{3cm}|  }
 \hline
 \multicolumn{4}{|c|}{\textbf{$q^{2}= 7.0 - 8.0$} GeV$^{2}$} \\
 \hline
 \textbf{Angular Observables}& \textbf{SM} &\textbf{$Z^{\prime}$ Scenario 1}&\textbf{$Z^{\prime}$ Scenario 2}\\
 \hline
 $\langle I^{\rho}_{1s}\rangle$&$0.2483^{+0.0003}_{-0.0003}$&$0.2448^{+0.0005}_{-0.0005}$&  $0.2406^{+0.0005}_{-0.0005}$\\
 \hline
 $\langle I^{\rho}_{1c}\rangle$& $0.6708^{+0.0004}_{-0.0004}$ &  $0.6753^{+0.0006}_{-0.0006}$ &  $0.6820^{+0.0007}_{-0.0007}$ \\
 \hline
$\langle I^{\rho}_{2s}\rangle$&$0.0826^{+0.0001}_{-0.0001}$&$0.0811^{+0.0002}_{-0.0002}$&  $0.0798^{+0.0002}_{-0.0002}$\\
 \hline
$\langle I^{\rho}_{2c}\rangle$&$-0.6632^{+0.0005}_{-0.0005}$&$-0.7230^{+0.0009}_{-0.0009}$& $-0.6815^{+0.0006}_{-0.0006}$\\
 \hline
  $\langle I^{\rho}_{3}\rangle$&$-0.0280^{+0.0068}_{-0.0068}$& $-0.0282^{+0.0065}_{-0.0065}$&  $-0.0278^{+0.0067}_{-0.0067}$\\
 \hline
 $\langle I^{\rho}_{4}\rangle$&$0.2437^{+0.0042}_{-0.0042}$& $0.2504^{+0.0044}_{-0.0044}$& $0.2480^{+0.0043}_{-0.0043}$\\
 \hline
$\langle I^{\rho}_{5}\rangle$&$-0.0357^{+0.0009}_{-0.0009}$&$-0.0126^{+0.0003}_{-0.0003}$& $-0.0239^{+0.0006}_{-0.0006}$\\
 \hline
$\langle I^{\rho}_{6s}\rangle$&$-0.2701^{+0.0035}_{-0.0035}$& $-0.2650^{+0.0029}_{-0.0029}$& $-0.2681^{+0.0034}_{-0.0034}$\\
 \hline 
 $\langle I^{\rho}_{6c}\rangle$ &$0$& $0$&  $0$\\
 \hline
\end{tabular}
\end{center}
\end{table}

\begin{table}[ht!]
\begin{center}
\caption{Predictions of averaged values of angular observables for the $B\to a_{1}(\to\rho_{\parallel}\pi)\mu^{+}\mu^{-}$ decay, in $q^2 = 0.1 - 1.0$ GeV$^{2}$ bin, for the SM as well as the NP scenarios (S1, S2) of $Z^{\prime}$ model listed in Table \ref{TZP1}. The errors presented mainly come from the uncertainties of the form factors.}\label{table:19}
\begin{tabular}{ |p{5cm}||p{3cm}|p{3cm}|p{3cm}|  }
 \hline
 \multicolumn{4}{|c|}{\textbf{$q^{2}= 0.1 - 1.0$} GeV$^{2}$} \\
 \hline
 \textbf{Angular Observables}& \textbf{SM} &\textbf{$Z^{\prime}$ Scenario 1}&\textbf{$Z^{\prime}$ Scenario 2}\\
 \hline
 $\langle\widehat{I}^{a_{1}}_{1s,\parallel}\rangle$   & $0.0414^{+0.0007}_{-0.0007}$    & $0.0190^{+0.0006}_{-0.0006}$&   $0.0251^{+0.0006}_{-0.0006}$\\
 \hline
$\langle\widehat{I}^{a_{1}}_{1c,\parallel}\rangle$&$0.2390^{+0.0009}_{-0.0009}$& $0.2737^{+0.0007}_{-0.0007}$&  $0.2643^{+0.0008}_{-0.0008}$\\
 \hline
$\langle\widehat{I}^{a_{1}}_{2s,\parallel}\rangle$&$0.0107^{+0.0002}_{-0.0002}$& $0.0048^{+0.0001}_{-0.0001}$&  $0.0064^{+0.0002}_{-0.0002}$\\
 \hline
$\langle\widehat{I}^{a_{1}}_{2c,\parallel}\rangle$&$-0.2027^{+0.0015}_{-0.0015}$& $-0.2211^{+0.0008}_{-0.0008}$&  $-0.2159^{+0.0010}_{-0.0010}$\\
 \hline
  $\langle\widehat{I}^{a_{1}}_{3,\parallel}\rangle$  &$0.0027^{+0.00001}_{-0.00001}$& $0.0024^{+0.00003}_{-0.00003}$&  $0.0026^{+0.000002}_{-0.000002}$\\
 \hline
 $\langle\widehat{I}^{a_{1}}_{4,\parallel}\rangle$&$-0.0218^{+0.0007}_{-0.0007}$& $-0.0183^{+0.0007}_{-0.0007}$&  $-0.0207^{+0.0007}_{-0.0007}$\\
 \hline
$\langle\widehat{I}^{a_{1}}_{5,\parallel}\rangle$&$0.0639^{+0.0002}_{-0.001}$& $0.0265^{+0.0003}_{-0.0003}$&  $0.0366^{+0.0003}_{-0.0003}$\\
 \hline
$\langle\widehat{I}^{a_{1}}_{6s,\parallel}\rangle$&$0.0133^{+0.0002}_{-0.0002}$& $0.0051^{+0.00002}_{-0.00002}$&  $0.0073^{+0.0001}_{-0.0001}$\\
 \hline 
 $\langle\widehat{I}^{a_{1}}_{6c,\parallel}\rangle$   &$0$& $0$&  $0$\\
 \hline
\end{tabular}
\end{center}
\end{table}

\begin{table}[ht!]
\begin{center}
\caption{Same as in Table \ref{table:19}, but for $q^2 = 1.0 - 2.0$ GeV$^{2}$ bin.}\label{table:20}
\begin{tabular}{ |p{5cm}||p{3cm}|p{3cm}|p{3cm}|  }
 \hline
 \multicolumn{4}{|c|}{\textbf{$q^{2}= 1.0 - 2.0$} GeV$^{2}$} \\
 \hline
 \textbf{Angular Observables}& \textbf{SM} &\textbf{$Z^{\prime}$ Scenario 1}&\textbf{$Z^{\prime}$ Scenario 2}\\
 \hline
 $\langle\widehat{I}^{a_{1}}_{1s,\parallel}\rangle$  &$0.0112^{+0.0002}_{-0.0002}$& $0.0038^{+0.00003}_{-0.00003}$&  $0.0052^{+0.0004}_{-0.0004}$\\
 \hline
 $\langle\widehat{I}^{a_{1}}_{1c,\parallel}\rangle$ &$0.2743^{+0.0002}_{-0.0002}$& $0.2855^{+0.00007}_{-0.00007}$&  $0.2533^{+0.0001}_{-0.0001}$\\
 \hline
$\langle\widehat{I}^{a_{1}}_{2s,\parallel}\rangle$&  $0.0036^{+0.00005}_{-0.00005}$& $0.0012^{+0.00001}_{-0.00001}$&  $0.0017^{+0.00001}_{-0.00001}$\\
 \hline
$\langle\widehat{I}^{a_{1}}_{2c,\parallel}\rangle$&   $-0.2640^{+0.0005}_{-0.0005}$& $-0.2702^{+0.00004}_{-0.00004}$&  $-0.2692^{+0.0002}_{-0.0002}$\\
 \hline
  $\langle\widehat{I}^{a_{1}}_{3\parallel}\rangle$   &$0.0004^{+0.0001}_{-0.0001}$& $-0.0002^{+0.0002}_{-0.0002}$&  $0.0016^{+0.0002}_{-0.0002}$\\
 \hline
 $\langle\widehat{I}^{a_{1}}_{4\parallel}\rangle$ &$0.0103^{+0.0012}_{-0.0012}$& $0.0135^{+0.0011}_{-0.0011}$&  $0.0116^{+0.0011}_{-0.0011}$\\
 \hline
$\langle\widehat{I}^{a_{1}}_{5\parallel}\rangle$&$0.0408^{+0.0008}_{-0.0008}$& $0.0124^{+0.0005}_{-0.0005}$&  $0.0193^{+0.0007}_{-0.0007}$\\
 \hline
$\langle\widehat{I}^{a_{1}}_{6s,\parallel}\rangle$&$0.0136^{+0.0002}_{-0.0002}$& $0.0029^{+0.0002}_{-0.0002}$&  $0.0053^{+0.0002}_{-0.0002}$\\
 \hline 
 $\langle\widehat{I}^{a_{1}}_{6c,\parallel}\rangle$  &$0$& $0$&  $0$\\
 \hline
\end{tabular}
\end{center}
\end{table}

\begin{table}[ht!]
\begin{center}
\caption{Same as in Table \ref{table:19}, but for $q^2 = 2.0 - 3.0$ GeV$^{2}$ bin.}\label{table:21}
\begin{tabular}{ |p{5cm}||p{3cm}|p{3cm}|p{3cm}|  }
 \hline
 \multicolumn{4}{|c|}{\textbf{$q^{2}= 2.0 - 3.0$} GeV$^{2}$} \\
 \hline
 \textbf{Angular Observables}& \textbf{SM} &\textbf{$Z^{\prime}$ Scenario 1}&\textbf{$Z^{\prime}$ Scenario 2}\\
 \hline
 $\langle\widehat{I}^{a_{1}}_{1s,\parallel}\rangle$   &$0.0111^{+0.0003}_{-0.0003}$& $0.0072^{+0.0004}_{-0.0004}$&  $0.0076^{+0.0004}_{-0.0004}$\\
 \hline $\langle\widehat{I}^{a_{1}}_{1c,\parallel}\rangle$&$0.2733^{+0.0004}_{-0.0004}$& $0.2792^{+0.0006}_{-0.0006}$&  $0.2786^{+0.0005}_{-0.0005}$\\
 \hline
$\langle\widehat{I}^{a_{1}}_{2s,\parallel}\rangle$&$0.0037^{+0.0001}_{-0.0001}$& $0.0024^{+0.0001}_{-0.0001}$&  $0.0025^{+0.0001}_{-0.0001}$\\
 \hline
$\langle\widehat{I}^{a_{1}}_{2c,\parallel}\rangle$&$-0.2675^{+0.0002}_{-0.0002}$& $-0.2705^{+0.0005}_{-0.0005}$&  $-0.2705^{+0.0004}_{-0.0004}$\\
 \hline
  $\langle \widehat{I}^{a_{1}}_{3\parallel}\rangle$  &$-0.0035^{+0.0003}_{-0.0003}$& $-0.0040^{+0.0003}_{-0.0003}$&  $-0.0037^{+0.0003}_{-0.0003}$\\
 \hline
 $\langle\widehat{I}^{a_{1}}_{4\parallel}\rangle$ &$0.0305^{+0.0014}_{-0.0014}$& $0.0329^{+0.0012}_{0.0012}$&  $0.0314^{+0.0013}_{-0.0013}$\\
 \hline
$\langle\widehat{I}^{a_{1}}_{5\parallel}\rangle$&$0.0255^{+0.0011}_{-0.0011}$& $0.0043^{+0.0006}_{-0.0006}$&  $0.0090^{+0.0007}_{-0.0007}$\\
 \hline
$\langle\widehat{I}^{a_{1}}_{6s,\parallel}\rangle$&$0.0088^{+0.0005}_{-0.0005}$& $-0.0014^{+0.0005}_{-0.0005}$&  $0.0001^{+0.0001}_{-0.0001}$\\
 \hline 
 $\langle\widehat{I}^{a_{1}}_{6c,\parallel}\rangle$   &$0$& $0$&  $0$\\
 \hline
\end{tabular}
\end{center}
\end{table}

\begin{table}[ht!]
\begin{center}
\caption{Same as in Table \ref{table:19}, but for $q^2 = 3.0 - 4.0$ GeV$^{2}$ bin.}\label{table:22}
\begin{tabular}{ |p{5cm}||p{3cm}|p{3cm}|p{3cm}|  }
 \hline
 \multicolumn{4}{|c|}{\textbf{$q^{2}= 3.0 - 4.0$} GeV$^{2}$} \\
 \hline
 \textbf{Angular Observables}& \textbf{SM} &\textbf{$Z^{\prime}$ Scenario 1}&\textbf{$Z^{\prime}$ Scenario 2}\\
 \hline
 $\langle\widehat{I}^{a_{1}}_{1s,\parallel}\rangle$   &$0.0158^{+0.0006}_{-0.0006}$& $0.0135^{+0.0007}_{-0.0007}$&  $0.0134^{+0.0007}_{-0.0007}$\\
 \hline $\langle\widehat{I}^{a_{1}}_{1c,\parallel}\rangle$&$0.2666^{+0.0008}_{-0.0008}$& $0.2702^{+0.0010}_{-0.0010}$&  $0.2702^{+0.0001}_{-0.0001}$\\
 \hline
$\langle\widehat{I}^{a_{1}}_{2s,\parallel}\rangle$&$0.0053^{+0.0002}_{-0.0002}$& $0.0044^{+0.0002}_{-0.0002}$&  $0.0044^{+0.0002}_{-0.0002}$\\
 \hline
$\langle\widehat{I}^{a_{1}}_{2c,\parallel}\rangle$&$-0.2625^{+0.0007}_{-0.0007}$& $-0.2642^{+0.0009}_{-0.0009}$&  $-0.2647^{+0.0009}_{-0.0009}$\\
 \hline
  $\langle\widehat{I}^{a_{1}}_{3\parallel}\rangle$  &$-0.0076^{+0.0004}_{-0.0004}$& $-0.0082^{+0.0004}_{-0.0004}$&  $-0.0078^{+0.0003}_{-0.0003}$\\
 \hline
 $\langle\widehat{I}^{a_{1}}_{4\parallel}\rangle$ &$0.0448^{+0.0014}_{-0.0014}$& $0.0467^{+0.0012}_{-0.0012}$&  $0.0455^{+0.0013}_{-0.0013}$\\
 \hline
$\langle\widehat{I}^{a_{1}}_{5\parallel}\rangle$&$0.0155^{+0.0012}_{-0.0012}$& $-0.0011^{+0.0007}_{-0.0007}$&  $0.0021^{+0.0008}_{-0.0008}$\\
 \hline
$\langle\widehat{I}^{a_{1}}_{6s,\parallel}\rangle$&$0.0030^{+0.0009}_{-0.0009}$& $0.0060^{+0.0007}_{-0.0007}$&  $0.0048^{+0.0008}_{-0.0008}$\\
 \hline 
 $\langle\widehat{I}^{a_{1}}_{6c,\parallel}\rangle$ &$0$& $0$&  $0$\\
 \hline
\end{tabular}
\end{center}
\end{table}

\begin{table}[ht!]
\begin{center}
\caption{Same as in Table \ref{table:19}, but for $q^2 = 4.0 - 5.0$ GeV$^{2}$ bin.}\label{table:23}
\begin{tabular}{ |p{5cm}||p{3cm}|p{3cm}|p{3cm}|  }
 \hline
 \multicolumn{4}{|c|}{\textbf{$q^{2}= 4.0 - 5.0$} GeV$^{2}$} \\
 \hline
 \textbf{Angular Observables}& \textbf{SM} &\textbf{$Z^{\prime}$ Scenario 1}&\textbf{$Z^{\prime}$ Scenario 2}\\
 \hline
 $\langle\widehat{I}^{a_{1}}_{1s,\parallel}\rangle$ &$0.0224^{+0.0009}_{-0.0009}$& $0.0209^{+0.0010}_{-0.0010}$&  $0.0206^{+0.0009}_{-0.0009}$\\
 \hline
 $\langle\widehat{I}^{a_{1}}_{1c,\parallel}\rangle$&$0.2575^{+0.0012}_{-0.0012}$& $0.2599^{+0.0013}_{-0.0013}$&  $0.2603^{+0.0010}_{-0.0010}$\\
 \hline
$\langle\widehat{I}^{a_{1}}_{2s,\parallel}\rangle$&$0.0075^{+0.0003}_{-0.0003}$& $0.0069^{+0.0003}_{-0.0003}$&  $0.0068^{+0.0003}_{-0.0003}$\\
 \hline
$\langle\widehat{I}^{a_{1}}_{2c,\parallel}\rangle$&$-0.2545^{+0.0010}_{-0.0010}$& $-0.2554^{+0.0012}_{-0.0012}$&  $-0.2561^{+0.0012}_{-0.0012}$\\
 \hline
  $\langle\widehat{I}^{a_{1}}_{3\parallel}\rangle$&$-0.0121^{+0.0005}_{-0.0005}$& $-0.0126^{+0.0005}_{-0.0005}$&  $-0.0123^{+0.0005}_{-0.0005}$\\
 \hline
 $\langle\widehat{I}^{a_{1}}_{4\parallel}\rangle$&$0.0561^{+0.0013}_{-0.0013}$& $0.0576^{+0.0012}_{-0.0012}$&  $0.0567^{+0.0012}_{-0.0012}$\\
 \hline
$\langle\widehat{I}^{a_{1}}_{5\parallel}\rangle$&$0.0082^{+0.0013}_{-0.0013}$& $-0.0051^{+0.0007}_{-0.0007}$&  $0.0029^{+0.0009}_{-0.0009}$\\
 \hline
$\langle\widehat{I}^{a_{1}}_{6s,\parallel}\rangle$&$-0.0032^{+0.0012}_{-0.0012}$& $-0.0109^{+0.0008}_{-0.0008}$&  $-0.0104^{+0.0001}_{-0.0001}$\\
 \hline 
 $\langle\widehat{I}^{a_{1}}_{6c,\parallel}\rangle$   &$0$& $0$&  $0$\\
 \hline
\end{tabular}
\end{center}
\end{table}

\begin{table}[ht!]
\begin{center}
\caption{Same as in Table \ref{table:19}, but for $q^2 = 5.0 - 6.0$ GeV$^{2}$ bin.}\label{table:24}
\begin{tabular}{ |p{5cm}||p{3cm}|p{3cm}|p{3cm}|  }
 \hline
 \multicolumn{4}{|c|}{\textbf{$q^{2}= 5.0 - 6.0$} GeV$^{2}$} \\
 \hline
 \textbf{Angular Observables}& \textbf{SM} &\textbf{$Z^{\prime}$ Scenario 1}&\textbf{$Z^{\prime}$ Scenario 2}\\
 \hline
$\langle\widehat{I}^{a_{1}}_{1s,\parallel}\rangle$&$0.0301^{+0.0011}_{-0.0011}$& $0.0291^{+0.0011}_{-0.0011}$&  $0.0286^{+0.0013}_{-0.0013}$\\
 \hline
 $\langle\widehat{I}^{a_{1}}_{1c,\parallel}\rangle$ &$0.2471^{+0.0014}_{-0.0014}$& $0.2487^{+0.0015}_{-0.0015}$&  $0.2493^{+0.0015}_{-0.0015}$\\
 \hline
$\langle\widehat{I}^{a_{1}}_{2s,\parallel}\rangle$&$0.0100^{+0.0004}_{-0.0004}$& $0.0096^{+0.0004}_{-0.0004}$&  $0.0095^{+0.0038}_{-0.0038}$\\
 \hline
$\langle\widehat{I}^{a_{1}}_{2c,\parallel}\rangle$&$-0.2448^{+0.0013}_{-0.0013}$& $-0.2452^{+0.0015}_{-0.0015}$&  $-0.2460^{+0.0015}_{-0.0015}$\\
 \hline
  $\langle\widehat{I}^{a_{1}}_{3\parallel}\rangle$ &$-0.0169^{+0.0006}_{-0.0006}$& $-0.0174^{+0.0006}_{-0.0006}$&  $-0.0170^{+0.0006}_{-0.0006}$\\
 \hline
 $\langle\widehat{I}^{a_{1}}_{4\parallel}\rangle$&$0.0655^{+0.0011}_{-0.0011}$& $0.0666^{+0.0010}_{-0.0010}$&  $0.0659^{+0.0011}_{-0.0011}$\\
 \hline
$\langle\widehat{I}^{a_{1}}_{5\parallel}\rangle$&$0.0025^{+0.0013}_{-0.0013}$& $-0.0082^{+0.0007}_{-0.0007}$&  $0.0068^{+0.0009}_{-0.0009}$\\
 \hline
$\langle\widehat{I}^{a_{1}}_{6s,\parallel}\rangle$&$-0.0096^{+0.0014}_{-0.0014}$& $-0.0158^{+0.0010}_{-0.0010}$&  $-0.0162^{+0.0013}_{-0.0013}$\\
 \hline 
 $\langle\widehat{I}^{a_{1}}_{6c,\parallel}\rangle$&$0$& $0$&  $0$\\
 \hline
\end{tabular}
\end{center}
\end{table}

\begin{table}[ht!]
\begin{center}
\caption{Same as in Table \ref{table:19}, but for $q^2 = 6.0 - 7.0$ GeV$^{2}$ bin.}\label{table:25}
\begin{tabular}{ |p{5cm}||p{3cm}|p{3cm}|p{3cm}|  }
 \hline
 \multicolumn{4}{|c|}{\textbf{$q^{2}= 6.0 - 7.0$} GeV$^{2}$} \\
 \hline
 \textbf{Angular Observables}& \textbf{SM} &\textbf{$Z^{\prime}$ Scenario 1}&\textbf{$Z^{\prime}$ Scenario 2}\\
 \hline $\langle\widehat{I}^{a_{1}}_{1s,\parallel}\rangle$&$0.0386^{+0.0012}_{-0.0012}$& $0.0379^{+0.0013}_{-0.0013}$&  $0.0374^{+0.0013}_{-0.0013}$\\
 \hline
 $\langle\widehat{I}^{a_{1}}_{1c,\parallel}\rangle$&$0.2357^{+0.0016}_{-0.0016}$& $0.2368^{+0.0017}_{-0.0017}$&  $0.2374^{+0.0017}_{-0.0017}$\\
 \hline
$\langle\widehat{I}^{a_{1}}_{2s,\parallel}\rangle$&$0.0128^{+0.0004}_{-0.0004}$& $0.0125^{+0.0004}_{-0.0004}$&  $0.0124^{+0.00004}_{-0.00004}$\\
 \hline
$\langle\widehat{I}^{a_{1}}_{2c,\parallel}\rangle$&$-0.2338^{+0.0015}_{-0.0015}$& $-0.2340^{+0.0017}_{-0.0017}$&  $-0.2348^{+0.0016}_{-0.0016}$\\
 \hline
  $\langle\widehat{I}^{a_{1}}_{3,\parallel}\rangle$ &$-0.0220^{+0.0006}_{-0.0006}$& $-0.0224^{+0.0006}_{-0.0006}$&  $-0.0221^{+0.0007}_{-0.0007}$\\
 \hline
 $\langle\widehat{I}^{a_{1}}_{4,\parallel}\rangle$&$0.0734^{+0.0010}_{-0.0010}$& $0.0743^{+0.0009}_{-0.0009}$&  $0.0737^{+0.0012}_{-0.0012}$\\
 \hline
$\langle\widehat{I}^{a_{1}}_{5,\parallel}\rangle$&$-0.0021^{+0.0012}_{-0.0012}$& $-0.0106^{+0.0007}_{-0.0007}$&  $-0.0091^{+0.0008}_{-0.0008}$\\
 \hline
$\langle\widehat{I}^{a_{1}}_{6s,\parallel}\rangle$&$-0.0162^{+0.0016}_{-0.0016}$& $-0.0209^{+0.0011}_{-0.0011}$&  $-0.0220^{+0.0013}_{-0.0013}$\\
 \hline 
 $\langle\widehat{I}^{a_{1}}_{6c,\parallel}\rangle$   &$0$& $0$&  $0$\\
 \hline
\end{tabular}
\end{center}
\end{table}

\begin{table}[ht!]
\begin{center}
\caption{Same as in Table \ref{table:19}, but for $q^2 = 7.0 - 8.0$ GeV$^{2}$ bin.}\label{table:26}
\begin{tabular}{ |p{5cm}||p{3cm}|p{3cm}|p{3cm}|  }
 \hline
 \multicolumn{4}{|c|}{\textbf{$q^{2}= 7.0 - 8.0$} GeV$^{2}$} \\
 \hline
 \textbf{Angular Observables}& \textbf{SM} &\textbf{$Z^{\prime}$ Scenario 1}&\textbf{$Z^{\prime}$ Scenario 2}\\
 \hline
 $\langle\widehat{I}^{a_{1}}_{1s,\parallel}\rangle$&$0.0478^{+0.0013}_{-0.0013}$& $0.0201^{+0.0008}_{-0.0008}$&  $0.0467^{+0.0013}_{-0.0013}$\\
 \hline
 $\langle\widehat{I}^{a_{1}}_{1c,\parallel}\rangle$&$0.2233^{+0.0017}_{-0.0017}$& $0.2241^{+0.0018}_{-0.0018}$&  $0.2248^{+0.0018}_{-0.0018}$\\
 \hline
$\langle\widehat{I}^{a_{1}}_{2s,\parallel}\rangle$&$0.0159^{+0.0004}_{-0.0004}$& $0.0157^{+0.0004}_{-0.0004}$&  $0.0155^{+0.0004}_{-0.0004}$\\
 \hline
$\langle\widehat{I}^{a_{1}}_{2c,\parallel}\rangle$&$-0.2218^{+0.0016}_{-0.0016}$& $-0.2218^{+0.0017}_{-0.0017}$&  $-0.2227^{+0.0018}_{-0.0018}$\\
 \hline
  $\langle\widehat{I}^{a_{1}}_{3,\parallel}\rangle$&$-0.0274^{+0.0006}_{-0.0006}$& $-0.0278^{+0.0006}_{-0.0006}$&  $-0.0275^{+0.0006}_{-0.0006}$\\
 \hline
 $\langle\widehat{I}^{a_{1}}_{4,\parallel}\rangle$&$0.0801^{+0.0008}_{-0.0008}$& $0.0808^{+0.0007}_{-0.0007}$&  $0.0804^{+0.0007}_{-0.0007}$\\
 \hline
$\langle\widehat{I}^{a_{1}}_{5,\parallel}\rangle$&$-0.0058^{+0.0012}_{-0.0012}$& $-0.0126^{+0.0006}_{-0.0006}$&  $-0.0125^{+0.0008}_{-0.0008}$\\
 \hline
$\langle\widehat{I}^{a_{1}}_{6s,\parallel}\rangle$&$-0.0229^{+0.0017}_{-0.0017}$& $-0.0260^{+0.0011}_{-0.0011}$&  $-0.0226^{+0.0017}_{-0.0017}$\\
 \hline 
 $\langle\widehat{I}^{a_{1}}_{6c,\parallel}\rangle$ &$0$& $0$&  $0$\\
 \hline 
\end{tabular}
\end{center}
\end{table}

\begin{table}[ht!]
\begin{center}
\caption{Predictions of averaged values of angular observables for the $B\to a_{1}(\to\rho_{\perp}\pi)\mu^{+}\mu^{-}$ decay, in $q^2 = 0.1 - 1.0$ GeV$^{2}$ bin, for the SM as well as the NP scenarios (S1, S2) of $Z^{\prime}$ model listed in Table \ref{TZP1}. The errors presented mainly come from the uncertainties of the form factors.}\label{table:28}
\begin{tabular}{ |p{5cm}||p{3cm}|p{3cm}|p{3cm}|  }
 \hline
 \multicolumn{4}{|c|}{\textbf{$q^{2}= 0.1 - 1.0$} GeV$^{2}$} \\
 \hline
 \textbf{Angular Observables}& \textbf{SM} &\textbf{$Z^{\prime}$ Scenario 1}&\textbf{$Z^{\prime}$ Scenario 2}\\
 \hline
 $\langle\widehat{I}^{a_{1}}_{1s,\perp}\rangle$   & $0.3423^{+0.0002}_{-0.0002}$    & $0.3574^{+0.0002}_{-0.0002}$&   $0.3533^{+0.0002}_{-0.0002}$\\
 \hline
 $\langle\widehat{I}^{a_{1}}_{1c,\perp}\rangle$&$0.1010^{+0.0018}_{-0.0018}$& $0.0465^{+0.0013}_{-0.0013}$& $0.0613^{+0.0092}_{-0.0092}$\\
 \hline
$\langle\widehat{I}^{a_{1}}_{2s,\perp}\rangle$&$-0.2345^{+0.0020}_{-0.0020}$& $-0.2641^{+0.0012}_{-0.0012}$&  $-0.2558^{+0.0015}_{-0.0015}$\\
 \hline
$\langle\widehat{I}^{a_{1}}_{2c,\perp}\rangle$&$0.0260^{+0.0005}_{-0.0005}$ & $0.0117^{+0.0004}_{-0.0004}$ &  $0.0156^{+0.0004}_{-0.0004}$ \\
 \hline
  $\langle\widehat{I}^{a_{1}}_{3,\perp}\rangle$  &$0.0008^{+0.00003}_{-0.00003}$& $0.0009^{+0.00003}_{-0.00003}$&  $0.0001^{+0.00003}_{-0.00003}$\\
 \hline
 $\langle\widehat{I}^{a_{1}}_{4,\perp}\rangle$&$0.0267^{+0.0008}_{-0.0008}$& $0.0224^{+0.0009}_{-0.0009}$&  $0.0252^{+0.0009}_{-0.0009}$\\
 \hline
$\langle\widehat{I}^{a_{1}}_{5,\perp}\rangle$&$-0.0780^{+0.0002}_{-0.0002}$&$-0.0324^{+0.0003}_{-0.0003}$& $-0.0447^{+0.0003}_{-0.0003}$\\
 \hline
$\langle\widehat{I}^{a_{1}}_{6s,\perp}\rangle$& $0.0162^{+0.0002}_{-0.0002}$& $0.0063^{+0.00002}_{-0.00002}$&  $0.0089^{+0.0001}_{-0.0001}$\\
 \hline 
 $\langle\widehat{I}^{a_{1}}_{6c,\perp}\rangle$ &  $0.0324^{+0.0005}_{-0.0005}$&  $0.0126^{+0.00005}_{-0.00005}$&  $0.0178^{+0.0001}_{-0.0001}$\\
 \hline
\end{tabular}
\end{center}
\end{table}

\begin{table}[ht!]
\begin{center}
\caption{Same as in Table \ref{table:28}, but for $q^2 = 1.0 - 2.0$ GeV$^{2}$ bin.}\label{table:29}
\begin{tabular}{ |p{5cm}||p{3cm}|p{3cm}|p{3cm}|  }
 \hline
 \multicolumn{4}{|c|}{\textbf{$q^{2}= 1.0 - 2.0$} GeV$^{2}$} \\
 \hline
 \textbf{Angular Observables}& \textbf{SM} &\textbf{$Z^{\prime}$ Scenario 1}&\textbf{$Z^{\prime}$ Scenario 2}\\
 \hline
 $\langle\widehat{I}^{a_{1}}_{1s,\perp}\rangle$  &$0.3486^{+0.00002}_{-0.00002}$& $0.3531^{+0.00005}_{-0.00005}$&  $0.3533^{+0.00002}_{-0.00002}$\\
 \hline
 $\langle\widehat{I}^{a_{1}}_{1c,\perp}\rangle$ &$0.0272^{+0.0004}_{-0.0004}$ &$0.0092^{+0.00008}_{-0.00008}$ &$0.0126^{+0.0001}_{-0.0001}$\\
 \hline
$\langle\widehat{I}^{a_{1}}_{2s,\perp}\rangle$& $-0.3179^{+0.0007}_{-0.0007}$&$-0.3284^{+0.00003}_{-0.00003}$&  $-0.3266^{+0.0002}_{-0.0002}$\\
 \hline
$\langle\widehat{I}^{a_{1}}_{2c,\perp}\rangle$&    $0.0088^{+0.0001}_{-0.0001}$ & $0.0030^{+0.00003}_{-0.00003}$ &   $0.0041^{+0.0002}_{-0.0002}$ \\
 \hline
  $\langle\widehat{I}^{a_{1}}_{3,\perp}\rangle$   &$0.0030^{+0.00009}_{-0.00009}$& $0.0031^{+0.00009}_{-0.00009}$&  $0.0031^{+0.0008}_{-0.0008}$\\
 \hline
 $\langle\widehat{I}^{a_{1}}_{4,\perp}\rangle$ &$-0.0126^{+0.0015}_{-0.0015}$&$-0.0165^{+0.0014}_{-0.0014}$& $-0.0141^{+0.0011}_{-0.0011}$\\
 \hline
$\langle\widehat{I}^{a_{1}}_{5,\perp}\rangle$&$-0.0498^{+0.0010}_{-0.0010}$& $-0.0151^{+0.0006}_{-0.0006}$&  $-0.0236^{+0.0008}_{-0.0008}$\\
 \hline
$\langle\widehat{I}^{a_{1}}_{6s,\perp}\rangle$& $0.0167^{+0.0002}_{-0.0002}$& $0.0035^{+0.0003}_{-0.0003}$&  $0.0065^{+0.0003}_{-0.0003}$\\
 \hline 
 $\langle\widehat{I}^{a_{1}}_{6c,\perp}\rangle$  & $0.0333^{+0.0004}_{-0.0004}$&  $0.0070^{+0.0005}_{-0.0005}$&  $0.0131^{+0.0006}_{-0.0006}$\\
 \hline
\end{tabular}
\end{center}
\end{table}

\begin{table}[ht!]
\begin{center}
\caption{Same as in Table \ref{table:28}, but for $q^2 = 2.0 - 3.0$ GeV$^{2}$ bin.}\label{table:30}
\begin{tabular}{ |p{5cm}||p{3cm}|p{3cm}|p{3cm}|  }
 \hline
 \multicolumn{4}{|c|}{\textbf{$q^{2}= 2.0 - 3.0$} GeV$^{2}$} \\
 \hline
 \textbf{Angular Observables}& \textbf{SM} &\textbf{$Z^{\prime}$ Scenario 1}&\textbf{$Z^{\prime}$ Scenario 2}\\
 \hline
 $\langle\widehat{I}^{a_{1}}_{1s,\perp}\rangle$   &$0.3473^{+0.0002}_{-0.0002}$& $0.3497^{+0.0002}_{-0.0002}$&  $0.3494^{+0.0002}_{-0.0002}$\\
 \hline
 $\langle\widehat{I}^{a_{1}}_{1c,\perp}\rangle$&$0.0271^{+0.0006}_{-0.0006}$ &$0.0176^{+0.0010}_{-0.0010}$&  $0.0185^{+0.0009}_{-0.0009}$\\
 \hline
$\langle\widehat{I}^{a_{1}}_{2s,\perp}\rangle$&$-0.3220^{+0.0003}_{-0.0003}$& $-0.3274^{+0.0008}_{-0.0008}$& $-0.3272^{+0.0006}_{-0.0006}$\\
 \hline
$\langle\widehat{I}^{a_{1}}_{2c,\perp}\rangle$&$0.0090^{+0.0002}_{-0.0002}$&$0.0058^{+0.0003}_{-0.0003}$& $0.0061^{+0.0003}_{-0.0003}$\\
 \hline
  $\langle\widehat{I}^{a_{1}}_{3,\perp}\rangle$  &$0.0055^{+0.0001}_{-0.0001}$& $0.0055^{+0.0001}_{-0.0001}$&  $0.0055^{+0.0001}_{-0.0001}$\\
 \hline
 $\langle\widehat{I}^{a_{1}}_{4,\perp}\rangle$ &$-0.0372^{+0.0017}_{-0.0017}$&$-0.0401^{+0.0015}_{-0.0015}$& $-0.0383^{+0.0016}_{-0.0016}$\\
 \hline
$\langle\widehat{I}^{a_{1}}_{5,\perp}\rangle$&$-0.0312^{+0.0013}_{-0.0013}$& $-0.0052^{+0.0008}_{-0.0008}$&  $-0.0109^{+0.0009}_{-0.0009}$\\
 \hline
$\langle\widehat{I}^{a_{1}}_{6s,\perp}\rangle$& $0.0107^{+0.0007}_{-0.0007}$&  $-0.0017^{+0.0006}_{-0.0006}$&  $0.0069^{+0.0006}_{-0.0006}$\\
 \hline 
 $\langle\widehat{I}^{a_{1}}_{6c,\perp}\rangle$   & $0.0215^{+0.0013}_{-0.0013}$& $-0.0034^{+0.0011}_{-0.0011}$&  $0.0014^{+0.0013}_{-0.0013}$\\
 \hline
\end{tabular}
\end{center}
\end{table}

\begin{table}[ht!]
\begin{center}
\caption{Same as in Table \ref{table:28}, but for $q^2 = 3.0 - 4.0$ GeV$^{2}$ bin.}\label{table:31}
\begin{tabular}{ |p{5cm}||p{3cm}|p{3cm}|p{3cm}|  }
 \hline
 \multicolumn{4}{|c|}{\textbf{$q^{2}= 3.0 - 4.0$} GeV$^{2}$} \\
 \hline
 \textbf{Angular Observables}& \textbf{SM} &\textbf{$Z^{\prime}$ Scenario 1}&\textbf{$Z^{\prime}$ Scenario 2}\\
 \hline
 $\langle\widehat{I}^{a_{1}}_{1s,\perp}\rangle$   &$0.3448^{+0.0003}_{-0.0003}$& $0.3463^{+0.0003}_{-0.0003}$&  $0.3462^{+0.0003}_{-0.0003}$\\
 \hline
 $\langle\widehat{I}^{a_{1}}_{1c,\perp}\rangle$ &$0.0386^{+0.0014}_{-0.0014}$ &$0.0328^{+0.0018}_{-0.0018}$& $0.0326^{+0.0017}_{-0.0017}$\\
 \hline
$\langle\widehat{I}^{a_{1}}_{2s,\perp}\rangle$&$-0.3141^{+0.0011}_{-0.0011}$&$-0.3172^{+0.0014}_{-0.0014}$& $-0.3178^{+0.0013}_{-0.0013}$\\
 \hline
$\langle\widehat{I}^{a_{1}}_{2c,\perp}\rangle$&$0.0128^{+0.0005}_{-0.0005}$& $0.0108^{+0.0006}_{-0.0006}$& $0.0108^{+0.0005}_{-0.0005}$\\
 \hline
  $\langle\widehat{I}^{a_{1}}_{3,\perp}\rangle$  &$0.0081^{+0.0002}_{-0.0002}$&$0.0081^{+0.0002}_{-0.0002}$& $0.0081^{+0.0002}_{-0.0002}$\\
 \hline
 $\langle\widehat{I}^{a_{1}}_{4,\perp}\rangle$ &$-0.0547^{+0.0017}_{-0.0017}$&$-0.0570^{+0.0015}_{-0.0015}$&  $-0.0556^{+0.0016}_{-0.0016}$\\
 \hline
$\langle\widehat{I}^{a_{1}}_{5,\perp}\rangle$&$-0.0190^{+0.0015}_{-0.0015}$&$-0.0013^{+0.0009}_{-0.0009}$& $-0.0027^{+0.0011}_{-0.0011}$\\
 \hline
$\langle\widehat{I}^{a_{1}}_{6s,\perp}\rangle$& $0.0037^{+0.0011}_{-0.0011}$& $-0.0073^{+0.0008}_{-0.0008}$&   $0.0051^{+0.0009}_{-0.0009}$\\
 \hline 
 $\langle\widehat{I}^{a_{1}}_{6c,\perp}\rangle$ & $0.0074^{+0.0022}_{-0.0022}$& $-0.0147^{+0.0016}_{-0.0016}$&   $0.0116^{+0.0019}_{-0.0019}$\\
 \hline
\end{tabular}
\end{center}
\end{table}

\begin{table}[ht!]
\begin{center}
\caption{Same as in Table \ref{table:28}, but for $q^2 = 4.0 - 5.0$ GeV$^{2}$ bin.}\label{table:32}
\begin{tabular}{ |p{5cm}||p{3cm}|p{3cm}|p{3cm}|  }
 \hline
 \multicolumn{4}{|c|}{\textbf{$q^{2}= 4.0 - 5.0$} $GeV^{2}$} \\
 \hline
 \textbf{Angular Observables}& \textbf{SM} &\textbf{$Z^{\prime}$ Scenario 1}&\textbf{$Z^{\prime}$ Scenario 2}\\
 \hline
 $\langle\widehat{I}^{a_{1}}_{1s,\perp}\rangle$ &$0.3418^{+0.0004}_{-0.0004}$& $0.3428^{+0.0004}_{-0.0004}$&  $0.3429^{+0.0005}_{-0.0005}$\\
 \hline
 $\langle\widehat{I}^{a_{1}}_{1c,\perp}\rangle$&$0.0548^{+0.0021}_{-0.0021}$& $0.0510^{+0.0024}_{-0.0024}$& $0.0502^{+0.0023}_{-0.0023}$\\
 \hline
$\langle\widehat{I}^{a_{1}}_{2s,\perp}\rangle$& $-0.3017^{+0.0016}_{-0.0016}$&  $-0.3034^{+0.0019}_{-0.0019}$&  $-0.3044^{+0.0019}_{-0.0019}$\\
 \hline
$\langle\widehat{I}^{a_{1}}_{2c,\perp}\rangle$&$0.0182^{+0.0007}_{-0.0007}$& $0.0168^{+0.0008}_{-0.0008}$&  $0.0166^{+0.0008}_{-0.0008}$\\
 \hline
  $\langle\widehat{I}^{a_{1}}_{3,\perp}\rangle$&$0.0109^{+0.0002}_{-0.0002}$&$0.0109^{+0.0002}_{-0.0002}$& $0.0109^{+0.0002}_{-0.0002}$\\
 \hline
 $\langle\widehat{I}^{a_{1}}_{4,\perp}\rangle$&$-0.0685^{+0.0016}_{-0.0016}$&$-0.0703^{+0.0014}_{-0.0014}$& $-0.0691^{+0.0014}_{-0.0014}$\\
 \hline
$\langle\widehat{I}^{a_{1}}_{5,\perp}\rangle$&$-0.0100^{+0.0015}_{-0.0015}$& $-0.0062^{+0.0009}_{-0.0009}$& $-0.0035^{+0.0011}_{-0.0011}$\\
 \hline
$\langle\widehat{I}^{a_{1}}_{6s,\perp}\rangle$& $-0.0039^{+0.0014}_{-0.0014}$&  $-0.0133^{+0.0010}_{-0.0010}$&  $-0.0127^{+0.0012}_{-0.0012}$\\
 \hline 
 $\langle\widehat{I}^{a_{1}}_{6c,\perp}\rangle$&   $-0.0078^{+0.0029}_{-0.0029}$&  $-0.0265^{+0.0020}_{-0.0020}$&   $-0.0253^{+0.0024}_{-0.0024}$\\
 \hline
\end{tabular}%
\end{center}
\end{table}

\clearpage
\bibliographystyle{refstyle}
\bibliography{references}
\end{document}